\documentclass[11pt,onecolumn,a4paper,accepted=2024-07-23]{quantumarticle}
\pdfoutput=1
\usepackage[font={small}]{caption}
\usepackage{subfigure}
\usepackage{graphicx}
\usepackage[numbers,sort&compress]{natbib}
\usepackage{float}
\usepackage[normalem]{ulem}
\usepackage{mathrsfs}

\usepackage{multicol}
\usepackage{cancel}
\usepackage{mathtools}
\usepackage{hyperref}

\usepackage{xcolor}
\usepackage{simplewick}

\usepackage{amsmath}
\usepackage{amsfonts}
\usepackage{amssymb}
\usepackage{graphicx}
\usepackage{simplewick}
\usepackage{hyperref}
\usepackage{multirow}
\usepackage{enumitem}
\usepackage{physics}
\usepackage{booktabs}
\usepackage{xcolor}
\usepackage{dsfont}
\usepackage{subfigure}

\def\({\left(} \def\){\right)}
\def\[{\left[} \def\]{\right]}

\newcommand{\eg}{{\it e.g.,}\ }
\newcommand{\ie}{{\it i.e.,}\ }

\def\Tr{{\rm Tr}}

\def\le{\left(}
\def\ri{\right)}

\makeatletter
\DeclareRobustCommand{\cev}[1]{%
  \mathpalette\do@cev{#1}%
}
\newcommand{\do@cev}[2]{%
  \fix@cev{#1}{+}%
  \reflectbox{$\m@th#1\vec{\reflectbox{$\fix@cev{#1}{-}\m@th#1#2\fix@cev{#1}{+}$}}$}%
  \fix@cev{#1}{-}%
}
\newcommand{\fix@cev}[2]{%
  \ifx#1\displaystyle
    \mkern#23mu
  \else
    \ifx#1\textstyle
      \mkern#23mu
    \else
      \ifx#1\scriptstyle
        \mkern#22mu
      \else
        \mkern#22mu
      \fi
    \fi
  \fi
}

\makeatother

\newcommand{\be}{\begin{equation}}
	\newcommand{\ee}{\end{equation}}
\newcommand{\bea}{\begin{equation} \begin{aligned}}
	\newcommand{\eea}{\end{aligned} \end{equation}}
\renewcommand{\th}{\theta}

\def\le{\left(}
\def\ri{\right)}

\def\tr           {\mathop{\rm tr}}

\def\Im           {{\rm Im\hskip0.1em}}

\renewcommand{\eqref}[1]{(\ref{#1})}

\usepackage[T1]{fontenc}

\begin{document}

\title[accepted=2024-07-23]{The Complexity of Being Entangled}

\author{Stefano Baiguera}
\email{baiguera@post.bgu.ac.il}
\orcid{0000-0003-3798-9038}
\author{Shira Chapman}
\orcid{0000-0002-9624-5488}
\email{schapman@bgu.ac.il}
\affiliation{Department of Physics, Ben-Gurion University of the Negev, \\ David Ben Gurion Boulevard 1, Beer Sheva 84105, Israel}
\author{Giuseppe Policastro}
\email{giuseppe.policastro@ens.fr}
\orcid{0000-0002-5026-7796}
\affiliation{Laboratoire de Physique  de l'\'Ecole Normale Sup\'eri{e}ure, \\
CNRS, PSL  Research University  and Sorbonne Universit\'es, \\
24 rue Lhomond, 75005 Paris, France}
\author{Tal Schwartzman}
\email{taljios@gmail.com}
\affiliation{Department of Physics, Ben-Gurion University of the Negev, \\ David Ben Gurion Boulevard 1, Beer Sheva 84105, Israel}

\begin{abstract}
Nielsen's approach to quantum state complexity relates the minimal number of quantum gates required to prepare a state to the length of  geodesics computed with a certain norm on the manifold of unitary transformations. For a bipartite system, we investigate binding complexity, which corresponds to norms in which gates acting on a single subsystem are free of cost. 
We reduce the problem to the study of geodesics on the manifold of Schmidt coefficients, equipped with an appropriate metric.  
Binding complexity is closely related to other quantities such as distributed computing and quantum communication complexity, and has a proposed holographic dual in the context of AdS/CFT.
For finite dimensional systems with a Riemannian norm, we find an exact relation between binding complexity and the minimal Rényi entropy. 
We also find analytic results for the most commonly used non-Riemannian norm (the so-called $F_1$ norm)
and  provide lower bounds for the associated notion of state complexity ubiquitous in quantum computation and holography.
We argue that our results are valid for a large class of penalty factors assigned to generators acting across the subsystems. We demonstrate that our results can be borrowed to study the usual complexity (not-binding) for a single spin for the case of the $F_1$ norm which was previously lacking from the  literature. Finally, we derive bounds for multi-partite binding complexities and the related (continuous) circuit complexity where the circuit contains at most $2$-local interactions. 
\end{abstract}

\maketitle

\setcounter{tocdepth}{1}

\newpage

\section{Introduction}
\label{sec:intro}

Computational circuit complexity has been established as a useful tool of quantum computing theory for some time, see \eg  \cite{watrous2008quantum,Aaronson:2016vto}.
Heuristically, complexity estimates the number of unitary gates (or time) required by an optimal circuit to produce a certain target state (typically entangled) starting from a reference state (typically factorized). 
As such, it is natural to ask whether there is a relation between the degree of entanglement and the complexity in producing a certain state \cite{Balasubramanian:2018hsu}.
In this work, we will show that under certain circumstances there is a relation between these two measures.

\paragraph{Circuit complexity.} Circuit complexity plays a crucial role in classifying computational problems and in establishing the advantages of quantum over classical computation. 
From a practical point of view, finding the optimal circuit for a given task is an important goal for the compiler of any quantum computer. 

In concrete circuits built in laboratories, quantum operations are usually performed using discrete elementary unitary operations (called \textit{gates})  which act on one or two qubits at the same time. 
While it is not possible to cover all the unitary group with a finite set of gates, any unitary operator can be built with arbitrary precision by selecting a universal set of gates \cite{nielsen_chuang_2010}.
The implementation of those gates in a quantum circuit can vary in terms of difficulty, time, or errors.
A proper definition of circuit complexity should take these variations into account. 
For generic circuits, finding the complexity is a hard problem.

Nielsen showed that lower and upper bounds can be obtained by looking at a continuous version of the problem \cite{Nielsen1,Nielsen3,Nielsen_2006}. 
The idea is to consider the unitary group, generate a continuous trajectory by a time-dependent Hamiltonian, and define a norm $F$ which assigns different costs (called \emph{penalty factors}) to the various directions along the tangent space of the group manifold.
The shortest circuit is given by a geodesic in the space of unitaries, and the complexity will be the length of this geodesic, measured using the notion of distance induced by the norm with penalty factors. 
A similar definition of complexity can be introduced in the space of states by performing a quotient over the stabilizer at each point along the trajectory. 
A closely related notion to Nielsen's approach is the framework of quantum optimal control, which studies the optimal manipulation of quantum dynamics under certain constraints \cite{carlini:2006,werschnik:2007,koch:2022gll}.

\paragraph{Complexity in holography.} 
Quantum complexity has gained much attention in holography after the observation that entanglement entropy in a conformal field theory (CFT) is insufficient to probe certain regions behind the horizons of dual black holes in Anti-de Sitter (AdS) space \cite{Hartman:2013qma,Susskind:2014moa,Freivogel:2015,Balasubramanian:2014sra}. 
While entanglement entropy is unable to capture the long-time linear growth of the interior volume of black holes, it was proposed that quantum computational complexity could be the quantity dual to the volume of the Einstein-Rosen bridge (this is the so-called complexity=volume conjecture) \cite{Susskind:2014rva,Stanford:2014jda}.\footnote{See, however, \cite{Belin:2018bpg,Rabinovici:2023yex} for alternative boundary interpretations of the bulk volume of the Einstein-Rosen bridge.} 
Later on, the volume conjecture was supplemented by other holographic proposals for complexity \cite{Brown:2015lvg,Brown:2015bva,Couch:2016exn,Belin:2021bga,Belin:2022xmt,Jorstad:2023kmq,Erdmenger:2021wzc,Erdmenger:2022lov,Chandra:2022pgl}, and their properties were compared by investigating their behavior in various backgrounds, see \eg \cite{Lehner:2016vdi,Barbon:2015ria,Chapman:2016hwi,Cai:2016xho,Carmi:2017jqz,Chapman:2018dem,Chapman:2018lsv,Chapman:2018bqj,Braccia:2019xxi,Sato:2019kik,Bernamonti:2019zyy,Bernamonti:2020bcf,Chapman:2021eyy,Auzzi:2021nrj,Baiguera:2021cba,Auzzi:2021ozb,Emparan:2021hyr,Jorstad:2022mls,Auzzi:2023qbm,Anegawa:2023wrk,Anegawa:2023dad,Baiguera:2023tpt,Aguilar-Gutierrez:2023zqm,Aguilar-Gutierrez:2023pnn,Akhavan:2019zax,Omidi:2020oit}. 
In parallel, efforts were made to put computational complexity on firmer ground from a quantum mechanical perspective and develop tools for calculating it in various setups.
This led to several investigations in quantum mechanics (QM), quantum field theory (QFT), and CFT \cite{Caputa:2017yrh,Jefferson:2017sdb,Chapman:2017rqy,Khan:2018rzm,Bhattacharyya:2018wym,Chapman:2018hou,Camargo:2018eof,Ge:2019mjt,Brown:2019whu,Balasubramanian:2019wgd,Chapman:2019clq,Caputa:2018kdj,Auzzi:2020idm,Caginalp:2020tzw,Flory:2020eot,Flory:2020dja,Chagnet:2021uvi,Basteiro:2021ene,Brown:2021rmz,Balasubramanian:2021mxo,Brown:2022phc,Erdmenger:2022lov,10.21468/SciPostPhys.16.2.041,Craps:2022ese}. For a comprehensive review, we refer the reader to \cite{Chapman:2021jbh}.

\paragraph{Binding Complexity.} In this work we will investigate \emph{Nielsen's binding complexity} (referred to, from now on, as binding complexity) \cite{Balasubramanian:2018hsu}, a specific version of Nielsen's geometric complexity.
Given a partition of a system into multiple parts, binding complexity assigns non-vanishing cost only to operations that are non-local, \ie that act on multiple subsystems.\footnote{Nonetheless, we will always require the gates to act at most on two subsystems at the same time.}
Contrarily, it is free to act with operations that are local (\ie act within a single subsystem).
While these requirements may seem unrealistic, we will find that the geodesics minimizing binding complexity do not present abrupt fluctuations along the free local directions during the time evolution. As a consequence, these geodesics would be a good approximation to the geodesics of an implementable circuit where the local operations have a small but non-zero cost.

There are several motivations to consider binding complexity. 
First of all, some of the main challenges for large-scale quantum computers (with many qubits) are to maintain coherence, \ie minimize the noise throughout the computation, and implement efficiently the necessary gates. 
Therefore, a prominent paradigm for future quantum computing is \emph{distributed computing} \cite{meter2008arithmetic,beals2013efficient,Caleffi:2022wxp}, where the task of computation is distributed between multiple small quantum computers or nodes, and in each such computer, the processing is coherent to a good degree. In such a model, it is necessary to minimize the amount of gates between the different computers, which are often assumed to be less reliable or slower. This minimization is precisely the notion of binding complexity. 
Distributed computing is also related to 
\textit{quantum communication complexity}, which measures the minimal information that distant parties have to exchange to perform a computational task \cite{buhrman:2010}. It was argued in \cite{Balasubramanian:2018hsu} that quantum communication complexity is upper bounded by the binding complexity.
Secondly, we will show that binding complexity for a bipartite system leads to a geometric distance on the space of states which is fully determined by the entanglement spectrum. This has an operational interpretation in terms of the minimal length of a circuit required to turn a non-separable state into a separable one (see \eg  \cite{rudnicki:2021}, which is related to the case of the Riemannian norm discussed in section \ref{ssec:F2norm}). 
Finally, binding complexity has been proposed to be dual to the interior volume of multi-boundary wormholes \cite{Balasubramanian:2018hsu}. 
Thus one may hope to apply the methods developed in this work to unveil new features of the wormhole interior in multi-boundary wormhole geometries \cite{Zolfi:2023bdp}. 

In \cite{Balasubramanian:2018hsu}, binding complexity was explored for Gaussian unitaries and states of a bosonic theory. Rough bounds for the binding complexity were also obtained in \cite{Zhang:2021xwx} using a discrete set of gates. 
Bounds on the minimal number of non-local gates required to optimize a quantum circuit were related in \cite{Eisert:2021}  to a sum (or maximum) of the entanglement entropy over all partitions. We will discuss the relation of \cite{Eisert:2021} to our results in section \ref{sec:geometrically_local_complexity}.

In the present work, we compute binding complexity for bipartite finite dimensional systems and unveil its precise relation to the entanglement spectrum.
One of the main novelties of our approach is that we will study the geometry of binding complexity using the so-called $F_1$ norm, which heuristically counts the number of gates required to perform a certain unitary operation, and find precise analytic results in this case. 
This quantity has received less attention in the literature because it is not a Riemannian norm, therefore certain analytic tools are not available in this case. 
In addition, we show that the binding complexity computed with the Riemannian $F_2$ norm is related to the minimal Rényi entropy of the reduced density matrix of the state.  
We argue that the $F_1$ binding complexity is, to some extent, independent of the penalty factors associated with the non-local generators acting between the two parties, up to a proportionality factor coming from the smallest of such penalty factors. 
We incidentally show that the methods developed to study the binding complexity of two qutrits can also be used to compute the $F_1$ (\emph{not} binding) complexity of a single qubit, which was missing from the previous Nielsen's complexity literature. Finally, we give lower bounds for multipartite binding complexity and demonstrate further inequalities for some more traditional notions of complexity using at most 2-local gates between different parts. The lower bounds are in terms of the bipartite binding complexity of various bipartitions of the state. A detailed summary of our results can be found at the beginning of section \ref{sec:discussion}.

\paragraph{Outline.}
The paper is organized as follows.
We begin in section \ref{sec:complexity_geometry} with a review of Nielsen's geometric approach to  quantum computational complexity, and the definition of binding complexity.
In section \ref{sec:SON_picture}, we show that the computation of binding state complexity reduces to a problem defined on the manifold of Schmidt coefficients.
In section \ref{sec:general_results}, using this general technique, we compute 
the norm and the binding complexity for bipartite quantum-mechanical systems with generic dimensions. 
We apply the general method to study the binding complexity for two qutrits in section \ref{ssec:example_2qutrits} and show that our results can be borrowed to study the usual notion of state complexity for a single qubit. 
In section \ref{sec:geometrically_local_complexity} we derive bounds on some measures of complexity (and binding complexity) in the case of multipartite systems.
We summarize the main results and discuss future directions in section \ref{sec:discussion}.
The appendices contain technical results. We derive a chain of inequalities for the cost functions in appendix \ref{ssec:bounds_norms}, which are required for the derivation of section \ref{sec:SON_picture}.  
In appendix \ref{app:ContSchmi}, we consider the limit of a large number of Schmidt coefficients.
In appendix \ref{app:Pauli}, we derive one of the bounds on the  $F_1$ complexity in the Pauli basis mentioned in section \ref{ssec:Pauli}.
Appendix \ref{app:ssec:Pauli_result} contains an alternative derivation of certain results presented in section \ref{ssec:F2norm} by using the Euler-Arnold formalism.

\section{Complexity geometry}
\label{sec:complexity_geometry}

We briefly review Nielsen's geometric approach to complexity  \cite{Nielsen1,Nielsen_2006,Nielsen3} and set up the relevant notation.
The idea of this approach is to replace circuits composed of discrete gates with continuous paths in a Lie group manifold.
Complexity is then defined as the length of the shortest trajectory in a geometry defined over this manifold, once a notion of distance is introduced.
While most of the discussions could apply to a generic Lie group, we will focus in the following on the special unitary group.

\subsection{Unitary and state complexities}

Given the special unitary group $\mathrm{SU}(N)$, we will be interested in finding the optimal way to construct a target unitary $U_T$ starting from the identity by means of a time-dependent Hamiltonian,
which can be expanded in terms of a basis of Hermitian  generators $\lbrace T_I \rbrace $ as follows
\be
H(t) = \sum_I Y^I (t) T_I \, ,
\label{eq:expansion_Ham}
\ee
where $Y^I$ are called \emph{velocities} and describe the tangent vector to a trajectory in the group manifold.
Different ways to construct the target unitary are parameterized as curves in the space of unitary transformations with a path parameter $t \in[0,1]$ such that $U(t=0) = \mathbf{1} $ and $U(t=1)=U_T$.
The unitary at a point on the curve is expressed as
\be
U(t) = \mathcal{\cev{P}} \exp \le -i \int_0^t dt' H(t') \ri \, ,
\label{eq:generic_path_unitaries}
\ee
where $\mathcal{\cev{P}}$ denotes the path ordering such that the circuit is built from right to left.
The time evolution of a generic curve is determined by the Schr\"{o}dinger equation $\dot{U}(t) = -i H(t) U(t)$.

We define \emph{unitary complexity} as the minimization of a \emph{cost function} $F[\vec{Y}]$ introduced on the tangent space: 
\be
\mathcal{\mathcal{C}}_F [U] = \min_{\lbrace Y^I : \, U(0)=\mathbf{1}, \, U(1)=U_T \rbrace} \int_0^1 dt \, F[Y^I (t)] \, .
\label{eq:unitary_complexity}
\ee
The path in eq.~\eqref{eq:unitary_complexity} is subject to the boundary conditions imposing that it connects the identity with the target unitary.\footnote{In general the trajectory in the space of unitaries will run along the time interval $t\in[0,t_f].$ If the cost function $F[\vec{Y}]$ is positive homogeneous of degree 1, we can always redefine the parameter $t$ in such a way that  $t_f=1$.\label{footnotehomog}} 
The cost function should satisfy certain properties so that it defines a notion of distance in the space of unitaries.
A large class of cost functions that we will study in this work is given by
\be
F_{p,\vec{q}} [\vec{Y}] = \le \sum_I q_I |Y^I|^p \ri^{\frac{1}{p}} \, ,
\label{eq:cost_function_Finsler}
\ee
where $q_I \geq 0$ are called \emph{penalty factors}. 
Roughly speaking, penalties parametrize the difficulty of moving along the various directions of the group manifold and are supposed to model systems in which certain gates are more difficult to realize than others.

When $p=2$, the cost function \eqref{eq:cost_function_Finsler} reduces to the distance induced by the Riemannian metric
\be
ds^2 =  M_{IJ} \rho^I \rho^J  \, , \qquad
\rho^I = \frac{i}{\mathcal{N}} \tr \le dU U^{\dagger} T^I \ri \, , \qquad
M_{IJ} = \mathrm{diag} \le q_1 , \dots , q_{N^2-1} \ri \, ,
\label{eq:Riem_metric}
\ee
where $\rho^I$ are right-invariant forms, $M_{IJ}$ is the matrix containing the penalties and we have assumed that the generators are normalized according to $\tr(T^I T^J) = \mathcal{N}\delta^{IJ}$ (that is $\rho^I=Y^I dt$).
When $M_{IJ}=\delta_{IJ},$ the expression \eqref{eq:Riem_metric} corresponds to the Cartan-Killing metric on the unitary group, which is bi-invariant and has positive curvature.
In the presence of non-trivial penalty factors, the metric is still right-invariant and therefore describes a homogeneous space, but now the curvature can also be negative \cite{Brown:2019whu}.
Indeed, negative curvature is required to reproduce geodesic deviation \cite{Brown:2017jil} and the switchback effect \cite{Stanford:2014jda,Brown:2016wib}, which are the standard features of complexity. 
When $p=1,$ the cost function \eqref{eq:cost_function_Finsler} has the physical interpretation of counting the number of infinitesimal gates being used. 
For these reasons, the norms with $p=1,2$ received special attention in the literature, see, \eg  \cite{Jefferson:2017sdb,Chapman:2017rqy,Brown:2019whu,Balasubramanian:2019wgd,Auzzi:2020idm,Chagnet:2021uvi,Basteiro:2021ene,Brown:2021rmz,Balasubramanian:2021mxo,Brown:2022phc,Erdmenger:2022lov,10.21468/SciPostPhys.16.2.041,Craps:2022ese}, and we shall also focus on them here. Another property of the case $p=1$ is that the cost function is not smooth, and so we are not able to use the Euler Lagrange equations or calculus of variations to find geodesic trajectories. This makes its analysis more complicated, see \eg \cite{Nielsen1} for a discussion on this point. In this paper, we develop efficient techniques to overcome this difficulty.

Let us now assume that the special unitary group $\mathrm{SU}(N)$ acts on an $N$-dimensional Hilbert space $\mathcal{H}$, where $\ket{\psi_R}$ is a given reference state and $\ket{\psi_T}$ is a target state which we wish to construct. 
The \emph{state complexity} is defined as the minimum of the unitary complexities computed for all the special unitary operators that connect the reference state to the target state:
\be
\mathcal{C}^{\rm state}_F [\ket{\psi_T},\ket{\psi_R}] = 
\min_{ \lbrace U \in G :~~ \ket{\psi_T} = U \ket{\psi_R} \rbrace} \mathcal{C}_F [U] \, .
\label{eq:def_compl_state}
\ee
Recall that the space of quantum states is defined as the space of rays in the Hilbert state, \ie vectors differing by a phase are identified. Elements of the special unitary group that leave a state  invariant define an equivalence class 
\be
V \ket{\psi} = e^{i \phi} \ket{\psi} \quad \Rightarrow \quad
U'=U V \sim U \, .
\label{eq:def_stabilizer}
\ee
We call $V$ the elements of the \emph{stabilizer} of the state $\ket{\psi}$, which 
corresponds to the maximal subgroup $\mathrm{SU}(N-1) \times \mathrm{U}(1)$ of $\mathrm{SU}(N)$.\footnote{The stabilizer is defined as the subgroup of a group $G$ that leaves invariant an element $x \in X$, where $X$ is a set on which the group acts.} 
The equivalence class in eq.~\eqref{eq:def_stabilizer} defines a map from the unitary space to the quotient  
\be
\pi : \mathrm{SU}(N) \rightarrow \mathbb{CP}^{N-1} \equiv  \frac{\mathrm{SU}(N)}{\mathrm{SU}(N-1) \times \mathrm{U}(1)} \, .
\label{eq:quotient_map}
\ee
The procedure \eqref{eq:def_compl_state}  induces a norm for the state complexity that depends only on the coordinates of the projective space.
While the concept of complexity on a state requires the notion of unitary complexity, the opposite is not true: we can define the complexity of a unitary without reference to a particular representation.

For a given trajectory in the Hilbert space, the minimization over the stabilizer group can be done locally, and thus \eqref{eq:cost_function_Finsler} induces a norm 
$F^{\rm state} [\ket{\psi(t)},|\dot{ \psi} (t)\rangle]$ on the tangent space of $\mathcal{H}$.  The Schr\"{o}dinger equation $|\dot{ \psi} (t)\rangle= -iH\ket{\psi(t)}$ constrains  $2(N-1)$ real degrees of freedom out of the $N^2-1$ degrees of freedom of $H$. Those are the degrees of freedom that control the evolution in the Hilbert space. Minimizing over the remaining $(N-1)^2$ degrees of freedom, which belong to the tangent space of the stabilizer group with respect to $\ket{\psi(t)}$, gives the induced norm
\be
F^{\rm state} [\ket{\psi(t)},|\dot{ \psi} (t)\rangle] \equiv
\min_{\mathrm{ stab} \, \ket{\psi (t)} } F[\vec{Y}] \, .
\label{eq:def_Fstate}
\ee
With this definition, we can re-express the distance between states as 
\begin{equation}
\label{eq:statemin}
    \mathcal{C}^{\rm state}_F [\ket{\psi_T},\ket{\psi_R}] =\min_{{\footnotesize 
    \begin{array}{c}
 |\psi(t)\rangle \\
|\psi(0)\rangle=\ket{\psi_R}  \\
|\psi(1)\rangle=\ket{\psi_T} 
\end{array}}
    }  \int_0^1 dt \, F^{\rm state} [\ket{\psi(t)},|\dot{ \psi} (t)\rangle] \, ,
\end{equation}
where now we can study optimal trajectories directly in terms of a geometry on the space of states.
When state and unitary complexity are defined using the cost function \eqref{eq:cost_function_Finsler} with $p=2$, the projection $\pi$ in eq.~\eqref{eq:quotient_map} is a  Riemannian submersion \cite{Auzzi:2020idm}.
The induced metric obtained on $\mathbb{CP}^{N-1}$ is only left-invariant for trivial penalty factors  \cite{Brown:2019whu}, in which case it is proportional to the Fubini-Study metric. 

\subsection{Binding complexity}
\label{ssec:binding_complexity}

Let us now assume that the physical system is split into $n$ subsystems, \ie the Hilbert space decomposes as
\be \label{multiH}
\mathcal{H} = \mathcal{H}_{A_1} \otimes \dots \otimes \mathcal{H}_{A_n} \, .
\ee
This setting is relevant to compare complexity with multiparty entanglement.
Heuristically, in a discrete model we define \emph{binding complexity} as the minimal number of gates, acting on more than one subsystem at the same time, that are needed to build the target unitary $U_T$ starting from the identity \cite{Balasubramanian:2018hsu}. 
At the same time, we will require gates to be \emph{bi-local}, meaning that they can act at most on two subsystems at the same time. 
This is a special case of the unitary complexity defined above with specific choices of penalties.
Let us focus on the case where the Hilbert space factorizes into two parts (we  return to the general case in section \ref{sec:geometrically_local_complexity})
\be
\mathcal{H} = \mathcal{H}_A\otimes \mathcal{H}_{B} \, , \qquad
N_A \equiv \mathrm{dim} (\mathcal{H}_A) \leq \mathrm{dim} (\mathcal{H}_B) \equiv N_B  \, .
\label{eq:splitting_Hilbert_spaces}
\ee
Each subspace admits a basis of Hermitian traceless generators $\lbrace T^A_a, T^B_i  \rbrace $ for the Lie subalgebras $\mathfrak{su}(N_A)$ and $\mathfrak{su}(N_B),$ respectively.\footnote{We will usually denote the generators of the special unitary group in subsystem $A$ with indices $a,b \in \lbrace 1, \dots, N_A^2 -1 \rbrace $ from the beginning of the alphabet, and the generators in $B$ with indices $i,j\in \lbrace 1, \dots, N_B^2 -1 \rbrace$ from the middle part of the alphabet. 
Capital indices $I,J \in \lbrace 1, \dots, (N_A N_B)^2 -1 \rbrace $ will collectively refer to the special unitary group $\mathrm{SU}(N_A N_B)$ of the full system.
We will also denote with a sub(super)script $A,B$ the subsystem where the local generators act.\label{foot:ind}}
We refer to those as the \emph{local} generators since they only act on one of the two sides of the system, and we will call local unitary
(LU) the transformations they span (in the sense of acting locally within each part). 
We choose a set that is orthogonal with respect to the trace in the fundamental representation
\begin{equation} \label{GenNormaliz}
	\Tr(T^A_a T^A_b)  = \mathcal{N}_A \, \delta_{ab} \, , \qquad
	\Tr(T^B_i T^B_j) = \mathcal{N}_B \, \delta_{ij} \, ,
\end{equation}
where $\mathcal{N}_{A(B)}$ are normalization constants. We do not assume that the generators are normalized, since we will later want to use the Pauli and Gell-Mann bases for $\mathrm{SU}(N_{A(B)})$ which oftentimes appear in the literature unnormalized.

The unitary operators that act on the entire system belong to the group $\mathrm{SU}(N_A N_B)$, which also entails operations that can entangle the two parts.
The latter are spanned by generators of the form $T^A_a \otimes T^B_i$, which we call \emph{non-local}.\footnote{In the nomenclature of \cite{Balasubramanian:2018hsu,Zhang:2021xwx} these are the \emph{relevant} or \emph{straddling} gates.}
The full Hamiltonian of the system can be decomposed as
\be
H(t) = Y^a_A (t) T^A_a \otimes \mathbf{1}^B +Y^i_B (t) \mathbf{1}^A \otimes T^B_i + Y^{ai} (t) T^A_a \otimes T^B_i \, .
\label{eq:decomposition_Hamiltonian}
\ee
The splitting of the velocities into the subsets $Y^I \equiv \lbrace Y^a_A, Y^i_B, Y^{ai} \rbrace$ implies a similar decomposition of the penalty factors into the corresponding subsets $q_I \equiv \lbrace q^A_a, q^B_i , q_{ai} \rbrace,$ respectively. Explicitly, the cost function \eqref{eq:cost_function_Finsler} becomes
\be
F_{p,\vec{q}} [\vec{Y}] = \le \sum_a q_a^A |Y^a_A|^p
+ \sum_i q_i^B |Y^i_B|^p 
+ \sum_{a,i} q_{ai} |Y^{ai}|^p
\ri^{\frac{1}{p}} \, .
\label{eq:cost_function_Finsler_subsystem}
\ee
We consider the case where it is free to act on each subsystem separately, and therefore only the non-local generators contribute to the unitary complexity. The corresponding penalty factors are\footnote{Choosing  vanishing penalty factors along certain directions in the group manifold has other interesting applications. For instance, in \cite{Craps:2023ivc} such a choice was used to relate Nielsen complexity to Krylov complexity.}
\be
q^A_a = q^B_i = \varepsilon \rightarrow 0 \, , \qquad
q_{ai} \geq 1 \, .
\label{eq:penalties_binding}
\ee
We have chosen $q_{ai} \geq 1$ rather than any other finite value without loss of generality.
This choice of penalty factors describes a framework where the implementation of gates that entangle two subsystems is much harder than performing operations on each of them separately, as we have discussed in the introduction.

Nielsen's binding complexity ($\mathcal{BC}$) is defined as the unitary complexity \eqref{eq:unitary_complexity} with penalty factors \eqref{eq:penalties_binding}, \ie\footnote{In reference \cite{Balasubramanian:2018hsu}, Nielsen's binding complexity corresponds to what we call the homogeneous norm defined by the cost function \eqref{eq:Fpenalties_flat}. In this work, we extend the definition of binding complexity to include any choice of penalties for the non-local generators.}
\be
\mathcal{B C}_{F} [U] = \lim_{q^A_a = q^B_i  \rightarrow 0} \mathcal{C}_F [U] \, .
\label{eq:def_binding_complexity}
\ee
We will focus on two norms of this kind:
\begin{itemize}
    \item The \emph{homogeneous} case assigns unit penalty factors to all the non-local generators
    \begin{equation}
F_{p,\text{hom}}[\vec{Y}] =  \le \sum_{a,i}|Y^{ai}|^p\ri^{\frac{1}{p}}
\, .
\label{eq:norm_hom_after_limit}
\end{equation}
\item The \emph{universal-set} case  is defined by the cost function
\be
F_{p,\text{set}}[\vec{Y}]  = |Y^{1 1}|  \, ,
\label{eq:norm_set_after_limit}
\ee
where the set $\vec Y$ does not include non-local gates other than $Y^{11}$. This corresponds to the scenario where all the non-local generators are forbidden, except for a single one (with velocity $Y^{1 1}$) which, together with the local generators, can be shown to form a universal set of gates (the subscript \textit{set} stands for universal set of gates).
In this way, it is still possible to reach all the unitary operators in the group $\mathrm{SU}(N_A N_B)$.
We immediately see that $F_{p, \text{set}}$ is independent of $p$; therefore, it is sufficient to study one special case to get information about all the other values of $p$.
\end{itemize}

\noindent
It is possible to recover these norms as formal limits of the cost function \eqref{eq:cost_function_Finsler_subsystem} as follows:
\begin{subequations}
\be
    \label{eq:Fpenalties_flat}
F_{p,\text{hom}}[\vec{Y}] = \lim_{\varepsilon \to 0} \le \sum_{a} \varepsilon |Y^a_A|^p+\sum_{i} \varepsilon |Y^i_B|^p +\sum_{a,i}|Y^{ai}|^p \ri^{\frac{1}{p}} \, ,
\ee
 \be
F_{p>1,\text{set}}[\vec{Y}]  = \lim_{\substack{\varepsilon \to 0 \\ q_{ai} \to \infty}} \le \sum_{a} \varepsilon |Y^a_A|^p + \sum_{i} \varepsilon |Y^i_B|^p
+\sum_{(a,i) \ne (1,1)} q_{ai} |Y^{ai}|^p + |Y^{1 1}|^p \ri^{\frac{1}{p}} \, .
\label{eq:penalty_uni}
\ee
\end{subequations}
In the latter case, we relied on the results in \cite{Wang_2015}, where it was shown that the formal $q_{ai} \rightarrow \infty$ limit forces the geodesics in the group manifold to live in the subspace of the tangent bundle where the directions associated to the penalties $q_{ai}$ are excluded.\footnote{In reference \cite{Wang_2015}, the authors performed the limit $q_{ai} \rightarrow \infty$, while at the same time keeping $q_{ai} Y^{ai}$ finite (equivalently, they kept the Hamiltonian bounded in operator norm) in order to recover the quantum brachistochrone equation by starting from the Euler-Arnold equations \cite{AIF_1966__16_1_319_0}.
The application of this prescription gives $q_{ai} |Y^{ai}|^p \rightarrow 0$ in eq.~\eqref{eq:penalty_uni} for all $p>1$. 
In this work, we will not make explicit use of this limit, instead we will directly assume that certain directions in the tangent space of the group manifold are forbidden.
 \label{footnote:limit} }

By applying the general definition of state complexity \eqref{eq:def_compl_state} to the case with penalty factors \eqref{eq:penalties_binding}, we get the binding complexity in the space of states
\be
\mathcal{BC}^{\rm state}_F [\ket{\psi_T},\ket{\psi_R}] = 
\min_{ \lbrace U \in \mathrm{SU}(N) : ~~\ket{\psi_T} = U \ket{\psi_R} \rbrace} \mathcal{BC}_F [U] \, .
\label{eq:def_binding_compl_state}
\ee
Although the space of states has fewer coordinates than the unitary group manifold, it is not homogeneous and the minimization procedure quickly becomes difficult to perform with the increase of the dimensionality \cite{Caginalp:2020tzw}.
However, a special feature of the current setting is the decomposition of the Hilbert space into a product according to eq.~\eqref{eq:splitting_Hilbert_spaces}.
This fact, combined with the assignment of vanishing cost to local operators, will reduce the computation of binding state complexity to the evaluation of geodesics on the group manifold $\mathrm{SO}(N_A)$, as we will show in section \ref{sec:SON_picture}.

\section{The orthogonal group picture}
\label{sec:SON_picture}

Our main task is to compute the binding state complexity \eqref{eq:def_binding_compl_state} for a bipartite system \eqref{eq:splitting_Hilbert_spaces} and eventually relate this quantity to the entanglement spectrum.
We will show that the relevant group manifold for this analysis is the orthogonal group $\mathrm{SO}(N_A)$,  defined with respect to the maximal number of Schmidt coefficients $N_A$ characterizing the size of the smaller of the two subsystems.
This section is the main core of the work since it contains the general strategy that we will apply in order to study all the specific cases presented later. 

As we are interested in the state complexity, the norms expressed before should be minimized over the stabilizer group of the state at any given point in the trajectory in the manifold of unitaries and eventually expressed in terms of the degrees of freedom of the state.
Given the decomposition of the Hilbert space \eqref{eq:splitting_Hilbert_spaces}, a natural and convenient parametrization for any state along a trajectory in the quotient space is given by the Schmidt decomposition
\be
\ket{\psi(t)} = \sum_{m=1}^{N_A} \lambda_m(t) \ket{\phi_m(t) \chi_m(t)} \, ,
\label{eq:Schmidt_decomposition}
\ee
where $\lambda_m(t)$ (called \emph{Schmidt coefficients}) are real and positive numbers, and $\ket{\phi_m(t)}, \ket{\chi_m(t)}$ form orthonormal linearly-independent sets for the Hilbert spaces $\mathcal{H}_A$ and $\mathcal{H}_B$, respectively. With some abuse of notation, we will later refer to $(\phi_m,\chi_m)$ as \emph{basis parameters.}\footnote{Since $N_A\leq N_B$, generally the set $\ket{\chi_m}$ does not form a complete orthonormal basis for $\mathcal{H}_B$, but rather an incomplete one.}
Note that for the state to be normalized, the Schmidt coefficients should lie on a sphere, \ie $\sum_m \lambda_m^2=1$, and thus only $N_A-1$ of them are independent. 
Any entanglement monotone on a bipartite system, when evaluated on pure states, can be written as a function of the Schmidt coefficients \cite{Vidal:1998re}.

In the context of binding complexity, local operations (which act only on one subsystem) play a special role.
It is then convenient to introduce the following terminology: 
two states $\ket{\psi},\ket{\psi'}$ belonging to the Hilbert space \eqref{eq:splitting_Hilbert_spaces} are called local unitary equivalent (LUE), denoted as $\ket{\psi} \underset{\rm LUE}{\sim} \ket{\psi'}$, if there exists a local unitary transformation mapping one into the other:
\be
\ket{\psi} \underset{\rm LUE}{\sim} \ket{\psi'}  \quad \Leftrightarrow \quad 
\ket{\psi'} = U_A \otimes U_B \ket{\psi} \, .
\label{eq:def_LUE}
\ee
This relation defines an equivalence class whose representative is denoted with $[\ket{\psi}]$.

Starting from the unitary group manifold $\mathrm{SU}(N_A N_B)$, we now argue that the evaluation of binding state complexity reduces to a problem defined in the special orthogonal space $\mathrm{SO}(N_A)$.
We refer to this setting as the $\mathrm{SO}(N_A)$ picture.
We will show in sections \ref{ssec:reduction_set} and \ref{ssec:relation_orthogonal} that the problem eventually depends only on the Schmidt coefficients.
Then we summarize the main steps of the procedure in section \ref{ssec:summary_procedure} and concretely apply them to a system composed of two qubits in section \ref{ssec:example_2qubits}.

\subsection{Reduction to a minimal set of parameters}
\label{ssec:reduction_set}

\begin{figure}[ht]
    \centering
   \subfigure[]{\label{fig:LUE} \includegraphics[scale=0.3]{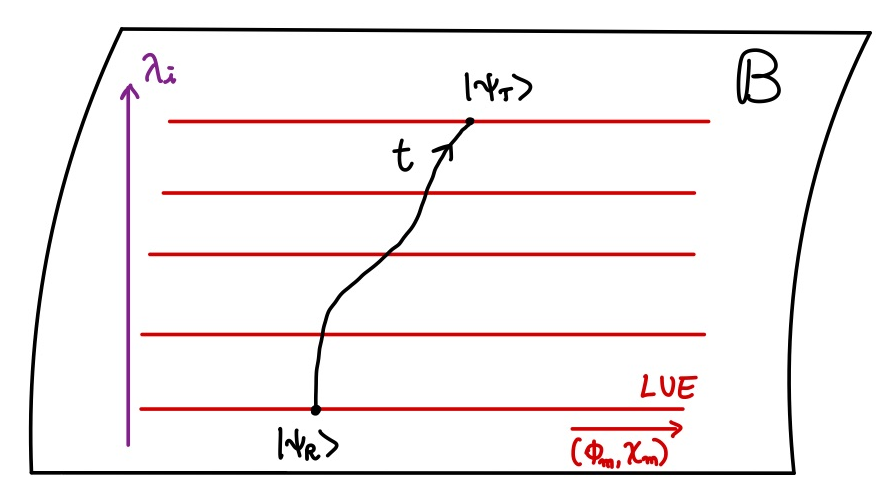} } \,\,
     \subfigure[]{\label{fig:LUE_distance} \includegraphics[scale=0.3]{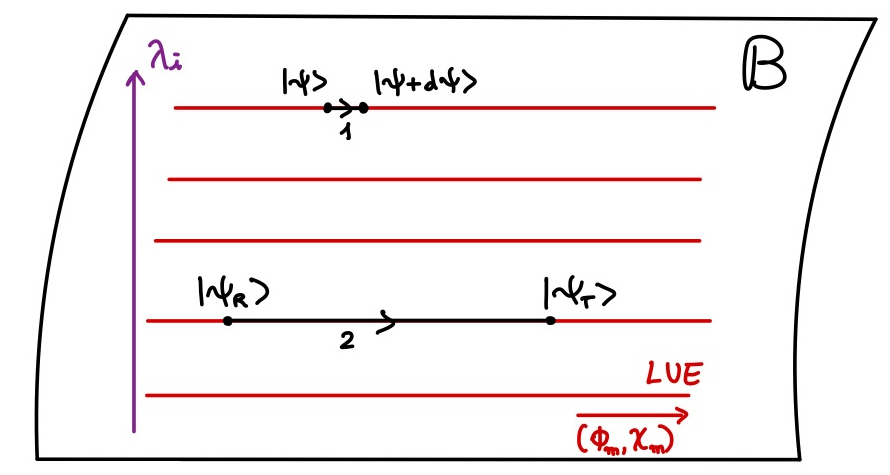} } 
    \caption{(a) Continuous path in the space of states $\mathcal{B}$ given by eq.~\eqref{eq:Schmidt_decomposition}.
    The vertical line represents different values of the Schmidt coefficients $\lambda_m$ which parametrize the equivalence classes of LUE states, while the horizontal axis corresponds to variations of the parameters $(\phi_m, \chi_m)$ of the basis.
    (b) The red lines represent the set of LUE states defined in eq.~\eqref{eq:def_LUE}. Those are regions where the distance between two states vanishes, either when the trajectory is infinitesimal (path 1) or when there is a finite variation of the parameters of the  basis (path 2).}
\end{figure}

We consider reference and target states defined by the Schmidt decompositions
\be
\ket{\psi_R} = \sum_{m=1}^{N_A} \lambda_m \ket{\phi_m \chi_m} \, , \qquad
\ket{\psi_T} = \sum_{m=1}^{N_A} \bar{\lambda}_m \ket{\bar{\phi}_m \bar{\chi}_m} \, ,
\label{eq:ref_tar_states}
\ee
and we study the binding state complexity $\mathcal{BC}^{\rm state}_F[\ket{\psi_T}, \ket{\psi_R}]$ 
induced by any $F$-norm of the form \eqref{eq:cost_function_Finsler_subsystem} with penalty factors \eqref{eq:penalties_binding}. 
A trivial consequence of the zero cost assigned to local generators is that the complexity of moving between LUE states vanishes
\be
\ket{\psi_R} \underset{\rm LUE}{\sim}  \ket{\psi_T} \quad \Rightarrow \quad
\mathcal{BC}^{\rm state}_F[\ket{\psi_T}, \ket{\psi_R}] = 0 \, ,
\label{eq:distance_LUE_finite}
\ee
because we can connect the states using a path generated by local operators only, which have zero cost.
We depict a generic trajectory in the space of states given by eq.~\eqref{eq:Schmidt_decomposition} in fig.~\ref{fig:LUE}. 
In the figure, we represent classes of LUE states as horizontal red lines. 
Those are null directions, along which the distance vanishes.
It is important to remark that this statement is not only valid infinitesimally, but even when the parameters $(\phi_m,\chi_m)$ of the basis undergo a finite change, see fig.~\ref{fig:LUE_distance}.

\begin{figure}[ht]
    \centering
    \includegraphics[scale=0.35]{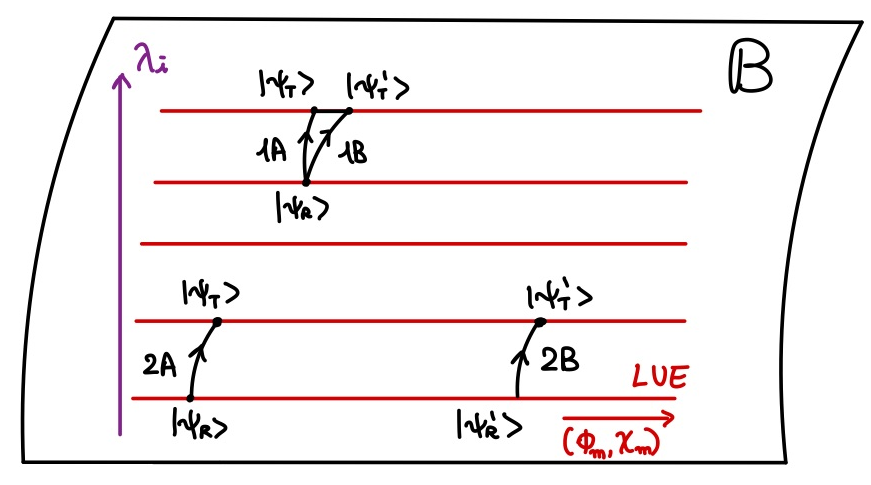} 
    \caption{
    Label 1 refers to infinitesimal paths, while label 2 to trajectories with finite variations of the parameters $(\phi_m,\chi_m)$ of the basis.
 Consider reference and target states with infinitesimally close Schmidt coefficients $\lambda_m$ and $ \lambda_m + d\lambda_m$. 
 The length of the infinitesimally close trajectories $1A,1B$ is the same.
 Instead, the length of the paths $2A,2B$ is in general different.
 Therefore the metric is not invariant under translations along $(\phi_m,\chi_m)$. }
    \label{fig:LUE_statements}
\end{figure}

Consider now the induced norm $F^{\rm state}$ on the Hilbert space, introduced in eq.~\eqref{eq:def_Fstate}.
Along a generic path in the space of states defined by the Schmidt decomposition \eqref{eq:Schmidt_decomposition}, we schematically denote the dependence of this norm on the parameters of the basis as\footnote{In writing the expression in this form (where it only depends on the parameters of the states) we have already performed the minimization over the stabilizer in eq.~\eqref{eq:def_Fstate}.}
\be
 F^{\rm state} [\ket{\psi(t)},|\dot{ \psi} (t)\rangle]  = 
 g (\lambda_m, \phi_m, \chi_m, \dot{\lambda}_m, \dot{\phi}_m, \dot{\chi}_m) \, .
 \label{eq:norm_g}
\ee
Let us focus on any $F$--norm with penalty factors \eqref{eq:penalties_binding}.
Since we assign vanishing penalty factors to local generators, which control the change in $(\phi_m$, $\chi_m)$, the norm \eqref{eq:norm_g} cannot depend on the direction of the tangent vector in that subspace:
\be
F^{\rm state} [\ket{\psi(t)},|\dot{ \psi} (t)\rangle] 
= g (\lambda_m, \phi_m, \chi_m, \dot{\lambda}_m) \, .
\label{eq:distance_g}
\ee
However, notice that the line element depends on the parameters $(\phi_m,\chi_m)$, implying that there is no translation invariance along these directions.
Pictorially, this corresponds to the fact that the paths $2A$ and $2B$ in fig.~\ref{fig:LUE_statements} have different lengths, while the infinitesimally close paths $1A, 1B$ have the same length.

Binding state complexity, as defined in eq.~\eqref{eq:def_binding_compl_state}, requires minimizing the distance among all the possible trajectories connecting the reference and target states. 
In particular, the identity \eqref{eq:distance_g} implies that the parameters $(\phi_m,\chi_m)$ in the Schmidt decomposition \eqref{eq:Schmidt_decomposition} become non-dynamical degrees of freedom in the geodesic minimization.
For this reason, the norm we obtain over the space of states, as per \eqref{eq:statemin}, can be minimized locally in terms of the parameters of the basis $\phi_m(t)$ and $\chi_m(t)$ to give a norm just in terms of the Schmidt coefficients $\lambda_m(t)$ and their differentials.\footnote{This is similar to partially solving the equations of motion for certain non-dynamical fields in QFT, whose momenta do not appear in the action, to obtain an on-shell action for the remaining fields.}

We shall denote this norm as $\mathcal{BF}$. More formally, for any $F$-norm with vanishing cost associated with local generators, we define 
\be 
\label{BFdef}
\mathcal{BF}[\vec{\lambda},\dot{\vec{\lambda}}]  =  \min_{\ket{\phi_n,\chi_n} } F^{\rm state} [\ket{\psi(t)} , |\dot{\psi}(t) \rangle]=
\min_{\ket{\phi_n \chi_n} \atop \rm{stab}~ |\psi(t)\rangle}
F[\vec{Y}] \, ,
\ee
where $\min_{\ket{\phi_n \chi_n} }$ is a minimization over the non-dynamical degrees of freedom in the possible bases in the Schmidt decomposition of the state $|\psi(t)\rangle$. The combined minimization on the right-hand side amounts, essentially, to minimizing over all control parameters $\vec Y$ and all basis parameters 
with the constraint of a prescribed change $\dot {\vec\lambda}(t)$ of the Schmidt coefficients. The last term in the definition \eqref{BFdef} describes the procedure we are going to follow in the next sections.

Using the definition \eqref{BFdef}, the complexity 
\eqref{eq:def_binding_compl_state} (see also eq.~\eqref{eq:statemin}) can be expressed as 
\be
\begin{aligned}
\mathcal{BC}^{\rm state}_F  [\ket{\psi_T}, \ket{\psi_R} ] 
 = \min_{{\footnotesize 
    \begin{array}{c}
 \vec{\lambda}(t) \\
\vec{\lambda}(0)=\vec{\lambda}_R  \\
\vec{\lambda}(1)=\vec{\lambda}_T 
\end{array}}} \int_0^1 dt \, \mathcal{BF} [\vec{\lambda},\dot{\vec{\lambda}}]
 \, ,
 \end{aligned}
 \label{eq:explicit_binding_state_compl}
\ee
where $\vec{\lambda}_R$ and $\vec{\lambda}_T$ are the vectors of Schmidt coefficients of the reference and target states. 
We see that binding state complexity can be reduced to finding geodesics 
in the space of Schmidt coefficients. The boundary conditions for the reference and target states are given in terms of the equivalence classes introduced in eq.~\eqref{eq:def_LUE}.
In order to describe the trajectory in the full Hilbert space that takes us from the exact reference state to the exact target state, we will perform a finite LU transformation (of zero cost) at the beginning and at the end of the trajectory selected by the parameters of the basis that minimize the norm above. We discuss this further below.

\subsection{Relation to the orthogonal group}
\label{ssec:relation_orthogonal}

Given the definition of $\mathcal{BF}$ in eq.~\eqref{BFdef}, we study the dependence of the evolution of the Schmidt coefficients on the parameters of the basis and on the Hamiltonian. 
First, we find from the conservation of the norm under unitary time evolution that
\be
\begin{aligned}
& \bra{\phi_m(t) \chi_m(t)}  \partial_t \le \ket{\phi_n(t) \chi_n(t)} \ri + \partial_t ( \bra{\phi_n(t) \chi_n(t)}) \ket{\phi_m(t) \chi_m(t)} = \\
 & = \delta_{mn} \le \bra{\phi_m(t)}\ket{\partial_t \phi_n(t)}+ \bra{\partial_t\phi_n(t)}\ket{ \phi_m(t)}+ \bra{\chi_m(t)}\ket{\partial_t \chi_n(t)}+\bra{\partial_t\chi_n(t)} \ket{ \chi_m(t)} \ri = \\
  & = \partial_t \le \bra{\phi_n(t)}\ket{\phi_n(t)}+\bra{\chi_n(t)}\ket{\chi_n(t)} \ri =0 \, .
\end{aligned}
\ee
In terms of the state defined in eq.~\eqref{eq:Schmidt_decomposition}, the previous identity implies 
\be
2 \dot{\lambda}_n(t) =
 \bra{\phi_n(t)\chi_n(t)}\ket{\partial_t\psi(t)} +\bra{\partial_t\psi(t)}\ket{\phi_n(t)\chi_n(t)} \, .
\ee
Using the Schr\"{o}dinger equation and the general decomposition \eqref{eq:decomposition_Hamiltonian} of the Hamiltonian, we find 
\be
\begin{aligned}
 	\dot{\lambda}_n(t) & =  \, \text{Im}(\bra{\phi_n(t)\chi_n(t)}H(t)\ket{\psi(t)})=  \sum_m \lambda_m (t) \, \text{Im}(\bra{\phi_n(t)\chi_n(t)}H(t)\ket{\phi_m(t)\chi_m(t)}) = \\ 
 	& = \sum_m \lambda_m (t) \, \text{Im}(\bra{\phi_n(t)\chi_n(t)}Y^{a i}(t) T^A_a \otimes T^B_i\ket{\phi_m(t)\chi_m(t)}) \, .  
 \end{aligned}
 \label{eq:constraint_der_Schmidt}
 \ee
Note that local generators do not contribute to the change of the Schmidt coefficients, \ie it is necessary to use operations involving both sides of the system (that create non-trivial entanglement) to modify them.\footnote{This statement does not rely on the choice of penalty factors in the cost function, but it is simply a consequence of the time evolution of the Schmidt decomposition. }
In summary, $\mathcal{BF}$ is computed by minimizing $F[\vec{Y}]$ over all the parameters of the basis and over all the degrees of freedom of $H$ 
that respect eq.~\eqref{eq:constraint_der_Schmidt}.

We now proceed to show that the problem reduces from the group manifold  SU($N_A N_B$) to SO($N_A$).
The previous manipulations naturally related the Schmidt coefficients to a matrix $R$ such that
\begin{subequations}
\be
\dot{\lambda}_m = \sum_{m} R_{mn} \lambda_n \, , 
\label{eq:evol_Schmidt}
\ee
\be
 R_{mn} (t) \equiv  \, \text{Im}(\bra{\phi_m(t)\chi_m(t)}Y^{a i}(t) T^A_a \otimes T^B_i\ket{\phi_n(t)\chi_n(t)})  \, .
\label{eq:RcnstrainH}
\ee
\end{subequations}
Since $R_{mn}(t)$ is real and antisymmetric, it is a generator of rotations of the special orthogonal group $\mathrm{SO}(N_A)$ on the sphere of Schmidt coefficients.
In other words, the constraint \eqref{eq:RcnstrainH} identifies a set of $\mathrm{dim} (\mathrm{SO}(N_A))= \frac{1}{2}N_A (N_A-1)$ degrees of freedom, which are relevant for the evolution of the Schmidt coefficients in eq.~\eqref{eq:evol_Schmidt}.
Note however that this is less than the number of independent degrees of freedom in $Y_{ab}(t)$, which is $(N_A^2-1)(N_B^2-1)$, as long as $N_A\geq 2$. This means that multiple different values of the velocity vectors are associated with the same matrix $R_{mn}$.

As we mentioned earlier, around eq.~\eqref{BFdef}, our final goal is to obtain a norm in terms of the Schmidt parameters.
We can simplify the minimizations in eq.~\eqref{BFdef} by taking an intermediate step in which we minimize the norm $F$ over the velocities $\vec Y(t)$ and basis parameters $\phi_m(t),\chi_m(t)$ subject to the constraint that these parameters yield a fixed rotation generator $R_{mn}(t)$ at any point along the trajectory according to  eq.~\eqref{eq:RcnstrainH}. In this way, we obtain a norm $\mathcal{BF}(R)$ in terms of the parameters of the matrix $R$. By having to deal with $\mathcal{BF}(R)$, instead of a norm on the Hamiltonian of SU($N_A N_B$), we say that we have reduced the problem to SO($N_A$).\footnote{By referring to the reduction of the problem to $\mathrm{SO}(N_A)$, we mean that the matrix $R_{mn}$ defined in eq.~\eqref{eq:constraint_der_Schmidt} and all the relevant degrees of freedom belong to are anti-symmetric and therefore belong to the algebra of $\mathrm{SO}(N_A)$. However, it is worth emphasizing that $R_{mn}$ is defined through the velocities and the generators of the full $\mathrm{SU}(N_A N_B)$ group. The precise problem that we are solving will be summarized in section~\ref{ssec:summary_procedure} and the reduction to SO($N_A$) that we refer to is completed after step 3.}

We then perform an additional minimization over the set of matrices $R$ which generate a given change $\dot {\vec\lambda} $ in the Schmidt coefficients $\vec \lambda$ according to eq.~\eqref{eq:evol_Schmidt}. This step amounts to minimizing over the stabilizer group of $\vec \lambda$ within $\mathrm{SO}(N_A)$. 

To perform the minimization of the basis parameters as in \eqref{BFdef},
we can re-express $R_{mn}$ in terms of a specific Hilbert space basis $\ket{m},\ket{\tilde{m}}$ of our choice using $\ket{\phi_m\chi_m} = U_A\otimes U_B\ket{m\tilde{m}}$, where $U_A, U_B$ are unitary operators acting in the respective subsystems.
This implies 
\be\label{eq:Rmnfixedbase}
R_{mn}=  \Im{(\bra{m\tilde{m}}Y^{a i}(t) (\text{Ad}_{U^A})^{\,\, b}_{a} (\text{Ad}_{U^B})^{\,\, j}_{i} T^A_b \otimes T^B_j \ket{n\tilde{n}})} \, , \quad
(\text{Ad}_{U})^{\,\,I}_{ J} \, T_I = U^{\dagger} T_J  U \, .
\ee
The adjoint operators implement the LU transformations and can be reabsorbed into the velocities. In this way, the generator of rotations \eqref{eq:RcnstrainH} becomes
\be\label{eq:Rcnstrain}
\tilde{Y}^{bj} = Y^{ai} (\text{Ad}_{U^A})^{\,\,b}_{ a}(\text{Ad}_{U^B})^{\,\, j}_{i}  \, , \qquad
 R_{mn} = \text{Im}(\bra{m\tilde{m}}\tilde{Y}^{a i}(t) T^A_a \otimes T^B_i\ket{n\tilde{n}}).
 \ee
In terms of the parameters in eq.~\eqref{eq:Rcnstrain}, the cost function \eqref{eq:norm_hom_after_limit} becomes
\be
F_{p,\text{hom}}[\vec{Y}] =  \le \sum_{b,j} \left|\tilde{Y}^{ai} (\text{Ad}_{U^A})^{\,\,b}_{a}(\text{Ad}_{U^B})^{\,\,j}_{i} \,  \right|^p \ri^{\frac{1}{p}} \, .
\label{eq:Fp_flat_Ytilde}
\ee
Minimizing the above equation over all adjoint operators and all the $\tilde{Y}^{ai}$ that respect the constraint \eqref{eq:Rcnstrain} will give a norm in terms of the parameters of $R$, that is:
\begin{equation}\label{R-norm}
   \mathcal{BF}_{p, \rm hom}(R_{mn}) = \min_{\tilde Y,\mathrm{Ad}_U} \sum_{b,j} \left(\left|\tilde Y^{ai} (\text{Ad}_{U^A})^{\,\,b}_{a}(\text{Ad}_{U^B})^{\,\,j}_{i}  \right|^p\right)^{\frac{1}{p}}.
\end{equation}
The final result $\mathcal{BF}_{p, \rm hom}(\vec \lambda, \dot{\vec\lambda})$ is then obtained by minimizing \eqref{R-norm}  over all matrices $R$ satisfying \eqref{eq:evol_Schmidt}.

We conclude this section by stating an important matrix inequality which will be useful for the evaluation of binding state complexity for the $F_{1, \rm hom}$ norm:\footnote{The index $a$ goes from $1$ to $N_A^2-1$, even if $\tilde Y$ is rectangular, see appendix \ref{ssec:bounds_norms}. Since the nuclear norm is invariant under rotations, $\tilde Y$ can be replaced in the rightmost-hand side by any $O_A \tilde Y O_B$, where $O$ are orthogonal matrices, yielding other equally valid inequalities.}
\be
  \min_{\mathrm{Ad}_U} \sum_{b,j} \left|\tilde Y^{ai} (\text{Ad}_{U^A})^{\,\,b}_{a}(\text{Ad}_{U^B})^{\,\,j}_{i}  \right| \geq  
 \sum_a \sigma_a (\tilde Y)  \geq  \sum_a |\tilde Y^{aa}|  \, ,
 \label{eq:chain_relations_F1normT}
\ee
where $\sum_a\sigma_a(\tilde Y)$ is the nuclear norm, which sums the singular values of the matrix $\tilde Y$.
The technical proof of this result is reported in Appendix \ref{ssec:bounds_norms}. 
When further minimized over all $\tilde Y$ with the appropriate boundary conditions \eqref{eq:Rcnstrain},  the first term on the left-hand side becomes the  $\mathcal{BF}_{1,\rm{hom}}(R_{mn})$ norm in terms of the matrix $R$
and the  inequality \eqref{eq:chain_relations_F1normT} provides the following lower bound 
\begin{equation}\label{eq:ineq2}
\mathcal{BF}_{1, \rm hom}(R_{mn})\geq \min_{\tilde Y:~ R_{mn}\text{ is fixed}}\sum_a  |\tilde Y^{aa}|.
\end{equation}
We will find in section \ref{ssec:F1norm} that 
this inequality is actually saturated, at least when considering the Gell-Mann basis for the generators of the unitary group, see eq.~\eqref{eq:gell}.  
In this way, we will be able to derive exact results for the binding state complexity with the $F_{1,\rm{hom}}$ norm.

\subsection{Summary of the procedure}
\label{ssec:summary_procedure}

Based on the previous analysis, we outline the steps that will be followed in the remainder of the paper to compute binding state complexity for a  bipartite system with Hilbert space \eqref{eq:splitting_Hilbert_spaces}:
\begin{enumerate}
    \item Specify a basis for the generators of $\mathrm{SU}(N_A), \mathrm{SU}(N_B)$ and define the reference and target states in the Hilbert space.
    \item Impose that the rotation matrix $R_{mn}$ is given by eq.~\eqref{eq:Rcnstrain}, and solve for as many velocities $\tilde{Y}^{ai}$ as the degrees of freedom encoded by this orthogonal matrix.\footnote{There always exists a solution to this equation if we allow a large enough set of non-local generators in our control Hamiltonian. However, the procedure may fail when the set of generators is not large enough. This will be the case for the universal-set of gates, which we discuss in detail in section \ref{app:1=2}. There,  we will use a different strategy than the one described here.}
    \item Plug the solution inside the desired cost function of the form \eqref{eq:cost_function_Finsler_subsystem} with penalties \eqref{eq:penalties_binding} and minimize over the remaining velocities $\tilde{Y}^{ai}$ and the adjoint of the local operators $U^A,U^B$ to obtain $\mathcal{BF}(R_{mn})$. 
    \item Solve the evolution equation for the Schmidt coefficients \eqref{eq:evol_Schmidt} and minimize over the remaining degrees of freedom inside the stabilizer of $\mathrm{SO}(N_A)$ to obtain $\mathcal{BF}(\vec\lambda, \dot{\vec \lambda})$.
    \item Minimize $\int_0^1\mathcal{BF}(\vec\lambda, \dot{\vec \lambda}) \, dt$ with boundary conditions $\vec{\lambda}(0)=\vec{\lambda}_R$, $\vec{\lambda}(1)=\vec{\lambda}_T$ to find the complexity.
\end{enumerate}

\subsection{Example: two qubits}
\label{ssec:example_2qubits}

Before diving into the general analysis, we exemplify our method in a simple setting: a system composed of two qubits, split into two subsystems of one qubit each. 
In the following, we specify the step numbers according to the list outlined in section \ref{ssec:summary_procedure} above. 
We will treat in turn the cases of the $F_{1,\text{hom}}$ and $F_{2,\text{hom}}$ norms. Steps 1 and 2 are actually independent of the norm used, except for the fact that the norm $F_{2,\text{hom}}$ does not depend on the choice of basis of generators of the unitary group.

\paragraph{Step 1.}
The basis of generators for the $\mathrm{SU}(2)$ group acting on each qubit separately is given by the Pauli matrices, therefore the Hamiltonian of the full system reads 
 \be
H(t) = Y^a_A (t) \, \sigma_a^A \otimes \mathbf{1}^B +Y^i_B (t) \, \mathbf{1}^A \otimes \sigma_i^B + Y^{ai} (t) \, \sigma_a^A \otimes \sigma_i^B \, .
\ee
From now until the end of the section, we will omit the superscripts referring to the subsystems $\lbrace A,B \rbrace,$ since the generators for the two copies are the same.
We denote the eigenstates of the Pauli matrix $\sigma_z$ by  $\lbrace \ket{1}\equiv \ket\uparrow,\ket{2}\equiv \ket\downarrow\rbrace$,\footnote{The  eigenstates of the matrix $\sigma_z$ are more commonly denoted by $\lbrace \ket{0},\ket{1} \rbrace$, but here we have used indices starting from $1$ to maintain consistency with the range of indices in eq.~\eqref{eq:ref_tar_states}.} and define 
$\lbrace \ket{\tilde{1}},\ket{\tilde{2}} \rbrace = \lbrace \ket{1},i\ket{2} \rbrace$  which will serve as a convenient basis for the Hilbert space of the second qubit.
We fix the reference state and allow for a generic Schmidt decomposition of the target state parametrized by a real angular coordinate $\theta$
\be
\ket{\psi_R} =  \ket{11} \, , \qquad
\ket{\psi_T} = \cos \theta \ket{ \bar{\phi}_1 \bar{\chi}_1 } + \sin \theta \ket{ \bar{\phi}_2 \bar{\chi}_2 } \, ,
\label{eq:ref_tar_2qubits}
\ee
where $\bar{\phi}_m, \bar{\chi}_m$ (with $m=1,2$) denote a fixed basis determined by the Schmidt decomposition, which is related to the computational basis $\ket{{m \tilde{m}}}$ by LU transformations.
These states fix the boundary conditions for any trajectory.
At each instant in the time evolution, the state can be expressed as
\be
\ket{\psi(t)} = \cos \theta(t) \,\ket{\phi_1(t)\chi_1(t)}+\sin \theta(t)\,\ket{\phi_2(t)\chi_2(t)} \, ,
\label{eq:generic_state_2qubit}
\ee
identifying the Schmidt coefficients $\lambda_1(t)=\cos  \theta(t) $ and $\lambda_2(t) = \sin \theta(t) $ in terms of a single parameter. 

\paragraph{Step 2.} 
As explained in section \ref{ssec:relation_orthogonal}, we identify the relation between $R$, the generator of rotations of the Schmidt coefficients, and the Hamiltonian. This determines the degrees of freedom in the Hamiltonian which are relevant for the problem. The $\mathrm{SO}(2)$ rotation generator has only one degree of freedom that we parametrize as $R =  v(t) \, i \sigma_y$ (here $\sigma_y$ is understood to act on the vector of Schmidt coefficients).
The constraint on the  control parameters of the Hamiltonian in terms of $R$ \eqref{eq:Rcnstrain} reads 
\be \label{eqR2qub}
\begin{aligned}
 R_{mn} = v(t) i (\sigma_y)_{mn} & =  \text{Im} \le \bra{\phi_m(t)\chi_m(t)}Y^{a i}(t) \sigma_a \otimes \sigma_i \ket{\phi_n(t)\chi_n(t)} \ri =
\\ 
& =  \text{Im} \le \bra{m\tilde{m}}\tilde{Y}^{a i}(t) \sigma_a \otimes \sigma_i \ket{n\tilde{n}} \ri  =
\\
& = (\tilde{Y}_{11}(t)-\tilde{Y}_{22}(t))(\delta_{m 1}\delta_{n 2}-\delta_{m 2}\delta_{n 1})  =
\\
& =  (\tilde{Y}_{11}(t)-\tilde{Y}_{22}(t)) i (\sigma_y)_{mn} \, .
\end{aligned}
\ee
In the first line, we used the definition in eq.~\eqref{eq:RcnstrainH}. 
In the second line, we redefined the matrix of velocities using an LU transformation on the basis elements according to eq.~\eqref{eq:Rcnstrain}, and we computed in the third line the non-vanishing elements.
Finally, in the last line, we recognized the appearance of the Pauli generator $\sigma_y$.

Binding state complexity requires to minimize the $F$-norm under the constraint
\be
\tilde{Y}_{11}(t)-\tilde{Y}_{22}(t) = v(t) \, .
\label{eq:constraint_2qubits}
\ee

\subsubsection*{$\mathcal{BF}_{1, \rm hom}$ norm} 

\paragraph{Step 3.}
Using the inequality \eqref{eq:ineq2}, we find
\be
\mathcal{BF}_{1,\text{hom}}(R_{mn})\geq \min_{\tilde{Y}} \le |\tilde{Y}_{11}| +|\tilde{Y}_{22}| +|\tilde{Y}_{33}| \ri = \min_{\tilde{Y}_{11},\tilde{Y}_{33}} \le |\tilde{Y}_{11}| +|-v(t)+\tilde{Y}_{11}| +|\tilde{Y}_{33}| \ri = |v(t)| \, .
\label{eq:inequalities_2qubits}
\ee
The lower bound, together with the constraint \eqref{eqR2qub}, can be attained, for instance, by the following Hamiltonian
\be
H(t) = -v(t) \, \sigma_x \otimes \sigma_x \,,
\ee
and by choosing $U^A$ and $U^B$ to  be the identity in eq.~\eqref{eq:Rcnstrain}.
We will, of course, require an additional LU transformation at $t = 1$ to reach the exact target state.
This shows that the minimization problem reduced from the group manifold SU($2$) to SO($2$), with $\mathcal{BF}_{1,\text{hom}}(R_{mn}) = |v(t)|$.

\paragraph{Steps 4 and 5.} Finally, we should minimize the norm over the directions of the stabilizer of the Schmidt vector $\vec{\lambda}(t)$ under the constraint \eqref{eq:evol_Schmidt}.
Since in this case $R$ and $\vec{\lambda}(t)$ depend on a single parameter, the minimization is trivial and we find $|v(t)| = |\dot \th(t)|$, where $\theta(t)$ was introduced in eq.~\eqref{eq:generic_state_2qubit}.
We then conclude that $\mathcal{BF}_{1,\text{hom}} =|\dot\theta (t)|$.
One can argue that the geodesic for this norm does not flip the sign of $\dot{\theta}(t)$,  since backtracking in $\theta$ will only elongate the integrated trajectory. Therefore a direct integration gives
\be	
\mathcal{BC}_{1,\rm hom} [\ket{\psi_T}, \ket{\psi_R}] = |\theta| \, , 
\text{\quad with $|\theta| \leq \frac{\pi}{4}$,}
\label{eq:binding_state_compl_2qubits}
\ee
where the restriction on the angle comes from the fact that we can always use a free local unitary transformation to rotate $\theta \to \theta +\frac{\pi}{2}$. 
We notice that the maximally entangled state, precisely achieved when $\theta=\pi/4$, is also maximally complex. 
Finally, we remark that related results were obtained in \cite{Cirac:2001} by evaluating the minimal time to entangle two qubits under certain constraints.

\paragraph{Trajectory in the Hilbert space.} 
Since the binding complexity 
\eqref{eq:def_compl_state} associated with the norm \eqref{eq:Fpenalties_flat} is invariant under time reparametrizations (see footnote \ref{footnotehomog}), we can actually choose $\theta(t)=\theta t,$ such that the velocity is simply a constant along the trajectory $v(t)=\theta$.
The trajectory of the state under the Hamiltonian evolution will first reach an intermediate state in the Hilbert space
\be
| \tilde{\psi}_T \rangle = \vec{P} e^{-i \int_0^1 dt' H (t')}  \ket{\psi_R}
= e^{-i H t}  \ket{\psi_R}\Big|_{t=1}  \, , \qquad
H = -\theta \, \sigma_x \otimes \sigma_x \, ,
\label{eq:optimal_Ham_qubits_F1norm}
\ee
where $ |\tilde{\psi}_T \rangle $ is related to the exact target state $\ket{\psi_T}$ in eq.~\eqref{eq:ref_tar_2qubits} by a LU transformation. 
Explicitly, the trajectory of the state generated by this Hamiltonian is given by
\be \label{traj2spin}
\ket{\psi(t)} = \cos (\theta t) \ket{11} + i \sin( \theta t) \ket{22}  \, , \qquad
\ket{\psi(0)} = \ket{\psi_R} \, , \qquad
\ket{\psi(1)} = | \tilde{\psi}_T \rangle \, .
\ee
To reach the exact target state, a final costless  (instantaneous) LU transformation is needed at $t=1$ such that $U\otimes V \ket{m\tilde{m}}=\ket{\bar{\phi}_m \bar{\chi}_m}$.

We observe that the Hamiltonian which generates the trajectories with minimal length is not unique. For example, another choice would be $H = a \, \sigma_x \otimes \sigma_y + b\, \sigma_y \otimes \sigma_x $, where $a$ and $b$ are positive and $a+b=\theta$.\footnote{At the price of being pedantic, let us comment that this Hamiltonian defines the values of $Y^{ai}(t)$, not those of $\tilde Y^{ai}(t)$ in eq.~\eqref{eqR2qub}.}
The trajectories obtained with this class of Hamiltonians will differ from the path in eq.~\eqref{traj2spin} by the basis used for the Schmidt decomposition and will require a different LU transformation at $t=1$ to reach the exact target state.

\subsubsection*{$\mathcal{BF}_{2, \rm hom}$ norm}
\paragraph{Step 3.} To obtain the complexity according to the $F_{2,\text{hom}}$ norm, see eq.~\eqref{eq:Fpenalties_flat}, we use  the following chain of identities: 
\bea
\mathcal{BF}_{2,\text{hom}}  = 
\min_{\tilde{Y}} \sqrt{\tilde{Y}_{ai}\tilde{Y}_{ai}}  = \min_{\tilde{Y}_{11}} \sqrt{\tilde{Y}_{11}^2+(\tilde{Y}_{11}-v(t))^2} = \frac{1}{\sqrt{2}} \, |v(t)| \, .
\label{eq:cost_F2hom_2qubits}
\eea
The first step exploits the invariance of the norm under orthogonal transformations $Y\to O^A\tilde{Y} O^B$.
The second step uses the constraint \eqref{eq:constraint_2qubits} and
involves a minimization over all the velocities except $\tilde{Y}_{11}$, which is minimized in the last step.
The minimum value is obtained when $\tilde{Y}_{11} = -\tilde{Y}_{22} = v(t)/2$.
Choosing the reference and target states in eq.~\eqref{eq:ref_tar_2qubits}, we find that a unitary connecting them with cost \eqref{eq:cost_F2hom_2qubits} is given by 
\be
| \tilde{\psi}_T \rangle = e^{-i H t}  \ket{\psi_R}\Big|_{t=1} \, , \qquad
H = -\frac{\theta}{2}  \le \sigma_x \otimes \sigma_x - \sigma_y \otimes \sigma_y \ri \, .
\ee
where again a final LU transformation is needed to go from $|\tilde{\psi}_T \rangle$ to $| \psi_T \rangle$.
\paragraph{Steps 4 and 5.} Steps 4 and 5 are trivial like in the previous case, because the problem has only one degree of freedom. We can again identify   $v(t)=-\theta$ and compute  the binding complexity
\be	
\mathcal{BC}^{\rm state}_{2,\rm hom} [\ket{\psi_T}, \ket{\psi_R}] = \int_0^1 dt \frac{1}{\sqrt{2}} \, |\theta| \, = \frac{1}{\sqrt{2}} \, |\theta| \, .
\label{eq:bind_complexity_F2_2qubits}
\ee
This result is smaller by a factor of $\sqrt{2}$ with respect to the complexity associated with the $F_{1, \rm hom}$ norm.

\section{Results for arbitrary qudits}
\label{sec:general_results}

We apply the strategy outlined in section \ref{ssec:summary_procedure} to study the general case of two subsystems described by the unitary groups $\mathrm{SU}(N_A)$ and $\mathrm{SU}(N_B)$.
One can interpret this setting as composed of two qudits.\footnote{Qudits are invariant under the $\mathrm{SU}(d)$ group and model atomic systems with $d$ excited states. Qutrits correspond to the case $d=3$.}
Alternatively, when $N_{A(B)} = 2^{n_{A(B)}},$ we can interpret the system as a spin chain split into two parts.
We will be following the general discussion in section \ref{ssec:reduction_set} and specifying the step numbers according to section \ref{ssec:summary_procedure}. The results for the $F_{1,\mathrm{hom}}$ norm depend on the choice of basis of generators used in decomposing the Hamiltonian \eqref{eq:expansion_Ham}. Therefore we conduct our analysis in the context of specific examples of bases. We first study the Gell-Mann basis of $\mathrm{SU}(N)$ where we have better analytic control in section \ref{ssec:F1norm}. We discuss the Pauli basis in section \ref{ssec:Pauli} where, in several cases, we can present bounds for the $F_{1,\mathrm{hom}}$ complexity. Our results for the $F_{2,\mathrm{hom}}$ norm, discussed in section \ref{ssec:F2norm}, are (up to an overall constant) independent of the basis of generators and are therefore valid for any basis. In section \ref{app:1=2}, we demonstrate that our results for the binding complexity using the $F_1$ norm are, to some extent, insensitive to the assignment of penalty factors to the non-local generators.

\paragraph{Step 1.}  The minimal set of parameters required to study binding state complexity is given by the Schmidt coefficients and their differentials.
We can then specify the reference and target states by the set of Schmidt coefficients
\be
\vec{\lambda}_{\rm ref} = ( 1, 0 , \dots , 0 ) \, , \qquad
\vec{\lambda}_{\rm tar} = (\bar{\lambda}_1 , \bar{\lambda}_2 , \dots , \bar{\lambda}_{N_A} ) \, , \qquad
\bar{\lambda}_1 \geq \bar{\lambda} _2 \geq \dots \geq \bar{\lambda}_{N_A} \, .
\label{eq:generic_ref_tar_Schmidt}
\ee
We will always take the reference state to be factorized, $\ket{\psi_{\rm ref}} = \ket{\psi_A}\otimes \ket{\psi_B}$, so that it has only one non-vanishing Schmidt coefficient equal to 1. For the target state, we can always assume (modulo a local unitary operation) that the coefficients are positive real numbers and ordered from largest to smallest as indicated in \eqref{eq:generic_ref_tar_Schmidt}. 

\paragraph{Choice of basis for the generators.}
\label{ssec:basis_generators}

Consider the bases $\lbrace T^A_a \rbrace$ and $\lbrace T^B_i \rbrace$ for the algebras $\mathfrak{su}(N_A)$ and $\mathfrak{su}(N_B),$ containing $N_A^2-1$ and $N_B^2-1$ generators, respectively.
The basis for the unitary algebra $\mathfrak{su}(N_A N_B)$ of the full system is built by taking the set of $(N_A N_B)^2 -1$ generators
\be
T^A_a \otimes \mathbf{1}^B \, , \qquad 
\mathbf{1}^A \otimes T^B_i \, , \qquad
 T^A_a \otimes T^B_i \, .
 \label{eq:total_basis_generators}
\ee
In the following, we will consider two possibilities for the basis $T^{A(B)}_{a(i)}$ of each subsystem: (1) the basis of generalized Gell-Mann matrices or (2) tensor products of Pauli matrices. These two possibilities coincide for the case of $\mathrm{SU}(2)$ discussed above when $N_A=N_B=2$. Let us start with the generalized Gell-Mann basis.

\subsection{$\mathcal{BF}_{1,\mathrm{hom}}$  norm with the Gell-Mann basis}
\label{ssec:F1norm}

The so-called Gell-Mann basis of generators of the special unitary group $\mathrm{SU}(N)$ is a natural generalization of the Pauli matrices to describe qudits, the information units of higher dimensional quantum computing \cite{Wang:2020}. 
The basis is defined by
	\be\label{eq:gell}
	\begin{aligned}
		T_{c(k,p)} \equiv  \begin{cases} 
		(T^x)_{k,p}  \equiv \ket{k}\bra{p} +\ket{p}\bra{k} & \text{if }  1 \leq k < p \leq N \\
	(T^y)_{k,p}  \equiv 		i\left(\ket{k}\bra{p} -\ket{p}\bra{k}\right)  &  \text{if } 1 \leq  p<k \leq N \\
		(T^z)_{k,p} \equiv	\sqrt{\frac{2}{k(k+1)}}(-k\ket{k+1}\bra{k+1}+\sum_{i=1}^{k}\ket{i}\bra{i}) & \text{if } 1 \leq  k=p < N \, ,
		\end{cases}
	\end{aligned}
	\ee
where $\ket{p}$ denotes a fixed basis in the Hilbert space, and
the three cases in the previous list generalize the Pauli matrices $\sigma_x, \sigma_y, \sigma_z$, respectively. 
The $N^2-1$ generators are labelled using the indices $k,p \in \lbrace 1, 2 ,\dots ,N\rbrace $, with the only constraint that we cannot have $k=p=N.$
The generalized Gell-Mann matrices form an orthogonal basis normalized as $\Tr (T_{c(k,p)} T_{c(m,n)} ) = 2 \delta_{k m} \delta_{p n} .$
It is sometimes convenient to label the basis using a single index, defined by
\be
c(k,p) = N (k-1) + p \, .
\label{eq:def_c_kp}
\ee
We build the basis for the algebra $\mathfrak{su}(N_A N_B)$ generating unitary transformations on the full system by combining the generators according to eq.~\eqref{eq:total_basis_generators}.
The bases of the Hilbert spaces of the systems $A$ and $B$ \eqref{eq:splitting_Hilbert_spaces} will be denoted with $\ket{p}, \ket{\tilde{p}}$, and the associated Gell-Mann matrices will take the form \eqref{eq:gell} with or without tilde, respectively. Of course, as implied by this notation, the bases $\ket{p}, \ket{\tilde{p}}$, used here to define the Gell-Mann matrices will be the same fixed bases used in eq.~\eqref{eq:Rmnfixedbase}-\eqref{eq:Rcnstrain} to define $\tilde Y$.

\paragraph{Step 2.} We now set up the constraint equations in the $\mathrm{SO}(N_A)$ picture (see section \ref{ssec:relation_orthogonal}) starting from the rotation matrix $R_{mn}$ defined in eq.~\eqref{eq:Rcnstrain}. 
Since $R_{mn}$ is antisymmetric, there is not any contribution of the diagonal generators (those with $m=n$), either in the subsystem $A$ or $B$ or both. Furthermore, since the expression \eqref{eq:Rcnstrain}
involves just the imaginary part, only non-local generators of the form $T^x \otimes T^y$ and $T^y \otimes T^x$ give rise to non-vanishing contributions.\footnote{Note that in the two qubits example in section \ref{ssec:example_2qubits}, we had, unlike here, a relative phase between the bases of the two systems, and therefore the generators that came into play were $\sigma_x\otimes \sigma_x$ and $\sigma_y\otimes \sigma_y$.}
Assuming $m>n$, we get\footnote{Assuming instead $n>m$ will give the same result up to an overall minus sign, as it should from the fact that $R_{mn}$ is anti-symmetric}
\be \label{eq:cnstGen}
R_{mn} = \text{Im} \le \bra{m\tilde{m}} \tilde{Y}_{c(a,b),c(i,j)} T^A_{c(a,b)} \otimes T^B_{c(i,j)}\ket{n\tilde{n}} \ri = \tilde{Y}_{c(m,n),c(n,m)} +\tilde{Y}_{c(n,m),c(m,n)} \, .
\ee
In this notation, the velocities $\tilde{Y}$ form a matrix with collective indices $c(a,b)$ and $c(i,j)$.
Notice that $ (T^y_{a,b})^A$ form a subset of $\frac{1}{2} N_A (N_A-1)$ Gell-Mann matrices that generate the orthogonal subgroup SO($N_A$) (they are angular momentum operators in Cartesian coordinates). 
Therefore, we can write
	\be
	R =  - i  \sum^{N_A}_{m>n=1} R_{mn}   (T^y_{m,n})^A  \, .
	\label{eq:decomposition_R_GellMann}
	\ee
We now perform for convenience an orthogonal transformation $O^B\in \mathrm{SO}(N_B^2-1)$ to define a new matrix of velocities $Y' = \tilde{Y} O^B$, such that the constraints \eqref{eq:cnstGen} are written only in terms of diagonal elements of $Y'$ as\footnote{Explicitly, the relevant matrix is $O^B_{c(i,j),c(k,l)}=\delta_{i,l}\delta_{j, k}$, where here the indices $i,j,k,l$ run over the range ${1,\ldots,N_B}$.}
\be
 Y'_{c(m,n), c(m,n)}+Y'_{c(n,m), c(n,m)} = R_{mn} \, , \qquad  (m>n) \, .
 \label{eq:constraints_GellMann_diag}
 \ee
The constraints \eqref{eq:constraints_GellMann_diag} are the starting point to study binding state complexity in a system composed of two qudits (or equivalently a spin chain when $N_A$ is a power of two) using the Gell-Mann basis.

\paragraph{Step 3.}
Using the inequalities \eqref{eq:chain_relations_F1normT}-\eqref{eq:ineq2} and the observation that the nuclear norm is invariant under an orthogonal transformation (applied separately on the left or right), \ie $\sum_a \sigma_a(\tilde{Y}) = \sum_a \sigma_a(Y')$, we find the following bound
\be
 \begin{aligned}
 	\mathcal{BF}_{1,\rm hom}(R_{mn}) & \geq  \min_{Y' :~ R_{mn}\text{ is fixed} } \sum_a \sigma_a (Y') \geq \min_{Y' :~ R_{mn}\text{ is fixed} } \sum_{m,n} |Y'_{c(m,n),c(m,n)}| \\
 &	\geq \min_{Y' :~ R_{mn}\text{ is fixed} } \sum_{m>n} |Y'_{c(m,n),c(m,n)}|+|Y'_{c(n,m),c(n,m)}| = \sum_{m>n} |R_{mn}| \, ,
 \end{aligned}
\ee 
where the velocities $Y'$ were defined below eq.~\eqref{eq:decomposition_R_GellMann} and the minimization is subject to the constraint \eqref{eq:constraints_GellMann_diag}.
In going from the first to the second line, we used a trivial bound based on excluding $m=n$ terms from the summation. In the last step, we performed the explicit minimization over $Y'$ using the constraints \eqref{eq:constraints_GellMann_diag}.  
One can check that the lower bound, together with the constraint \eqref{eq:cnstGen}, can be achieved by using, for instance, the optimal Hamiltonian\footnote{One can check that in the two-qubit case presented in section \ref{ssec:example_2qubits}, the Hamiltonian \eqref{eq:optimal_Ham_GellMann} degenerates to $H = \theta \sigma_x \otimes \sigma_y$, which connects the reference state $\ket{\psi_R} = \ket{11}$ to the generic target state $| \tilde{\psi}_T \rangle = \cos \theta \ket{11} + \sin \theta \ket{22},$ where we have chosen the basis $\ket{\tilde{1}}=\ket{1}$ and $\ket{\tilde{2}}=\ket{2}$. In particular, this $H$ is optimal because it gives the same binding state complexity as evaluated in eq.~\eqref{eq:binding_state_compl_2qubits}. This is a Hamiltonian in the class described below eq.~\eqref{traj2spin}.}
\be
H = \sum_{m>n} R_{mn} T^A_{c(m,n)} \otimes T^B_{c(n,m)} \, ,
\label{eq:optimal_Ham_GellMann}
\ee
and by choosing $U^A$ and $U^B$ to  be the identity in eq.~\eqref{eq:Rcnstrain}.
We will, of course, require an additional LU transformation at the end of the trajectory to reach the exact target state. 
Therefore the minimization yields
\be 
\label{eq:F1ofR}
 \mathcal{BF}_{1, \rm hom}(R_{mn}) =
 \sum_{m>n} |R_{mn} | \, .
 \ee
We note that $R_{mn}$ rotates a point along the plane spanned by the $m$ and $n$ Cartesian axes. This means that the problem now has the interpretation of the minimal sum of (absolute values of) rotation angles needed to reach a given point starting from the pole on a unit sphere, where the rotations are restricted to planes spanned by all pairs of  Cartesian axes. To illustrate this point further, for a unit sphere embedded in three dimensions, the equivalent quantity is the minimal sum of absolute values of rotation angles, where only rotations around the $x$, $y$ and $z$ axes are allowed.

\paragraph{Step 4.} To perform the minimization over the stabilizer group and find the state complexity, we need to solve eq.~\eqref{eq:evol_Schmidt}, which in this setting reads
	\be\label{eq:EoMtil}
	\sum_{m<n} R_{nm}  \lambda_m -\sum_{m>n} R_{mn} \lambda_m = \dot{\lambda}_n \, .
	\ee
There are $N_A-1$ independent equations (recall that the Schmidt coefficients live on the sphere) and a total of $\frac{1}{2} N_A (N_A-1)$ independent velocities $R_{mn}$.
Let us enumerate the different velocity components in terms of a single index $R_\alpha$  with $\alpha\in \{1,\dots,\frac{1}{2} N_A(N_A-1)\}$. 
We can, without loss of generality, solve the equations above for the first $N_A-1$ velocities and express $R_\alpha$ for $\alpha\in \{1, \dots, N_A-1\}$ in terms of the velocities $R_\beta$ for $\beta\geq N_A$ (if the equations cannot be solved for the first $N_A-1$ velocities, just enumerate them differently).
Inserting the solution back into the $F_{1,\rm hom}$ norm from eq.~\eqref{eq:F1ofR} gives
\be
\mathcal{BF}_{1, \rm hom} (\vec\lambda,\dot {\vec\lambda}) = \min_{\rm stab}\sum_{m>n} |R_{mn}|=
\min_{R_\beta} \le \sum_{\alpha=1}^{N_A-1} |R_\alpha (\lambda_m,\dot{\lambda}_m,R_\beta)|+ \sum_{\beta=N_A}^{N_A(N_A-1)/2} |R_\beta| \ri \, ,
\label{eq:piecewise_F1norm}
\ee
where the minimization ``${\rm stab}$'' refers to step 4 in section \ref{ssec:summary_procedure}, and allows us to express the norm in terms of the Schmidt coefficients.  To minimize over a \emph{single} $R_\beta$, we notice that the right-hand side of eq.~\eqref{eq:piecewise_F1norm} is a piece-wise linear positive function, so the minimum must be at the points where one of the arguments of the absolute values vanish, \ie 
\be\label{eq:somevarezero}
R_\beta  =0  \quad
\vee \quad  R_\alpha (\lambda_m,\dot{\lambda}_m, R_\beta)=0 \, .
\ee
Reiterating this process for all the $ R_\beta$-s amounts to setting any selection of $(N_A-1)(N_A-2)/2$ of the above absolute values to zero. Of course, we then have to take the absolute minimum among all these choices.
In other words, we should solve eq.~\eqref{eq:EoMtil} with only $N_A-1$ non-vanishing velocities, when the rest of the velocities are set to zero. Then we substitute the solution, in terms of $\vec \lambda$ and $\dot{\vec{\lambda}}$, into the sum over the absolute values of all velocities and we finally take the minimum over all choices of those $N_A-1$ non-vanishing velocities out of the $N_A(N_A-1)/2$ velocities.

Let us now show that the minimum of \eqref{eq:F1ofR} is obtained by setting to zero all the velocities other than  $R_{m1}$ in eq.~\eqref{eq:decomposition_R_GellMann} subject to the constraint \eqref{eq:EoMtil} with given $\vec \lambda$ and $\dot{\vec{\lambda}}$.
In other words, the following matrix minimizes the norm, at least for points on the sphere that are not too far from the reference state $\vec{\lambda}_{\rm ref}$ in eq.~\eqref{eq:generic_ref_tar_Schmidt}: 
\be
R_{\rm opt} =  - i \sum^{N_A}_{m>1}  w_m  (T^y_{m,1})^A  \, .
\label{eq:initial_optimal_gen_F1}
\ee
The requirement to solve \eqref{eq:EoMtil} fixes
\begin{equation}
\begin{aligned}
 & w_m \equiv R_{m1} = \frac{\dot \lambda_m}{\lambda_1} \, , \qquad (m>1) \, , &  \\
 & R_{mn}=0 \, ,  \qquad    (m>n>1) \, . &
\end{aligned}
    \label{eq:TheAmazingSolution}
\end{equation}
Let us first motivate this ansatz. At the start of the trajectory, all $(T^y)_{m,n}$ with $m,n>1$ are members of the stabilizer subgroup of rotations of the vector of Schmidt coefficients. Therefore, the use of these generators in $R$ will not influence the constraint \eqref{eq:EoMtil}, but they will increase the norm and so there is no point in using those generators in $R$. When we are slightly away from the start of the trajectory, these generators are still close to the stabilizer, and so are less efficient than $(T^y)_{m,1}$ in modifying the Schmidt coefficients. 
Intuitively, this suggests that the same velocities will also be set to zero for a region on the Schmidt sphere that is close enough to the reference state. 
We will prove that this is the optimal solution also away from the pole, as long as $\lambda_2+\lambda_3 \leq \lambda_1$ (assuming that the Schmidt coefficients are ordered in decreasing order).

Let us now prove that eq.~\eqref{eq:initial_optimal_gen_F1} gives the optimal matrix $R$ for the minimization of the norm.
We compare the norm computed using eq.~\eqref{eq:initial_optimal_gen_F1} to any other choice of $R$ subject to the constraint \eqref{eq:EoMtil}. 
A generic option for $R$ can be expressed as
\be
	R_{\rm gen} = -i  \sum^{N_A}_{m>n=1} \nu_{m,n}  (T^y_{m,n})^A \, ,
	\label{eq:generic_gen_F1}
	\ee
where $\nu_{m,n}$ form another set of velocities.
By comparing $ R_{\rm opt} \vec{\lambda} = R_{\rm gen} \vec{\lambda}$ defined in eqs.~\eqref{eq:initial_optimal_gen_F1} and \eqref{eq:generic_gen_F1}, we can express the velocities $w_m$ in terms of $\nu$ as 
\be
	 w_m = \frac{1}{\lambda_1} \le \sum_{n<m}  \nu_{m,n} \lambda_n -\sum_{n>m}  \nu_{n,m} \lambda_n \ri 
 \qquad m>1 \, .
	\label{eq:velocities_nu_v}
	\ee
The difference between the $F_{1, \rm hom}$ norm using the two different rotation matrices reads
	\be
	\sum^{N_A}_{m>n =1} |\nu_{m,n}| -  \sum^{N_A}_{ m >1} |w_m|  
	\geq
	\sum^{N_A}_{m>n>1} |\nu_{m,n}|\le 1 -\frac{\lambda_n+\lambda_m}{\lambda_1} \ri \, ,
	\ee
where we used the identity \eqref{eq:velocities_nu_v} and applied the triangle inequality to the second sum on the left-hand side following this replacement.
We see that for a region on the sphere in which $\lambda_n+\lambda_m \leq \lambda_1$, for any $m > n>1$, this expression is positive (since we take the Schmidt coefficients in a decreasing order, this region amounts to $\lambda_2 + \lambda_3 \leq \lambda_1$).
 Therefore the matrix $R_{\rm opt}$ proposed in eq.~\eqref{eq:initial_optimal_gen_F1} indeed minimizes the norm for this region on the sphere. The $\mathcal{BF}_{1, \rm hom}$ norm then takes the following form in terms of the Schmidt coefficients $\lambda_m$ and $\dot \lambda_m$: 
\be
\mathcal{BF}_{1, \rm hom} \, dt =  \sum^{N_A}_{m>1} | w_m|  dt = \sum^{N_A}_{m>1} \frac{ |d\lambda_m|}{\lambda_1} = \sum^{N_A}_{m>1} \frac{ |d\lambda_m|}{\sqrt{1-\sum^{N_A}_{n>1} \lambda_n^2}} \, ,
\label{eq:distance_F1norm_GellMann}
\ee
where $\lambda_1$ is the maximal Schmidt coefficient, and in the third equality we used the solution for the velocities $w_m$ given in eq.~\eqref{eq:TheAmazingSolution}.
This cost function defines an infinitesimal distance in the space of Schmidt coefficients, and the length of its geodesics computes the binding state complexity.

\paragraph{Step 5.}
Next, we compute the geodesics connecting the reference and target states \eqref{eq:generic_ref_tar_Schmidt} in this geometry.
Let us first show that for the shortest path, $d\lambda_m\geq 0$ for $m>1$ at all points along the path. We will prove this by contradiction. Let us assume that, for a trajectory parameterized by $\vec{\lambda}(t)$ with $\vec{\lambda}(0) = \vec{\lambda}_{\rm ref}$ defined in eq.~\eqref{eq:generic_ref_tar_Schmidt}, there exists some range of time $(t_2,t_4)$ during which $\dot\lambda_m(t) < 0$, while $\dot\lambda_m >0$ for $t<t_2$. Then there must exist two times $t_1$ and $t_3$ such that $t_1<t_2<t_3$,  for which $\lambda_m(t_1) = \lambda_m(t_3)\leq \lambda_m(t)$ for  \,\,$ t \in (t_1,t_3)$. The factor of $(1-\sum^{N_A}_{n>1} \lambda_n^2)^{-1/2}$ entering the norm \eqref{eq:distance_F1norm_GellMann} is a monotonically increasing function of the Schmidt coefficients, thus there is a shorter trajectory such that $\lambda_m(t)=\lambda_m(t_1)=\lambda_m(t_3)$ for $t \in [t_1,t_3]$, contradicting the assumption. 

Since the Schmidt coefficients are arranged in decreasing order, the possible candidates for the geodesics are trajectories in a region defined by $\lambda_1(t)\geq \lambda_2(t)\geq \dots \geq \lambda_{N_{A}}(t)$, and have, as shown above, $\dot \lambda_m \geq 0$ for all $m>1$. 
Consider the following piecewise smooth trajectory $\lambda^g_m(t)$: 
\be
\begin{aligned}
(1,0,...,0) & \to \le \sqrt{1-(N_A-1) \bar{\lambda}_{N_A}^2}, \bar{\lambda}_{N_A }, \dots , \bar{\lambda}_{N_A} \ri \\
& \rightarrow \le \sqrt{1-\bar{\lambda}_{N_A}^2-(N_A-2) \bar{\lambda}_{N_A -1}^2},\bar{\lambda}_{N_A-1}, \dots , \bar{\lambda}_{N_A-1}, \bar{\lambda}_{N_A} \ri \\
&  \to \dots  \to (\bar{\lambda}_1 , \bar{\lambda}_2 , \dots , \bar{\lambda}_{N_A} ) \, ,
\end{aligned}
\label{eq:piecewise_trajectory_GellMann}
\ee
where in the $p$-th step, the differentials satisfy $d\lambda_m = d\lambda_n$ for $m,n \in \lbrace 2, 3, \dots , N_A - p +1 \rbrace $ and $d\lambda_m=0$ for $m>N_A - p +1$.\footnote{Note that for states where certain Schmidt coefficients do not appear in the final state, these Schmidt coefficients will also be turned off along all the trajectory.}${}^{,}$\footnote{The specification of the relation between the different differentials, rather than the precise time dependence, is enough to find the complexity due to the reparametrization invariance of the integrated norm.}
We will show that $\lambda^g_m(t)$ is the geodesic.

First of all, let us explain how the trajectory works.
The first Schmidt coefficient is always fixed by the normalization condition $\sum_m \lambda_m^2 =1$. In the first step, all the other non-vanishing coefficients are increased at the same pace until the lowest coefficient reaches the target value $\bar{\lambda}_{N_A}$. 
After that, the last Schmidt coefficient stops evolving and we continue to increase the other ones at an equal rate until they reach the next-to-minimal value $\bar{\lambda}_{N_A-1}$.
Then also the second-to-last coefficient stops evolving, and we repeatedly apply this method until the target state is reached.
 
We now prove that the trajectory \eqref{eq:piecewise_trajectory_GellMann} is the geodesic by showing that the difference between the length of any other trajectory and $\vec{\lambda}^g(t)$ is positive. Let us consider  another trajectory $\vec \lambda(t)$ which we parametrize using the time $t= 1-\lambda^g_1 = 1-\lambda_1$.\footnote{Note that since the start and endpoints of the trajectory are the same and $\lambda_1$ is monotonic, see argument above \eqref{eq:piecewise_trajectory_GellMann},   we have the freedom to choose a parametrization such that $\lambda_1=\lambda_1^g$.} 

With this parametrization, $t \in [0, t_f\equiv 1-\bar{\lambda}_1]$ and $\lambda_{m>2}(0)=0$.
Let us compute the difference in the length of the two trajectories, \ie the cost of $\lambda_{m}(t)$ minus the cost of $\lambda^g_{m}(t)$. We use integration by parts to find 
\be
	\begin{aligned}
	\int_{0}^{t_f} dt \left[ \mathcal{BF}_{1,\rm hom}(\vec{\lambda},\dot{\vec{\lambda}}) - \mathcal{BF}_{1, \rm hom}(\vec{\lambda}^g,\dot{\vec{\lambda}}^g)  \right] & =	
	\int_{0}^{t_f} dt \left(\sum^{N_A}_{m=2}  \frac{\dot{\lambda}_m}{1-t} - \sum^{N_A}_{m=2} \frac{ \dot{\lambda}^g_m}{1-t} \right) \\
&   =\int_{0}^{t_f} dt  \frac{1}{(1-t)^2}\sum^{N_A}_{m=2}  (\lambda^g_m(t)-\lambda_m(t)).
	\end{aligned}
	\label{eq:comparison_GellMann}
\ee
Because of the parametrization that we chose, the trajectories satisfy 
\begin{itemize}
    \item the normalization condition
    \be
    \sum_{m=2} \lambda^g_m(t)\lambda^g_m(t)  = \sum_{m=2}\lambda_m(t)\lambda_m(t)=  1-(1-t)^2 \, .
    \label{eq:identity1_GellMann}
    \ee
    \item the inequality
    \be
    0\leq\lambda^g_m(t), \lambda_m(t)\leq \bar{\lambda}_m \, , \qquad
    (\mathrm{for} \,\, m>1) \, ,
    \label{eq:identity2_GellMann}
    \ee
    where  $\bar\lambda_m$ is the endpoint of the trajectory. 
    \item decreasing order
    \be
    \lambda^g_m(t) \leq \lambda^g_{m-1}(t)
    , \qquad \lambda_m(t)\leq \lambda_{m-1}(t) \, , 
    \label{eq:identity3_GellMann}
    \ee
    chosen for convenience and without loss of generality.
\end{itemize}

\noindent
Under these constraints, $\vec{\lambda}^g(t)$ maximizes the sum $\sum^{N_A}_{m=2}  \lambda^g_m(t)$ and so the difference between the lengths in eq.~\eqref{eq:comparison_GellMann} is always positive. To see why, consider the sum, with $\lambda_2$ replaced by the sphere constraint in eq.~\eqref{eq:identity1_GellMann},
\be
\sum^{N_A}_{m=2}\lambda_m =\sum^{N_A}_{m=3}\lambda_m + \sqrt{ 1-(1-t)^2 - \le \sum^{N_A}_{m=3}\lambda_m^2 \ri}.
\ee
Taking the gradient with respect to $\lambda_{m\geq 3}$ gives
\be
\partial_{\lambda_m} \left(\sum^{N_A}_{m=2}\lambda_m \right) = 1-\frac{\lambda_m}{\sqrt{ 1-(1-t)^2 - \le \sum^{N_A}_{m=3}\lambda_m^2 \ri}} = 1- \frac{\lambda_m}{\lambda_2}.
\ee
Since $\lambda_m \leq \lambda_2$, the gradient is non negative, and vanishes only when $\lambda_m=\lambda_2$. This shows that the maximum of the sum is when the Schmidt coefficients for $m \geq 3$ are at their extreme values $\lambda_m(t) = \text{min}(\lambda_{m-1}(t),\bar \lambda_m)$, which is exactly how the geodesic $\lambda^g_m(t)$ behaves. 

Computing the length of the geodesic using the norm \eqref{eq:distance_F1norm_GellMann}, we obtain the binding state complexity
\be
\begin{aligned}
	& \mathcal{BC}^{\rm state}_{1, \rm hom}  = \int_0^{\bar \lambda_{N_A}} \frac{(N_A-1)dx}{\sqrt{1-(N_A-1)x^2}} +     \int_{\bar \lambda_{N_A}}^{\bar \lambda_{N_A-1}} \frac{(N_A-2)dx}{\sqrt{1-\bar \lambda_{N_A}^2-(N_A-2)x^2}} \\
&	+ \int_{\bar \lambda_{N_A-1}}^{\bar \lambda_{N_A-2}} \frac{(N_A-3)dx}{\sqrt{1-\bar \lambda_{N_A}^2 -\bar \lambda_{N_A-1}^2  -(N_A-3)x^2}}  + \dots  =  \\
	& =	\sum_{m=1}^{N_A-1} \sqrt{m} \, \bigg[\mathrm{arcsin} \,  \le \frac{\bar \lambda_{m+1}\sqrt{m}}{\sqrt{\sum_{n=1}^{m+1}\bar \lambda_{n}^2}} \ri - \mathrm{arcsin} \, \le \frac{\bar \lambda_{m+2}\sqrt{m}}{\sqrt{\sum_{n=1}^{m+1}\bar \lambda_{n}^2}} \ri \bigg] \,, 
	\end{aligned}
\label{eq:result_stateF1_hom_GellMann}
	\ee
where $\bar \lambda_{N_A+1} \equiv 0$.
In the case of the maximally entangled state, the binding complexity reduces to $\sqrt{N_A-1} \, \mathrm{arccos} \, (1/\sqrt{N_A})$. For a generic state of a spin chain consisting of $n_A$ spins, $\mathcal{BC}^{\rm state}_{1, \rm hom}$ scales exponentially as $\sqrt{N_A}=2^{n_A/2}$, as we have checked numerically by considering random sequences of Schmidt coefficients sampled from a uniform distribution and then normalized.
Eq.~\eqref{eq:result_stateF1_hom_GellMann} is our main result in the case of the $F_{1,\mathrm{hom}}$ norm in the Gell-Mann basis.

A Hamiltonian that builds a trajectory realizing the minimal cost can be constructed as follows 
\be
H= \sum^{N_A}_{m>1} w_m(t) T^A_{c(m,1)} \otimes T^B_{c(1,m)} 
\, , \qquad
w_m \equiv \frac{\dot{\lambda}^g_m(t)}{\lambda^g_1(t)} \, ,
\label{eq:optimal_Ham_F1GellMann}
\ee
where we used eqs.~\eqref{eq:optimal_Ham_GellMann} and \eqref{eq:TheAmazingSolution} with $\vec{\lambda^g}(t)$ the trajectory of the geodesic.
This Hamiltonian preserves the basis for the Schmidt decomposition along the entire time evolution, and the state along the trajectory reads
\be \label{stateTraj}
\ket{\psi(t)} = \sum_{m=1}^{N_A} \lambda^g_m(t) \ket{mm} \, .
\ee
Finally, to move from the state $\ket{\psi(t=1)}$ to the exact target state, an additional costless LU transformation is needed. 

We notice that the trajectory built with the optimal Hamiltonian \eqref{eq:optimal_Ham_F1GellMann} does not have large fluctuations along the null directions (\ie the red lines containing LUE states in fig.~\ref{fig:LUE}), because a finite LU transformation is only performed as the final step at $t=1$ to move from an LUE target state to the exact one.
Therefore the results obtained in the limit of vanishing cost for the local generators represent a good approximation for a  more realistic setting, where one has a small but non-vanishing cost for any local gate.\footnote{See the discussion below eq.~\eqref{eq:circuit_sec5} regarding the $F_{1, \mathrm{set}}$ norm, for which there are large fluctuations in the null directions and approximating the optimal circuits with non-vanishing cost for local gates is more involved.}

The above results are valid for systems with a finite-dimensional Hilbert space such as qubits and qudits. Recently, the study of complexity in continuous variable systems has raised significant interest. These studies served to find definitions of complexity in quantum field theory as a first attempt towards making the holographic complexity conjectures more precise, see \eg \cite{Jefferson:2017sdb,Chapman:2017rqy,Guo:2018kzl}. The study of binding complexity in quantum field theories is outside the scope of this manuscript, but as a first step, we can consider the complexity for a very large number of Schmidt coefficients. In that case, the complexity formula \eqref{eq:result_stateF1_hom_GellMann} can be cast in the form of a continuous integral. We present a closed formula in this limit in appendix \ref{app:ContSchmi}. 

\subsection{$\mathcal{BF}_{1,\mathrm{hom}}$  norm with the Pauli basis}
\label{ssec:Pauli}

The tools developed in section \ref{sec:SON_picture} can also be applied to the case in which the generators of the unitary group are the generalized Pauli matrices. However, it turns out that the implementation of the constraints \eqref{eq:Rcnstrain} and the minimization to perform are technically harder, making it difficult to determine the geodesics in the space of states for the $F_{1,\mathrm{hom}}$ norm.
For these reasons, we leave an exhaustive analysis of the $\mathcal{BC}^{\rm state}_{1, \rm hom}$ with the Pauli basis for future work, and instead derive bounds on the complexity and exact results for special cases. 

\paragraph{Generalized Pauli basis.} The generalized Pauli matrices give a 
basis of generators of the unitary group $\mathrm{U}(N)$ which is natural in the case of a spin chain $N=2^n$.
They are defined as 
\begin{equation}
    T_{\vec \mu} = \sigma^{1}_{\mu_{1}} \otimes  \sigma^{2}_{\mu_{2}} \otimes \ldots \otimes \sigma^{n}_{\mu_{n}},
    \label{eq:definition_Pauli_matrix}
\end{equation}
where each factor of the tensor product is a covariant two-dimensional Pauli matrix $\sigma_{\mu} = (\sigma_0, \sigma_i),$ with $\sigma_0=\mathbf{1}$.
The generalized Pauli matrices are orthogonal and normalized as $\Tr (T_{\vec\mu} T_{\vec \nu} ) = 2^{n} \delta_{\vec \mu \vec \nu}$.
Assuming that the chain is split into two parts as in eq.~\eqref{eq:splitting_Hilbert_spaces}, the basis for the full system is given  by
\be
T_I =  T_{\vec\mu}^A \otimes T_{\vec \nu}^B = \sigma^{A,1}_{\mu_{1}} \otimes \ldots \sigma^{A,{n_A}}_{\mu_{n_A}} \otimes \sigma_{\nu_{1}}^{B,1} \otimes \ldots \otimes \sigma^{B,{n_B}}_{\nu_{n_B}} \, ,
\label{eq:full_Pauli_basis}
\ee 
The superscript $(A,\alpha)$ refers to the $\alpha$-th qubit in subsystem $A$, while $(B,\beta)$ to the $\beta$-th qubit in subsystem $B$.  The index $I$ runs over the Pauli matrices of the combined system, cf. footnote \ref{foot:ind}. As we are interested in the generators of $\mathrm{SU}(N_A N_B)$, we will exclude the case where all the matrices in the product are the two-dimensional identity matrix. The generators are normalized as $\Tr (T_I T_J) = 2^{n_A+ n_B}\delta_{IJ} $.
In the case of the Pauli basis, the constraints \eqref{eq:Rcnstrain} will take a more complicated form than those for the Gell-Mann basis in eq.~\eqref{eq:cnstGen} and they will involve, generically, a larger number of velocities.  
As a consequence, we won't be able to derive an exact result for $\mathcal{BF}_{1, \rm hom}$  except for the special case where the smaller subsystem consists of a single qubit, $N_A=2$; in the general case we can only prove bounds. Therefore we will not be specifying the step structure in this section.

\paragraph{Upper bounds.}

Consider the binding state complexity for a system composed of $n$ qubits, divided into two subregions with $n_A$ and $n_B$ qubits, respectively.
 We choose reference and target states described by the Schmidt coefficients in eq.~\eqref{eq:generic_ref_tar_Schmidt} (\ie the reference state is a product state). 
We take the path built with the Hamiltonian \eqref{eq:optimal_Ham_F1GellMann} (which was a geodesic for the geometry on the space of states defined by the choice of Gell-Mann matrices as generators), but
now we compute its length according to the $\mathcal{BF}^{\rm state}_{1}$ norm for the Pauli basis.
In appendix \ref{app:Pauli}, we show that the cost is the same as with the Gell-Mann basis norm, and as such, the result in eq.~\eqref{eq:result_stateF1_hom_GellMann} gives an upper bound. However, as we show next, this upper bound can be rather weak in certain circumstances, and we can improve it in specific cases.

\begin{figure}[ht]
    \centering \includegraphics[scale=0.4]{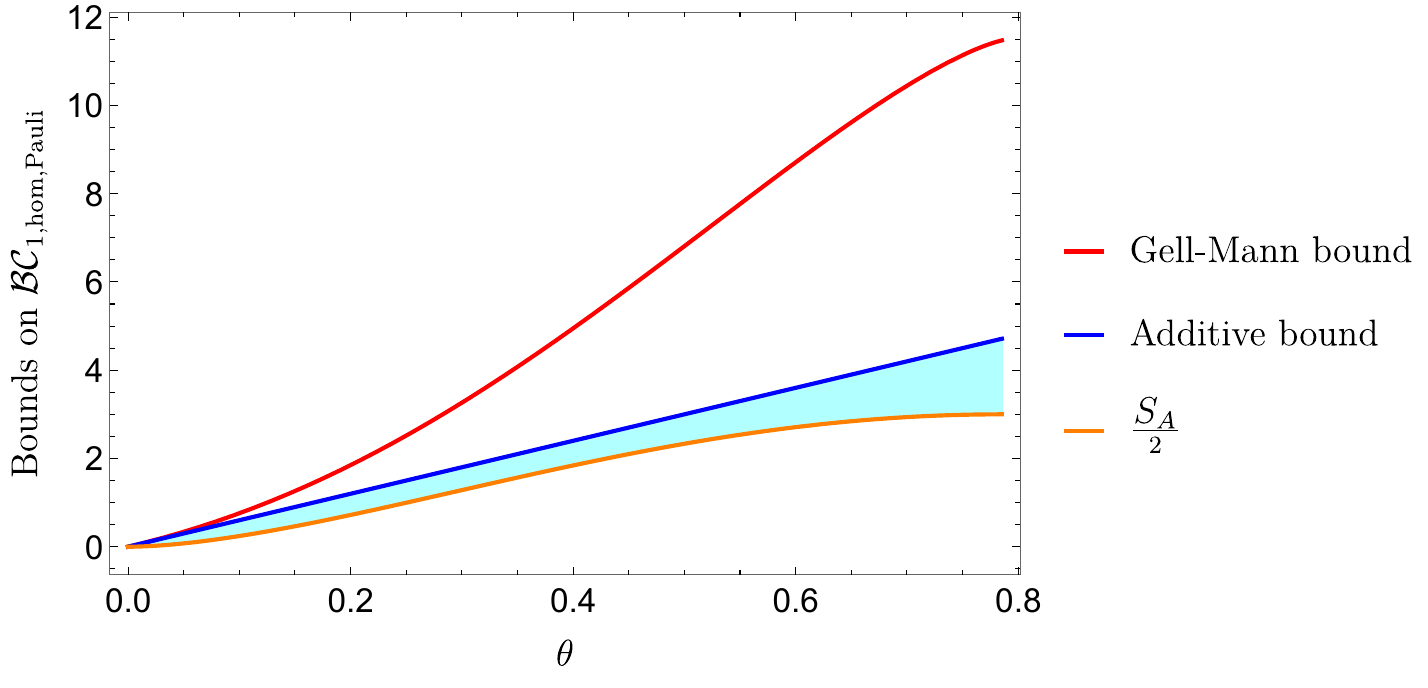}
    \caption{Comparison of the different bounds for the Pauli basis binding complexity of the state \eqref{eq:badbell} in the case of equal angles and 6 entangled qubit pairs, as a function of the angle $\theta$. Red: the bound coming from the Gell-Mann optimal trajectory. Blue: the bound from eq.~\eqref{PauliAddit} coming from additivity. Orange: one-half the entanglement entropy of the state. The allowed region for the binding complexity is between the blue and orange curves, according to the lower bound of \eqref{SlwrbndsDis2} when $c=2$.} \label{fig:PauliBoundCompare}
\end{figure}

Let us now consider the special case in which the two systems are of equal size $n_A=n_B$ and focus on target states which are LUE to a product of entangled qubit pairs across the two subregions, \ie\footnote{The discussion can be extended to $n_A < n_B$ by padding eq. \eqref{eq:badbell} with, for example, tensor products of spin up $\otimes \ket{\uparrow}$ for the Hilbert spaces of the remaining unentangled qubits of the system $B$.} 
\be\label{eq:badbell}
U_A\otimes U_B\ket{\psi_T} =  \otimes_i( \cos(\theta_i)\ket{\uparrow_{A_i}\uparrow_{B_i}} +  \sin(\theta_i)\ket{\downarrow_{A_i}\downarrow_{B_i}} )  \, .
\ee
The cost of a minimal circuit acting on two qubits (evaluated according to the $F_{1, \rm hom}$ norm) was given in eq.~\eqref{eq:binding_state_compl_2qubits}.
Extrapolating this result to a circuit $U_P (t)$ composed by the tensor product of the geodesics for each pair gives a total cost of $\sum_i |\theta_i|$.
Since the generators acting on each pair of qubits form a subalgebra of $\mathfrak{su}(N_A N_B)=\mathfrak{su}(N_A^2)$, and since these subalgebras commute with each other, it is reasonable to expect that this cost is the binding state complexity
\be \label{PauliAddit}
\mathcal{BC}^{\rm state}_{1, \rm hom} = \sum_i |\theta_i| \, ,
\ee
which would then be additive for factorized states.\footnote{See \cite{Nielsen1}, section IIIB, for a discussion on the additivity of geodesics lengths for tensor product target states in the context of Finsler metrics.} 
Let us compare this bound to the previous one from the Gell-Mann basis expression for positive angles $0<\theta_i<\pi/4$, where eq.~\eqref{eq:result_stateF1_hom_GellMann} can be applied. In a small angle expansion the two results coincide, while for large angles they differ significantly. This can be seen by the scaling with $n_A$ of the complexity bounds for the maximally entangled state, which is linear for the bound \eqref{PauliAddit}, but exponential for the bound following from the Gell-Mann basis complexity (in this case of course the above bound from additivity is much stronger). The bounds as a function of the angle for the case of 6 entangled qubit pairs with equal angles are plotted in figure \ref{fig:PauliBoundCompare}. We can see that they coincide for small angles, but deviate significantly for larger angles.

\paragraph{Exact Result.}
Let us consider yet another special case, in which the subsystem $A$ consists of a single qubit, and $B$ of an arbitrary number.
In this case, we can provide an exact result for the $\mathcal{BC}_{1,\text{hom}}^{\text{state}}$ complexity with the Pauli basis. Here, the target state is always LUE to a state in which there is non-vanishing entanglement only between $A$ and a single qubit in $B$.  
We can prove that all the other qubits (that we will call \textit{ancillary qubits}) in subregion $B$ do not give any advantage in minimizing a trajectory for the evaluation of binding state complexity according to the $\mathcal{BF}_{1, \rm hom}$ norm with the Pauli basis.
Since there is only one independent Schmidt coefficient (two in total), the equation of motion \eqref{eq:constraint_der_Schmidt} reduces to 
\begin{subequations}
\be
\dot{\lambda}(t) =  R_{12} \sqrt{1-\lambda(t)^2} \, ,
\ee
\be
R_{12}= \Im \le \bra{\phi_1(t)\chi_1(t)}Y^{\mu_1 \nu_1 \dots \nu_{n_B}}(t) \sigma_{\mu_1} \otimes \sigma_{\nu_1} \otimes \dots \otimes \sigma_{\nu_{n_B}} \ket{\phi_2(t)\chi_2(t)} \ri \, ,\label{eq:Pauli1nsystem}
\ee
\end{subequations}
where $\ket{\phi_m(t)}$ and $\ket{\chi_m(t)}$ are eigenstates for the subsystems $A$ and $B$, respectively. Eq.~\eqref{eq:Pauli1nsystem} provides a linear relation that can be solved for one of the $Y$ components in terms of the others (and of $R_{12}$). The norm $\sum |Y|$ can then be minimized (for a fixed value of $R_{12}$) using a similar procedure to the one explained around eq.~\eqref{eq:piecewise_F1norm}. That is, since the norm is a piecewise linear function, it will be minimized at points where the absolute values change signs. Therefore,  
the minimum of the norm will occur
when only one of the $Y^{\mu_1 \nu_{1} \dots \nu_{n_B}}(t)$ does not vanish. The norm will then be
\small
\be
\begin{aligned}
\mathcal{BF}_{1, \rm hom} & = \min_{{\footnotesize 
    \begin{array}{c}
\mu_1 \nu_{1} \dots \nu_{n_B} \\
U_A \otimes U_B
\end{array}}}
\left| \frac{\dot \lambda(t)}{\sqrt{1-\lambda(t)^2} \, \text{Im} \le \bra{
1\tilde 1
}U_A \otimes U_B \, \sigma_{\mu_1} \otimes \sigma_{\nu_1} \otimes \dots \otimes \sigma_{\nu_{n_B}} \, U^\dagger_A \otimes U^\dagger_B\ket{
2\tilde 2
} \ri \, } \right| = \\ & =\frac{|\dot \lambda(t)|}{\sqrt{1-\lambda(t)^2}} \, ,
\end{aligned}
\ee
\normalsize
\sloppy where the minimization is done over all possible choices of the single non-vanishing $Y^{\mu_1 \nu_{1} \dots \nu_{n_B}}$, and over all the possible LU transformations with respect to an arbitrary fixed orthonormal basis. 
We reach the final expression by observing that 
$\text{Im} \le \bra{1\tilde 1} U_A\otimes U_B \cdot \,\sigma_{\mu_1} \otimes \sigma_{\nu_1} \otimes \dots \otimes \sigma_{\nu_{n_B}} \cdot U_A^\dagger \otimes U_B^\dagger \ket{2 \tilde 2} \ri \, $
is upper bounded by the largest eigenvalue of $\sigma_{\mu_1} \otimes \sigma_{\nu_1} \otimes \dots \otimes \sigma_{\nu_{n_B}} $, which equals $1$.
This shows that the norm is the same as in the two-qubit case in section \ref{ssec:example_2qubits}, and that the presence of ancillary qubits did not improve the complexity. 

\paragraph{Lower bound.}
We can use the maximal entanglement rate studied in \cite{bravyi:2007,van:2013,Verstraete:2016} 
to give lower bounds for the binding state complexity. 
Let us consider a trajectory in the space of states generated by a Hamiltonian $H(t) =H_A(t)+H_B(t)+H_{AB}(t)$,  where $H_A$ and $H_B$ are composed by local generators in the subsystems $A$ and $B$, while  $H_{AB}$ is the interaction term between them. In such a case, an upper bound for the rate of change of the entanglement entropy $S_A$ between the two systems is given by \cite{Verstraete:2016}
\be
\frac{d S_A}{dt} \leq c \log {d} ||H_{AB}(t)|| \, .
\ee
Here $||*||$ is the operator norm (\ie the maximal singular value),  
$c$ is an $\mathcal{O}(1)$ constant (its optimal value has been numerically argued to be $2$ in reference \cite{Verstraete:2016}), $d$ is the dimension of the smaller subsystem out of $A$ and $B$ on which $H$ has support.\footnote{If the Hilbert space of the systems $A$ and $B$ can be factorised as $a\otimes\bar a$ and $b\otimes\bar b$, and $H_{AB}$ has the form of $\mathbf{1}_{\bar a} \otimes H_{ab} \otimes \mathbf{1}_{\bar b} $, then $d = \text{min}\left(\text{dim}(a),\text{dim}(b)\right)$.}
For instance, $d=N_A$ if $H$ contains interactions between any of the qubits, and $d=2$ if it only contains an interaction between a single pair of qubits. 
Since the operator norm of the generators, both in the Pauli and Gell-Mann bases, is equal to $1$, we can use the triangle inequality to get $||H_{AB}||=||\sum_{a,i} Y^{ai}T^A_a \otimes T^B_i||\leq \sum_{a,i} |Y^{ai}| \ ||T^A_a \otimes T^B_i|| = \sum_{a,i} |Y^{ai}|= \mathcal{BF}_{1, \rm hom}$. 
Integrating over time and remembering that the reference state is factorized, so it has zero entanglement entropy, gives
\be \label{Slwrbnds}
\frac{S_A}{c \log{d}} \leq \mathcal{BC}^{\rm state}_1[\ket{\psi_R},\ket{\psi_T}].
\ee
We note that this lower bound is valid for both choices of the basis for the unitary group and for any kind of interactions we allow in $H(t)$, as long as the value of $d$ is changed accordingly. The lower bound can be far from the binding complexity. For example, with the Gell-Mann basis, the binding complexity for the maximally entangled state grows as $\sqrt{N_A}$, while the entanglement entropy is $\log{N_A}$ and therefore for large $N_A$ the bound is far from being saturated.  On the other hand with the Pauli basis, the binding complexity for this state is upper bounded by $\frac{\pi}{4} \log (N_A)$, where $\log(N_A)$ is simply the number of spins, as can be seen from eq.~\eqref{PauliAddit}.
For the Pauli basis, the value of $d$ changes depending on whether we consider the $\mathcal{BC}^{\rm state}_{1, \rm hom}$ or $\mathcal{BC}^{\rm state}_{1, \rm set}$ norms. For $\mathcal{BC}^{\rm state}_{1, \rm hom}$, $d=N_A$, while for $\mathcal{BC}^{\rm state}_{1, \rm set}$, $d=2$ if the allowed interaction has support only on one spin on each side. In section \ref{app:1=2}, we will actually show that $\mathcal{BC}^{\rm state}_{1, \rm hom} = \mathcal{BC}^{\rm state}_{1, \rm set}$ and therefore for the Pauli basis,
\be \label{SlwrbndsDis2}
\frac{S_A}{c} \leq \mathcal{BC}^{\rm state}_{1,\text{Pauli}} [\ket{\psi_T}, \ket{\psi_R}].
\ee
We draw this lower bound in figure \ref{fig:PauliBoundCompare} and highlight in light blue the region between the most restrictive upper and lower bounds.

\subsection{$\mathcal{BF}_{2,\mathrm{hom}}$  norm in any basis}
\label{ssec:F2norm}

In this section, we compute the binding state complexity for a system of two qudits using the cost function $F_{2, \rm hom}$ in eq.~\eqref{eq:norm_hom_after_limit}. 
The $F_{2, \rm hom}$ norm can be expressed as
\be
F_{2, \rm hom} = \sqrt{\sum_{a,i}|Y^{ai}|^2} = \sqrt{\frac{\Tr(Q(H)^2)}{\mathcal{N}_A \mathcal{N}_B}} \, ,
\ee
where $Q(H) = \sum_{a,i} Y^{ai}T^A_a\otimes T^B_i$ is the non-local part of the Hamiltonian, while $\mathcal{N}_{A} = \Tr(T_a^{A}T_b^{A})\delta_{ab}$ and $\mathcal{N}_{B} = \Tr(T_i^{B}T_j^{B})\delta_{ij}$  are the normalization factors for the subsystems $A$ and $B$, respectively. 
Note that the only dependence of the $F_{2, \rm hom}$ norm on the basis of generators comes from an overall normalization, as long as the generators are orthogonal.   Therefore we can use the convenient Gell-Mann basis to compute the norm, and then rescale the result by an appropriate normalization in order to get a general expression valid for any basis.
The normalization of the Gell-Mann basis for arbitrary rank of the unitary group is $\mathcal{N}_A= \mathcal{N}_B = 2$.
Denoting the velocities of the Hamiltonian expanded in the Gell-Mann basis as $Y_G^{ai}$ and performing a convenient rescaling in terms of the normalization of the basis of generators, we find
\be
F_{2, \rm hom} = \sqrt{\frac{4}{\mathcal{N}_A \mathcal{N}_B }\sum_{a,i}|Y_G^{ai}|^2} \, .
\ee
The reference and target states are specified by the Schmidt coefficients in eq.~\eqref{eq:generic_ref_tar_Schmidt}. 
We consider, without loss of generality,  the factorized reference state $\ket{\psi_R}=\ket{1\tilde 1}$ defined in the basis for the Hilbert space chosen in eq.~\eqref{eq:gell}.
The constraint equations to get to the $\mathrm{SO}(N_A)$ picture were previously studied in eqs.~\eqref{eq:cnstGen}-\eqref{eq:constraints_GellMann_diag}. This concludes {\bf steps 1--2} of our procedure outlined in section \ref{ssec:summary_procedure}.

\paragraph{Step 3.} Since this norm is invariant under orthogonal transformations applied to the $Y$-s, we can use the velocities $Y'$ introduced below eq.~\eqref{eq:decomposition_R_GellMann} to find 
\bea \label{BF2hom}
\mathcal{BF}_{2, \rm hom} (R_{mn}) & = \frac{2}{\sqrt{\mathcal{N}_A \mathcal{N}_B }}\min_{Y'} \sqrt{ \sum_{m,n} \le {Y'}_{c(m,n),c(m,n)} \ri^2 } = \\
& = \frac{2}{\sqrt{\mathcal{N}_A \mathcal{N}_B }}\sqrt{ \sum_{m>n} \frac{R_{mn}^2}{2}} = \frac{1}{\sqrt{\mathcal{N}_A \mathcal{N}_B }}\sqrt{\text{Tr}(R^\text{T}R)} \, ,
\eea
where in the second equality we have explicitly performed the minimization over the redundant $Y'$ components. 

\paragraph{Step 4.} 
We define the orthonormal vectors 
\be
\hat r_1 \equiv \vec{\lambda}(t) \, , \qquad 
\hat r_2 \equiv \frac{\dot{\vec{\lambda}}(t)}{\left| \dot{\vec{\lambda}}(t) \right|} \, .
\label{eq:vectors_GramSchmidt_process_old}
\ee
and complete them to an orthonormal basis of $\mathbb{R}^{N_A}$. The frame associated with the new basis is related to the previous one by an orthogonal matrix $O$. In the rotated frame the constraint $R \vec \lambda = \dot{ \vec \lambda}$ fixes the following components of the matrix $\tilde{R}=O R O^T$
\be
\tilde{R}_{m,1} = - \tilde{R}_{1,m} = \delta_{m,2} \sqrt{\dot{\vec{\lambda}}(t)^2} \, .
\label{eq:rotation_GramSchmidt}
\ee
while all the other components of $\tilde R$ belong to the stabilizer group over which we have to minimize the norm to get the state complexity.
The minimum is achieved when the latter components are set to zero, giving 
\begin{equation}
    \begin{split}
\mathcal{BF}_{2, \rm hom}\left(\vec \lambda,\dot {\vec\lambda}\right) = &\frac{1}{\sqrt{\mathcal{N}_A \mathcal{N}_B }}\min_{\rm stab} \sqrt{\text{Tr}(R^\text{T}R)} = \frac{1}{\sqrt{\mathcal{N}_A \mathcal{N}_B }}\min_{\rm stab} \sqrt{\text{Tr}(\tilde{R}^\text{T}\tilde{R})} = \\
= &
\frac{1}{\sqrt{\mathcal{N}_A \mathcal{N}_B }} \sqrt{\tilde{R}_{1,2} \tilde{R}_{1,2}+\tilde{R}_{2,1}\tilde{R}_{2,1}} = 
\sqrt{\frac{2}{\mathcal{N}_A \mathcal{N}_B }} \sqrt{\dot{\vec{\lambda}}(t)^2} \, ,\label{eq:F2_hom_gennom}
\end{split}
\end{equation}
where the ${\rm ``stab"}$ minimization refers to step 4 in section \ref{ssec:summary_procedure}, and the second step in the first line is a consequence of the rotational invariance of the $F_{2, \rm hom}$ norm.
It is implicit that the vector $\vec{\lambda}$ is normalized, and so this defines the Euclidean metric on a unit sphere. 

\begin{figure}[ht]
    \centering
    \includegraphics[scale=0.1]{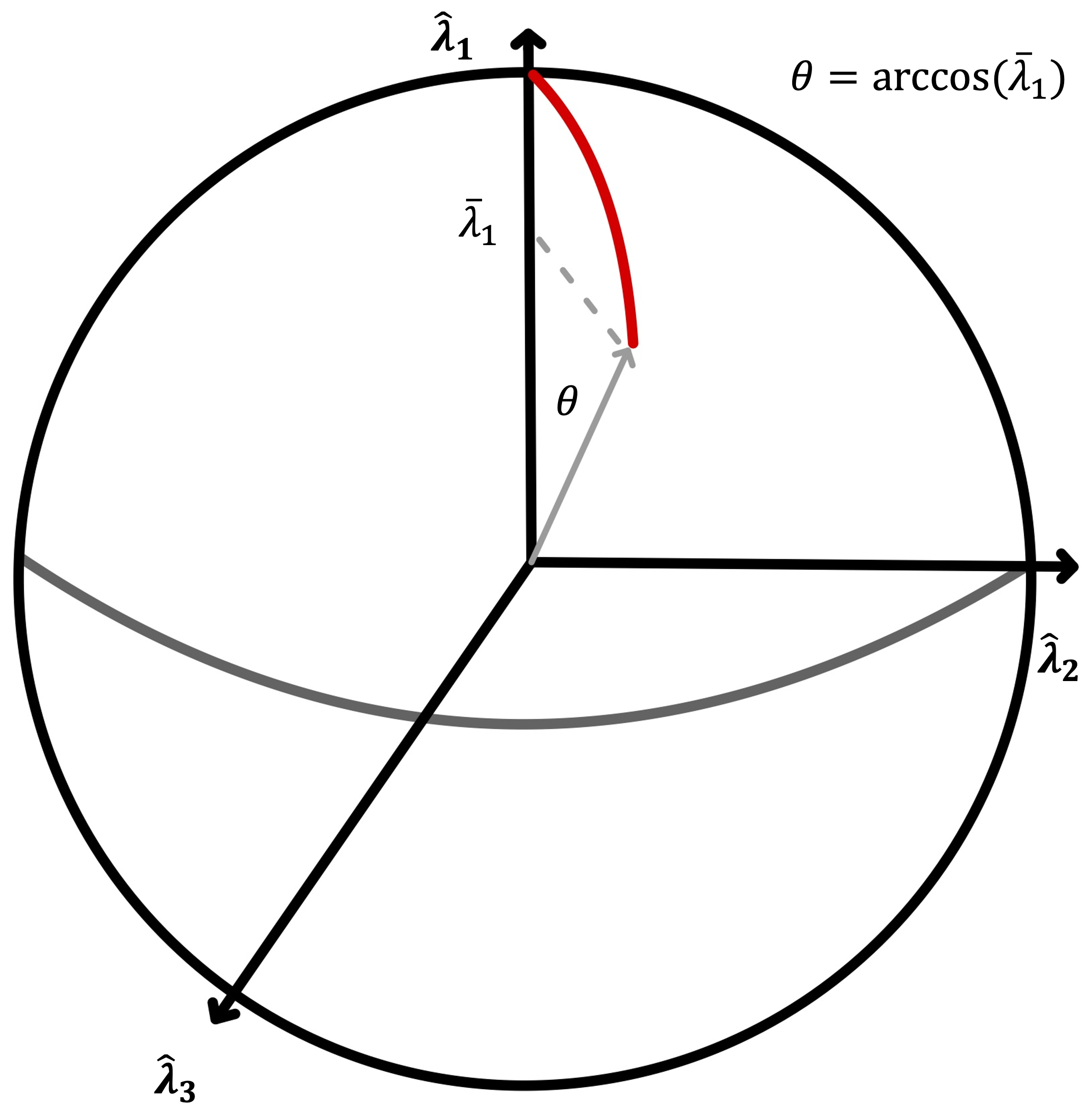}
    \caption{Unit sphere spanned by the vector of Schmidt coefficients, with the vertical axis representing the first direction. The red trajectory is the geodesic connecting the reference and target states in eq.~\eqref{eq:generic_ref_tar_Schmidt}.  }
    \label{fig:sphere_F2norm}
\end{figure}

\paragraph{Step 5.} 
The geodesics are the arcs and the binding state complexity is the arc length from the reference to the target state:
\be
\mathcal{BC}^{\rm state}_{2, \rm hom} = \sqrt{\frac{2}{\mathcal{N}_A \mathcal{N}_B }} \, \mathrm{arccos} (\bar{\lambda}_1) =\sqrt{\frac{2}{\mathcal{N}_A \mathcal{N}_B }} \, \mathrm{arccos} \le e^{-\frac{1}{2}S_{\infty}(\ket{\psi_T})} \ri \, .
\label{eq:bind_compl_GellMann}
\ee
Here $\bar{\lambda}_1$ is the maximal Schmidt coefficient of the target state defined in eq.~\eqref{eq:generic_ref_tar_Schmidt}, while $S_{\infty}(\ket{\psi_T})$ is the minimal Rényi entropy of the reduced density matrix of the target state $\ket{\psi_T}$. Eq.~\eqref{eq:bind_compl_GellMann} is our main result for the case of the $F_2$ norm. 
We note that up to a constant factor due to the normalization, this result is the Fubini-Study distance between the target state and the closest product state, which can be associated with the geometric measure of entanglement  \cite{Chieh:2003}.  Here, in the context of complexity, this distance naturally receives an operational meaning (see \cite{rudnicki:2021} for a related discussion). 
The geometry generated by the $\mathcal{BF}_{2, \rm hom}$ norm and the corresponding geodesic are depicted in fig.~\ref{fig:sphere_F2norm}.

For completeness, let us include the normalization constants for the different bases. For the Gell-Mann basis, the result reads
\be
\mathcal{BC}^{\rm state}_{2, \rm hom} = \frac{1}{\sqrt{2}} \, \mathrm{arccos} (\bar{\lambda}_1) = \frac{1}{\sqrt{2}} \, \mathrm{arccos} \le e^{-\frac{1}{2}S_{\infty}(\ket{\psi_T})} \ri \, ,
\label{eq:bind_compl_GellMann2}
\ee
while for the Pauli basis of a system with $n_A$ and $n_B$ spins the result is
\be
\mathcal{BC}^{\rm state}_{2, \rm hom} = \frac{1}{\sqrt{2^{n_a+n_b-1}}} \, \mathrm{arccos} (\bar{\lambda}_1) = \frac{1}{\sqrt{2^{n_a+n_b-1}}} \, \mathrm{arccos} \le e^{-\frac{1}{2}S_{\infty}(\ket{\psi_T})} \ri \, .
\label{eq:bind_compl_Pauli}
\ee
The two results coincide when $n_a=n_b=1$, as the two bases become the same.

The Schmidt coefficients follow a trajectory along the great circle connecting the Schmidt vectors of the reference and target states. A particular Hamiltonian that realizes this trajectory and has the cost $\mathcal{BC}^{\rm state}_{2, \rm hom}$ is 
	\begin{align}
		H = -i \, \frac{ \mathrm{arccos} \, (\bar{\lambda}_1)}{\sqrt{1-\bar{\lambda}_1^2}} 
			\le \sum_{m=1}^{N_A}  
			\bar{\lambda}_m ( \ket{m\tilde m} \bra{1\tilde 1} - \ket{1\tilde 1} \bra{m\tilde m} \ri
	 \, .
	 \label{eq:min_Ham_F2norm}
	\end{align}
With this Hamiltonian, the basis of the Schmidt decomposition will remain in the computational basis $\ket{i\tilde i}$ until $t=1$, and the state will be related to the exact target state by a final costless LU transformation $\ket{\psi_T} = e^{-i S_0} \sum_{m=1}^{N_A} 
\bar \lambda_m \ket{m\tilde m} = e^{-i S_0}e^{-i H}\ket{1\tilde 1}$, where $S_0$ is a time independent generator of LU transformations (\ie take eq.~\eqref{eq:decomposition_Hamiltonian} with $Y^{ai}=0$). 

Since the geometry associated with binding complexity contains null directions determined by LUE states (see the discussion in section \ref{ssec:reduction_set}), there is a degeneracy in the unitaries that take the reference state to the target state with minimal cost.
Specifically for the case of the $F_2$ norm, it is possible to find a continuous trajectory that does not require a final LU transformation.
We determine such geodesic in appendix \ref{app:geodesics_EA} by using the Euler-Arnold equations, and we show an alternative way to compute its cost \eqref{eq:bind_compl_GellMann} in appendix \ref{app:binding_compl_appendix}.

A possible extension of the $F_{2,\rm hom}$ norm to a multipartite system associates vanishing cost to each local transformation, and equal cost to any non-local generator. Related notions were considered in \cite{rudnicki:2021,Chieh:2003}.
We can speculate, extrapolating from the above bipartite case, that the state complexity of such generalization will be given in terms of the Fubini-Study distance to the closest multipartite pure factorized state.
Notice that the above-mentioned extension will differ from the multipartite notion of complexity that will be discussed in section \ref{sec:geometrically_local_complexity} because there, we will only allow non-local generators to act on a maximum of two subsystems at the same time.

\subsection{Insensitivity of $\mathcal{BF}_1$ binding complexity to penalty factors of non-local gates} 
\label{app:1=2}

We have obtained two results that give the binding state complexity for the $F_{1,{\rm hom}}$ and $F_{2,{\rm hom}}$ norms as two different functions of the Schmidt coefficients, see eqs.~\eqref{eq:result_stateF1_hom_GellMann} and \eqref{eq:bind_compl_GellMann}. One may be tempted to conjecture that by a suitable assignment of penalty factors, it would be possible to obtain an arbitrary function of the Schmidt coefficients.\footnote{We thank Yaron Oz for raising this question.} We will now show that this is not the case, at least for the  $\mathcal{BC}_1$ complexity. On the contrary, the
binding complexity computed according to the $\mathcal{BF}_1$ norm
is rather insensitive to the precise choice of penalty factors for the non-local gates.\footnote{The insensitivity to penalty factors does not apply to other $\mathcal{BC}_{p>1}$ complexities, as we explain below.}
More precisely, we will argue that $\mathcal{BC}_1$ does not depend on the choice of penalties assigned to the non-local generators, up to an overall normalization stemming from the value of the minimal penalty, as long as the generators forming the optimal trajectory all lie on the same adjoint orbit of the $\mathrm{SU}(N_A)\otimes \mathrm{SU}(N_B)$ action as the cheapest generator. This condition is satisfied for the Gell-Mann and Pauli bases. The logic behind this is that if LU transformations are free, one could use them together with the cheapest non-local gate to create any other non-local gate.

To prove this statement, let us consider the norm
	\be
F_{1,\text{int}}[\vec{Y}]  = \lim_{\substack{\varepsilon \to 0 \\ q_{ai} > 1}} \le \sum_{a} \varepsilon |Y^a_A| + \sum_{i} \varepsilon |Y^i_B|
+\sum_{(a,i) \ne (1,\tilde{1})} q_{ai} |Y^{ai}| + |Y^{11}| \ri  \, ,
\label{eq:penalty_intermediate}
\ee
which represents an intermediate case between the homogeneous norm \eqref{eq:Fpenalties_flat} and the norm for a universal set of gates \eqref{eq:penalty_uni}.
All three norms assign a vanishing cost to local operations, but the penalties $q_{ai}$ associated with non-local generators are 1 for $F_{1,\rm hom}$, $q_{ai}>1$ finite for $F_{1,\rm int}$ and $q_{ai} = \infty$ for $F_{1, \rm set}.$
Since complexity increases with the penalties, we conclude that the binding state complexities associated with the three cost functions satisfy
\be
\mathcal{BC}^{\rm state}_{1, \rm hom} \leq 
\mathcal{BC}^{\rm state}_{1, \rm int} \leq
\mathcal{BC}^{\rm state}_{1, \rm set} \, .
\label{eq:inequality_binding_complexities}
\ee
We are going to show that these bounds are saturated, \ie that all these complexities are the same as long as the conditions specified at the beginning of this subsection hold.

To make this argument formal, let us first notice that we can express a generic trajectory in the $\mathrm{SU}(N_A N_B)$ group manifold as a circuit of infinitesimal steps
\be \label{OptCircuit}
U = \mathcal{\cev{P}} \exp \le -i \int_0^1 dt' H(t') \ri \, = \lim_{M \to \infty} \prod_{m=1}^{M} \exp \le -\frac{i}{M} H\le \frac{m}{M} \ri \ri \, ,
\ee
where $H(t) = Y^a_A (t) T^A_a \otimes \mathbf{1}^B +Y^i_B (t) \mathbf{1}^A \otimes T^B_i + Y^{ai} (t) T^A_a \otimes T^B_i $.
 Let us now consider a different circuit that creates the same unitary, in which we break the gate $\exp \le -\frac{i}{M} H\le \frac{m}{M} \ri \ri$ into $(N_A N_B)^2-1$ consecutive gates
    \be \label{OptBrokCircuit}
    U = \lim_{M \to \infty} \prod_{m=1}^{M} \prod_{a} e^{ -\frac{i}{M} Y^{a}_A \le \frac{m}{M} \ri  T^A_a \otimes \mathbf{1}^B } \prod_{i} e^{ -\frac{i}{M} Y^{i}_B \le \frac{m}{M} \ri  \mathbf{1}^A \otimes T^B_i } \prod_{a, i} e^{ -\frac{i}{M} Y^{a i} \le \frac{m}{M} \ri  T^A_a \otimes T^B_i } \, ,
    \ee
where the product over $a$ and $i$ is from $1$ to $N_A^2-1$ and $N_B^2-1$ respectively, and we used the Baker-Campbell-Hausdorff formula in the limit $M\rightarrow \infty$.
Notice that the cost of the circuits \eqref{OptCircuit} and \eqref{OptBrokCircuit} is the same according to the $F_{1,\text{hom}}$ norm, but is a priori different using the $F_{p>1,\text{hom}}$ norm.\footnote{For this reason, the arguments presented in this subsection only apply to the $\mathcal{BF}_{1}$ norm.} This is because the cost of the circuit  \eqref{OptBrokCircuit} is $\sum_{m,a,i} |Y^{a i} \le \frac{m}{M} \ri|$ regardless of $p$, as the Hamiltonian is composed out of a single generator at any given time step.

We now assume that the circuit in \eqref{OptCircuit} is the optimal trajectory obtained using the $F_{1,\text{hom}}$ norm. Let us construct yet another circuit that creates the same unitary and has the same cost. This circuit will only use a single non-local generator $ T^A_{1} \otimes T^B_{1}$, thus its cost will be the same whether the norm is $F_{1,\text{hom}}$ or $F_{1,\text{set}}$.
To begin with, let us assume that for every non-local generator $T^A_a\otimes T^B_i$, there is an LU transformation such that
\be\label{eq:ourassumption}
{U^A_a}^\dagger T^A_{1} {U^A_a} \otimes {U^B_b}^\dagger T^B_{1} {U^B_b} =  T^A_a\otimes T^B_i.
\ee
Notice that this statement is about the individual systems, \ie there exists an LU transformation that takes a $T_1^{A/B}$ to any other generator $T_I^{A/B}$.
The construction is now straightforward: take the circuit in \eqref{OptBrokCircuit} and replace any occurrence of a non-local generator with the preferred non-local generator $T_1^A\otimes T_1^B$\footnote{It would be sufficient to show this only for the generators forming the optimal trajectory, but since we do not know the optimal trajectory in the Pauli basis, showing it for all generators will be the strategy there.}
 \be
    U = \lim_{M \to \infty} \prod_{m=1}^{M} \prod_{a} e^{ -\frac{i}{M} Y^{a}_A \le \frac{m}{M} \ri  T^A_a } \prod_{i} e^{ -\frac{i}{M} Y^{i}_B \le \frac{m}{M} \ri  T^B_i } \prod_{a, i} {U^A_a}^\dagger {U^B_i}^\dagger e^{ -\frac{i}{M} Y^{a i} \le \frac{m}{M} \ri  T^A_1 T^B_1 } {U^A_a} {U^B_i}\, ,
    \ee
where we have omitted the tensor product notation for convenience. 
Since the cost of the LU transformations is zero, the cost of this circuit will be the same regardless of whether we used the $F_{1,\text{hom}}$ or $F_{1,\text{set}}$ norm.
Due to the inequalities \eqref{eq:inequality_binding_complexities}, we conclude that for $p=1$ all these binding complexities are the same, in particular this is true for $\mathcal{BC}^{\rm state}_{1,\rm int}$, which assigns arbitrary penalties $q_{ai}>1$ to the non-local generators. 

We are left to discuss the validity of our assumption \eqref{eq:ourassumption}. For the Pauli basis, take, without loss of generality, the single allowed generator to be $\sigma^{A,1}_x \sigma^{B,1}_x$. 
We show below that we can create any generator out of the set of generalized Pauli basis with $\sigma^{A,1}_x \sigma^{B,1}_x$ and an LU transformation. Let us focus on a system of $n$ spins and ask whether there is a $U \in \text{SU}(2^n)$ such that ${U}^\dagger \sigma^{1}_x U = \sigma^{1}_{\mu_1} \sigma^{2}_{\mu_2} ... \sigma^{n_a}_{\mu_{n}}$, for any sequence of $\mu_i$s except for all zeros, where the superscript marks the site, and the subscript marks which Pauli matrix is being considered (cf. the text below eq.~\eqref{eq:definition_Pauli_matrix}).  
Since we have not made any assumption about the number of spins, this argument will be valid for both systems $A$ and $B$.  Let us define $\bar \sigma = \sigma^{2}_{\mu_2} ... \sigma^{n}_{\mu_{n}}$. If $\mu_1 = y \text{ or } z$, then
\be
e^{-i \frac{\pi}{4} \sigma^{1}_z \bar \sigma} \sigma^{1}_x \text{e}^{i \frac{\pi}{4} \sigma^{1}_z\bar \sigma} = \sigma^{1}_y \bar \sigma, \ \ \ \ \ 
e^{i \frac{\pi}{4} \sigma^{1}_y \bar \sigma} \sigma^{1}_x e^{-i \frac{\pi}{4} \sigma^{1}_y \bar \sigma} = \sigma^{1}_z \bar \sigma.
\ee
To reach $\sigma^{1}_x \bar \sigma$, we can first perform the transformation to $\sigma^{1}_y \bar \sigma$ and then complement it with an additional LU transformation $\text{e}^{i \frac{\pi}{4} \sigma^{1}_z } \sigma^{1}_y \bar \sigma \text{e}^{-i \frac{\pi}{4} \sigma^{1}_z} = \sigma^{1}_x \bar \sigma$. To reach $\mathbf{1}^{1} \bar \sigma$, we can use the swap $S= \text{exp} \le {-i}\frac{\pi}{4}(\sigma^{1}_x \sigma^{2}_x+\sigma^{1}_y \sigma^{2}_y+\sigma^{1}_z \sigma^{2}_z) \ri$, which gives $S^\dagger \sigma^{1}_x S=\sigma^{2}_x$, and then recursively use similar transformations on the subset of $n-1$ remaining spins to reach $\bar \sigma$ from $\sigma^{2}_x$.
This concludes the argument for the equivalence of $\mathcal{BC}^{\rm state}_{1, \rm hom} = 
\mathcal{BC}^{\rm state}_{1, \rm set}$ in the Pauli basis.

For the Gell-Mann basis, one cannot hope to create from one generator and an LU transformation any other generator, as the diagonal and off-diagonal generators have in general different eigenvalues. 
If the single non-local generator is a tensor product of two off-diagonal generators $ T^A_{c(n,m)} \otimes T^B_{c(k,p)}$ where $n \neq m, k \neq p$, then all the other off-diagonal generators can be reached with an LU transformation. To demonstrate that, we note the following relations for $T_{c(2,1)}$, where $p>k>2$:
\begin{subequations}
    \bea
e^{-i\frac{\pi}{2}T_{c(p,2)}} T_{c(2,1)}  e^{i\frac{\pi}{2}T_{c(p,2)}} = T_{c(p,1)} \, , \qquad
e^{-i\frac{\pi}{2}T_{c(2,p)}} T_{c(2,1)}  e^{i\frac{\pi}{2}T_{c(2,p)}} = T_{c(1,p)} \, , 
\\
e^{i\frac{\pi}{2}T_{c(p,1)}} T_{c(2,1)} e^{-i\frac{\pi}{2}T_{c(p,1)}} = T_{c(p,2)} \, , \qquad
e^{i\frac{\pi}{2}T_{c(1,p)}} T_{c(2,1)}  e^{-i\frac{\pi}{2}T_{c(n,p)}} = T_{c(2,p)} \, ,
\eea
\bea
e^{i\frac{\pi}{2}T_{c(2,p)}} e^{i\frac{\pi}{2}T_{c(1,k)}} T_{c(2,1)} & e^{-i\frac{\pi}{2}T_{c(1,k)}}e^{-i\frac{\pi}{2}T_{c(2,p)}} = T_{c(p,k)} \, , \\
e^{i\frac{\pi}{2}T_{c(2,p)}} e^{i\frac{\pi}{2}T_{c(k,1)}} T_{c(2,1)} & e^{-i\frac{\pi}{2}T_{c(k,1)}}e^{-i\frac{\pi}{2}T_{c(2,p)}} = T_{c(k,p)}.
\eea
\end{subequations}
These relations show that any off-diagonal generator can be reached from $T_{c(2,1)}$ by an LU transformation. This is not special to $T_{c(2,1)}$, since from the above identities we learn that all the off-diagonal generators are on the same adjoint orbit.

Fortunately, the geodesic circuit with the Gell-Mann basis, which is generated by the Hamiltonian in eq.~\eqref{eq:optimal_Ham_F1GellMann}, only uses off-diagonal generators. Thus, the previous argument is still valid if we limit the generality of the allowed single non-local generator such that it is off-diagonal. 
For the geodesic obtained using the Gell-Mann basis, the associated circuit in eq.~\eqref{OptBrokCircuit} will take the form 
    \be
    U = \lim_{M \to \infty} \prod_{m=1}^{M} \prod_{p>1}^{N_A} \exp \le -\frac{i}{M} R_{p1} \le \frac{m}{M} \ri T^A_{c(p,1)} \otimes T^B_{c(1,p)}  \ri \, .
    \ee   
For instance, if the allowed single non-local generator is $ T^A_{c(2,1)} \otimes  T^B_{c(2,1)}$, we can construct the circuit as 
\be
U = \lim_{M \to \infty} \prod_{m=1}^{M} \prod_{p>1}^{N_A} e^{-i\frac{\pi}{2} \left[ T^A_{c(p,2)}\otimes \mathbf{1}^B + \mathbf{1}^A \otimes T^B_{c(2,p)} \right]} e^{ -\frac{i}{M}  R_{p1}  \le \frac{m}{M} \ri T^A_{c(2,1)} \otimes T^B_{c(2,1)} } e^{i\frac{\pi}{2} \left[ T^A_{c(p,2)}\otimes \mathbf{1}^B + \mathbf{1}^A \otimes T^B_{c(2,p)} \right] } \, .
\label{eq:circuit_sec5}
  \ee
We still need to discuss an additional subtlety. The derivation given above applies when the local generators have exactly zero cost. However, in any realistic experimental set-up, the penalty assigned to local generators is $\varepsilon \ne 0$, in which case the circuit \eqref{eq:circuit_sec5} would have divergent cost in the limit $M \rightarrow \infty$. 
However, it is possible to remedy this issue by considering the unitary \eqref{eq:circuit_sec5}, but with finite $M$. If the additional cost and the error coming from the Trotterization of a Hamiltonian \cite{Lloyd:1996,Childs:2021} can be kept small, this will describe an efficient approximation for the target unitary. To estimate the error and cost of such a circuit, consider a simple case in which the Hamiltonian \eqref{OptCircuit} is constant. This is the form of the geodesic in the Gell-Mann basis if all but the largest Schmidt coefficients are equal. The Trotterization error can be estimated as $\mathcal{O}(\frac{\mathcal{BC}^2}{M})$ (compare to eq.~(16) in \cite{Childs:2021}),
and the extra cost to the binding complexity will be $ \mathcal{O}(M N_A \varepsilon)$. Depending on the engineer's budget, states with small enough $\mathcal{BC}$ could be reached to a good approximation. Furthermore, as long as $\varepsilon$ is small enough, the cost of the above circuit will provide a good estimate of the states' complexity.

\section{Binding complexity of two qutrits and complexity of a single qubit}
\label{ssec:example_2qutrits}

The simplest system on which one can ask questions of Nielsen's complexity is a single qubit.  
However, much less is known for the single-qubit complexity with an $F_1$ norm. 
It turns out that the results we obtained earlier in this paper for the binding complexity can be applied to this problem, too.
More specifically, we will show that the binding complexity for a system of two qutrits in the Gell-Mann basis is related to the $F_1$ Nielsen's complexity for a single qubit.

\begin{figure}[ht]
    \centering
    \includegraphics[scale=0.1]{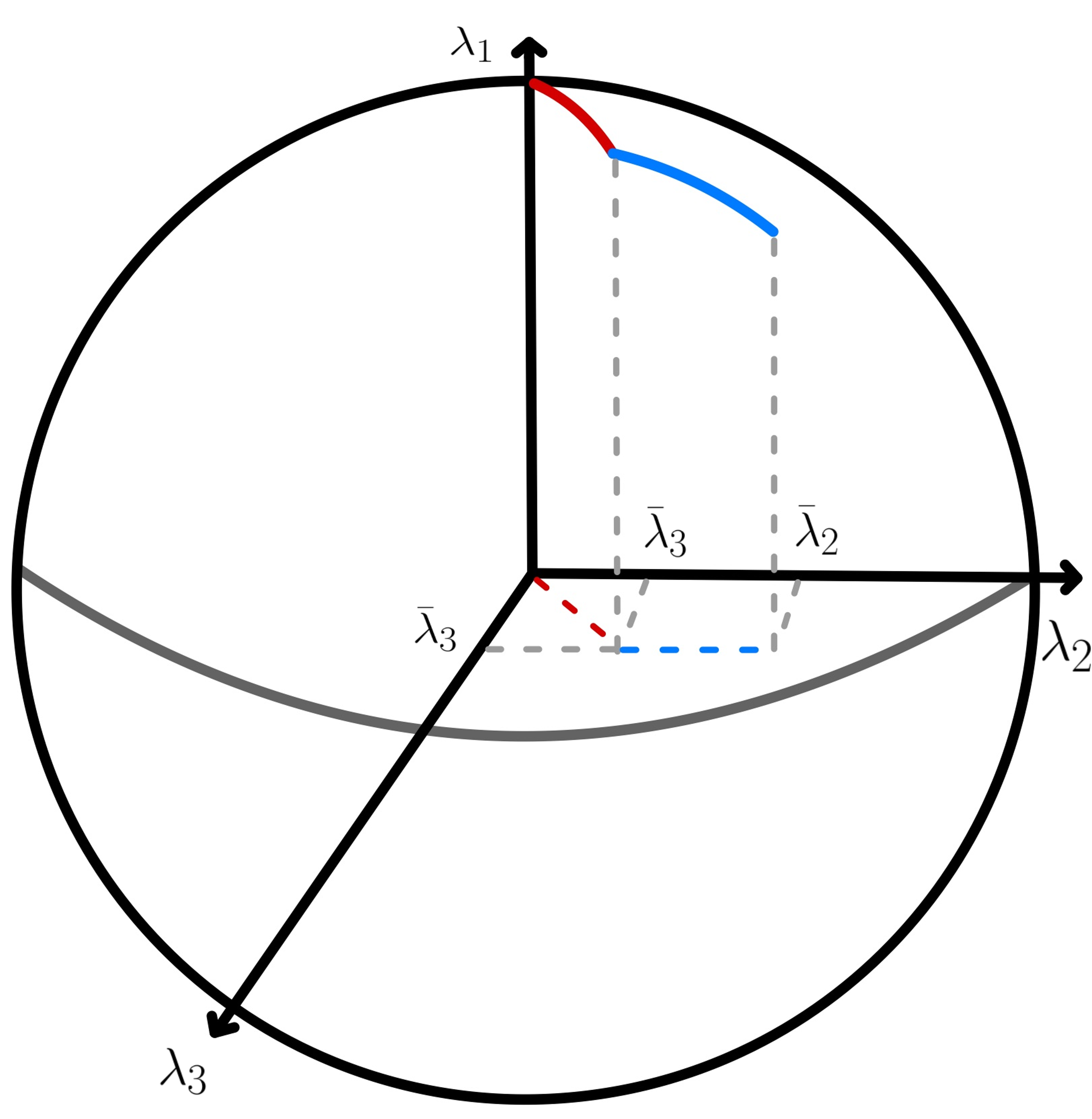}
    \caption{Piecewise trajectory \eqref{eq:piecewise_trajectory_GellMann} in the space of Schmidt coefficients for the case of two qutrits. Binding complexity is given by the sum of the lengths of the red and blue curves in this geometry.}
    \label{fig:F1norm}
\end{figure}

In the case of two qutrits, we expand the $\mathrm{SO}(3)$ generator $R$ introduced in eq.~\eqref{eq:RcnstrainH} in terms of the rotations around each axis 
	\be \label{Rso3}
	R(t) = v_1(t) L_1+v_2(t) L_2 +v_3(t) L_3 \, ,
	\ee
where $(L_i)_{jk} = \epsilon_{ijk}$ with $\epsilon$ the Levi-Civita symbol. 
The relation to the Gell-Mann generators introduced in eq.~\eqref{eq:gell} is $L_1 = i (T^y)_{32}, \ L_2 = - i (T^y)_{31}, \ L_3 = i (T^y)_{21}$. 
According to the result \eqref{eq:F1ofR}, 
in the Gell-Mann basis, the minimization yields the norm
\be \label{eq:F1qtr}
\mathcal{BF}_{1, \rm hom} (R_{mn})= |v_1(t)| + |v_2(t)|+|v_3(t)| \, .
\ee
When $\lambda_1 \geq \lambda_2+\lambda_3$, as we have shown in eq.~\eqref{eq:distance_F1norm_GellMann}, the metric can be written in terms of the Schmidt coefficients as  
	\be \label{qutritMet}
	\mathcal{BF}_{1, \rm hom} \, dt = \frac{|d\lambda_2|+|d\lambda_3|}{|\lambda_1|},
	\ee
and the optimal trajectory from $\vec{\lambda}(0)=(1,0,0)$ to $\vec{\lambda}(1)=(\bar{\lambda}_1,\bar{\lambda}_2,\bar{\lambda}_3)$, depicted in fig.~\ref{fig:F1norm}, has the following cost determined according to eq.~\eqref{eq:result_stateF1_hom_GellMann}:
	\be \label{BCqutrits}
	\mathcal{BC}^{\rm state}_{1,\rm hom} = \sqrt{2} \, \mathrm{arccos} \, \le \sqrt{1-2 \bar{\lambda}_3^2}\ri +\mathrm{arccos} \, \le \frac{\bar{\lambda}_3\bar{\lambda}_2+\bar{\lambda}_1 \sqrt{1-2\bar{\lambda}_3^2} }{1-\bar{\lambda}_3^2}\ri  \, .
	\ee
For the maximally entangled state, the complexity is $\sqrt{2} \, \mathrm{arccos} \, (1/\sqrt{3})$.

Since the $\mathfrak{so}(3)$ and $\mathfrak{su}(2)$ algebras are isomorphic, and since the nature of the complexity problem is completely algebraic, eq.~\eqref{eq:F1qtr} can also be interpreted as the unitary (\emph{not} binding) complexity norm for a single qubit using the basis of Pauli matrices. This is because a trajectory in SU($2$) can be written as
\be
U(t) = \mathcal{\cev{P}} \exp( -i \int_0^t dt' H(t')) \, , \qquad
H(t') = v_1(t') \sigma_1 +v_2(t') \sigma_2 +v_3(t') \sigma_3 \, ,
\ee 
for which the norm $F_1 = \sum_i|\Tr(\dot U dU \sigma_i)|/2$ is equal to \eqref{eq:F1qtr}. The density matrix of a single spin in a pure state can be written as 
\be
\rho = \ket{\phi}\bra{\phi} = \frac{1}{2}(\mathbf{1} + x_1 \sigma_1 + x_2 \sigma_2 +x_3 \sigma_3) \, , \qquad
\mathrm{where} \,\, |\vec{x}|=1 \, .
\ee
The vector $\vec x$ is real, while the condition $|\vec x|=1$ defines the Bloch sphere and ensures that the state is pure.  
The six points $x_i= \pm 1$ correspond to the eigenstates of the Pauli matrices $\sigma_i$. To some extent, as we discuss below, the complexity problems can be mapped to each other by identifying the vector of Schmidt coefficients $\vec \lambda$ with the position on the Bloch sphere $\vec x$. Starting from the equations of motion $\dot \rho = -i[H,\rho]$, one can find the state complexity norm by applying similar methods to those discussed in section \ref{ssec:F1norm} (see eq.~\eqref{eq:EoMtil} with the map $ R_{32} =2 v_1$, $ R_{31} =-2 v_2$, $R_{21} =2 v_3$). 
The minimization over the stabilizer group with respect to $\rho$ implies that either $v_1,v_2$ or $v_3$ vanish, giving
\be \label{F1spin}
F_1 dt = \frac{1}{2} \text{min} \le \frac{|dx_2|+|dx_3|}{|x_1|}, \frac{|dx_3|+|dx_1|}{|x_2|},\frac{|dx_1|+|dx_2|}{|x_3|} \ri.
\ee
Certain points on the sphere are related by LU transformations, which are free in the framework of binding complexity. Therefore, any reference and target states can be mapped to the region $\lambda_1\geq \lambda_2 \geq \lambda_3 \geq 0$. 
For the complexity of a single qubit, we do not have this freedom. Nevertheless, the norm in \eqref{F1spin} is invariant under a sign flip of each coordinate, and under permutations of the coordinates. 
Therefore, we can map any reference state which is an eigenstate of one of the Pauli matrices to $(x_1=1,x_2=0,x_3=0)$, and any target state to another state that has $\bar x_2\geq \bar x_3 \geq 0$. It is then possible to map the single qubit complexity to the binding complexity of two qutrits if the inequality $\bar x_1 \geq \bar x_2$ is also satisfied. 
Since we could only compute the binding complexity for two qutrits when $\bar \lambda_1 \geq \bar \lambda_2 +\bar \lambda_3 $, then the result in eq.~\eqref{BCqutrits} corresponds to twice the complexity of a single qubit in the regime $\bar x_1 \geq \bar x_2+\bar x_3$, provided that we identify $\bar \lambda_i=\bar x_i$, that is
	\be \label{BCqutrits2qubit}
	\mathcal{C}^{\rm state}_{1,\rm qubit} = \frac{1}{\sqrt{2}} \, \mathrm{arccos} \, \le \sqrt{1-2 \bar{x}_3^2}\ri +\frac{1}{2}\mathrm{arccos} \, \le \frac{\bar{x}_3\bar{x}_2+\bar{x}_1 \sqrt{1-2\bar{x}_3^2} }{1-\bar{x}_3^2}\ri  \, .
	\ee
This expression is directly applicable when the reference state is an eigenstate of the Pauli matrix  $\sigma_1$. More generally, when the reference state is an eigenstate of $\sigma_i$ for some given $i$, the state complexity of the single qubit is given by half of eq.~\eqref{BCqutrits}, with $\bar \lambda_1=|\bar x_i|$, $\bar \lambda_2= \max(|\bar x_j|,|\bar x_k|)$ and $\bar \lambda_3=\min(|\bar x_j|,|\bar x_k|)$, where $j,k$ label the other two directions in the Bloch sphere. This solution is valid as long as the inequality $|\bar x_i| \geq |\bar x_j|+|\bar x_k|$, defining two separate connected regions on the sphere (around the positive and negative $i$-th axis), is satisfied, and as long as the reference and target states both belong to the same region out of the two.

It is interesting to notice that the minimization over the stabilizer group, which yielded the norm for the state complexity, amounts to setting one of the velocities $v_i$ to zero (see eq.~\eqref{eq:somevarezero}). In the context of the complexity of a single qubit, this gives a trajectory in SU($2$) that does not use the $\sigma_i $ Pauli matrix associated with the velocity $v_i$. Thus, giving $\sigma_i$ a larger penalty will not change the results. It was argued in \cite{Brown:2019whu} that in certain experimental two-level systems the $\sigma_z$ gate is much harder to implement (see their discussion below eq.~(2.4)). In order to estimate the complexity, the authors therefore proposed to use a Riemannian $F_2$ norm with a very large penalty assigned to the $\sigma_z$ direction. In our results we see that we can alternatively use the $F_1$ cost function with reference state $\ket{\uparrow}$ and the same effect of suppressing the use of the $\sigma_z$ gate will be achieved, even without the use of penalty factors. We expect that a similar effect will come into play for larger systems and intend to investigate this issue further in the future.

\section{Geometrically local complexity and multipartite binding complexity}
\label{sec:geometrically_local_complexity}

\begin{figure}[ht]
    \centering
\subfigure[Local]{ \includegraphics[scale=0.2]{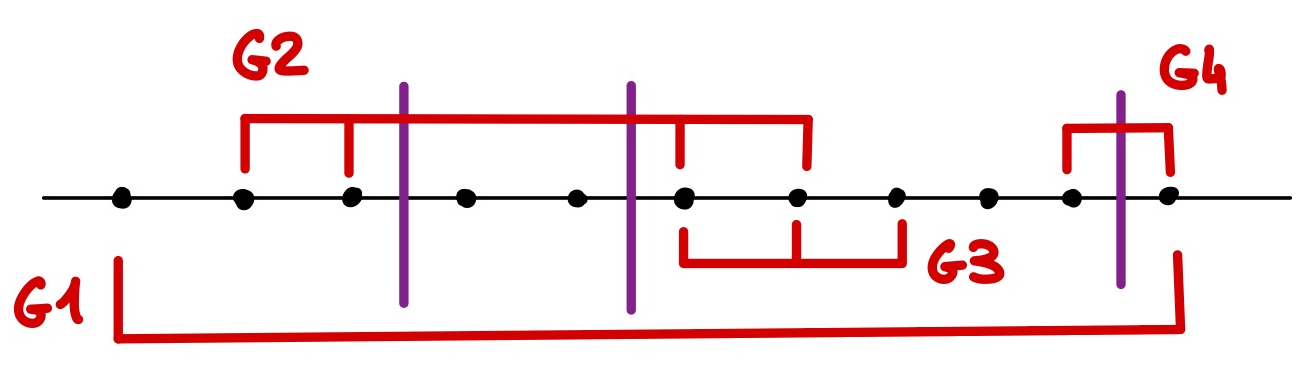} } \quad
\subfigure[Geometrically-local]{ \includegraphics[scale=0.2]{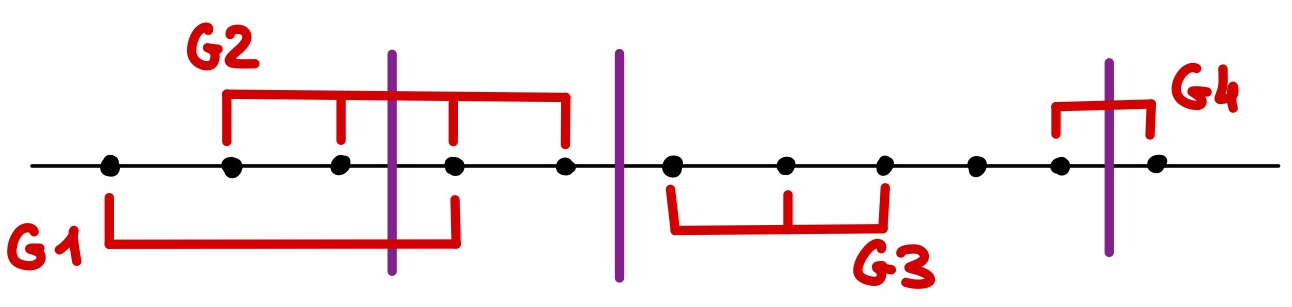} }
    \caption{Pictorial representation of a spin chain where the vertical purple lines represent a separation between parts. 
    (a) Circuit with gates acting on at most two parts of a system.
    (b) Circuit with gates acting only on nearest-neighbour parts or within each part. 
    Both circuits can be used to study local complexity \eqref{eq:Clocal}, but only case (b) is a circuit that can be used to study geometrically-local complexity \eqref{eq:Cg_hom}. When the penalties associated with operations within a subsystem vanish (in this example, gate $G_3$ for both circuits), the definitions reduce to multipartite binding complexity, see eqs.~\eqref{eq:def_multipartite_complexityA} and \eqref{eq:def_multipartite_complexity}, respectively. }
    \label{fig:geom_local_compl}
\end{figure}

We now consider a multipartite system with a factorizable Hilbert space of the form \eqref{multiH}.
As we mentioned below eq.~\eqref{multiH}, oftentimes one of the constraints imposed on complexity (even outside the context of binding complexity)
is that only generators between pairs of qubits are allowed. 
This type of locality constraint plays an important role in circuit complexity in realistic quantum computation scenarios, where many-qubit gates are typically harder to implement.
Motivated by these considerations, in this section, we study Nielsen's state complexity according to the $F_1$ norm in eq.~\eqref{eq:cost_function_Finsler}, using the following nomenclature adopted in \cite{Eisert:2021}:
\begin{itemize}
    \item \textit{Local complexity} $\mathcal{C}^{\rm state}_{1,l}$ allows gates acting on one part or any pair of parts of the system. 
    \item \textit{Geometrically-local complexity} $\mathcal{C}^{\rm state}_{1,g}$ allows gates acting only on nearest-neighbour parts or within each part.
\end{itemize}
These notions of complexity are depicted in fig.~\ref{fig:geom_local_compl}. 
As a side comment, let us mention that the authors of \cite{Fu:2018kcp} have noted that the notion of geometrically-local complexity seems incompatible with the holographic notions of complexity. 
On the other hand, many works have taken locality (as opposed to geometrical-locality) as a guiding principle in looking for a quantum mechanical dual of holographic complexity, see \eg \cite{Brown:2017jil,Balasubramanian:2018hsu,Brown:2019whu,Balasubramanian:2019wgd}. 

In \cite{Eisert:2021}, the authors derived lower bounds on the notion of local and geometrically-local complexity in terms of the entanglement entropies of different bipartitions of the target state.
We will now use the techniques developed in this work to derive new lower bounds, in terms of binding complexity, which improve the lower bounds of \cite{Eisert:2021}.

Let us define the set-up.
The system is split into $n$ identical parts, each of them has an $N$-dimensional Hilbert space on which the special unitary space $\mathrm{SU}(N)$ can act. Therefore, the possible circuits in the composite system belong to $\mathrm{SU}(N^n)$.\footnote{This applies, for example, to a subdivision of a spin chain to $n$ equal parties, where $N = 2^{\#\text{ spins}}$. } 
We denote the set of generators for each part as $T^{A}_i$, where the subscript $i$ runs over the list of generators $\lbrace 1, \dots, N^2 -1 \rbrace$ within each subsystem, while the superscript $A$ refers to the site on which they act (\ie a generator $T^{A}_i$ is the identity on all sites except for the $A$-th site).\footnote{The notation has changed from the previous sections in which $A$ denoted one of the two parties $A$ and $B$, and not a running index.} The full set of generators for the system is made by taking products of these $T_i^A$ for the different parts. 
Suppose that we construct our control Hamiltonian and unitary circuits from  gates which act on at most two parties, that is
\bea
 H_l(t)  & = \sum^{n}_{A=1}\sum^{N^2-1}_{i=1}Y^{A}_{i} T^{A}_{i} 
+ \sum^{n-1}_{A=1} \sum^{n}_{B=A+1}   \sum^{N^2-1}_{i,j=1}Y^{A B}_{i j} T^{A}_{i} T^{B}_j
\, , \qquad t\in [0,1], 
\\ 
H_g(t) & = \sum^{n}_{A=1}\sum^{N^2-1}_{i=1}Y^{A}_{i} T^{A}_{i} 
+  \sum^{n-1}_{A=1}  \sum^{N^2-1}_{i,j=1}Y^{A  A+1}_{i j} T^{A}_{i} T^{A+1}_j
\, , \qquad t\in [0,1],
\label{eq:control_Ham}
\eea
where the two Hamiltonians differ by the restriction of the non-local generators of $H_g(t)$ to nearest-neighbour-parts interactions.
We will denote our reference and target states as $\ket{\psi_R},\ket{\psi_T}$.
The notions of complexity introduced in this section correspond to 
\begin{itemize}
    \item The local complexity 
    \be
\mathcal{C}^{\rm state}_{1,l, \rm hom} [\ket{\psi_T},\ket{\psi_R}] = \min_{H_l(t)}
\int_0^1 dt \le \sum^{n}_{A=1}\sum^{N^2-1}_{i=1}|Y^{A}_{i}| +
 \sum^{n-1}_{A=1} \sum^{n}_{B=A+1}  
 \sum^{N^2-1}_{i,j=1} |Y^{A B}_{i j}| \ri \, .
\label{eq:Clocal}
\ee
\item The geometrically-local complexity
    \be
\mathcal{C}^{\rm state}_{1,g,\rm hom} [\ket{\psi_T},\ket{\psi_R}] = \min_{H_g(t)}
\int_0^1 dt \le \sum^{n}_{A=1}\sum^{N^2-1}_{i=1}|Y^{A}_{i}| +
 \sum^{n-1}_{A=1}  \sum^{N^2-1}_{i,j=1} |Y^{A,A+1}_{i j}| \ri \, .
\label{eq:Cg_hom}
\ee
\end{itemize}

\noindent
In both cases, the minimization is performed over all the control Hamiltonians of the form \eqref{eq:control_Ham} which connect the reference and target states. 
We also added the subscript ``hom'' referring to the homogeneous choice of penalties equal to unity for all the generators (including the local ones).

It is useful for the following to introduce also a multipartite notion of binding complexity $\mathcal{MBC}[\ket{\psi_T}, \ket{\psi_R}]$.
This can be defined as a special case of the (geometrically-) local complexity, when the penalty factors  assigned to local gates vanish: 
\begin{subequations}
\be\label{eq:def_multipartite_complexityA}
\mathcal{MBC}^{\rm state}_{1, l, \rm hom} [\ket{\psi_T},\ket{\psi_R}] = 
\min_{H_l(t)}
\int_0^1 dt 
 \sum^{n-1}_{A=1} \sum^{n}_{B=A+1}  \sum^{N^2-1}_{i,j=1} |Y^{A B}_{i j}|, 
\ee
\be
\mathcal{MBC}^{\rm state}_{1, g, \rm hom} [\ket{\psi_T},\ket{\psi_R}]  =  \min_{H_g(t)}
\int_0^1 dt 
\sum^{n-1}_{A=1}  \sum^{N^2-1}_{i,j=1} |Y^{A, A+1}_{i j}| \, . 
\label{eq:def_multipartite_complexity}
\ee
\end{subequations}
These notions of binding complexity are pictorially depicted in fig.~\ref{fig:geom_local_compl}. 
Since complexity decreases when we decrease the penalties, we immediately get
\be
\mathcal{MBC}^{\rm state}_{1,l, \rm hom} [\ket{\psi_T},\ket{\psi_R}] \leq \mathcal{C}^{\rm state}_{1,l, \rm hom} [\ket{\psi_T},\ket{\psi_R}] \, , \quad
\mathcal{MBC}^{\rm state}_{1,g, \rm hom} [\ket{\psi_T},\ket{\psi_R}] \leq \mathcal{C}^{\rm state}_{1,g, \rm hom} [\ket{\psi_T},\ket{\psi_R}] \, .
\label{eq:inequality1_sec6}
\ee

\vskip 5mm

\noindent
From now until the end of this section, we will omit the explicit dependence of the complexity on the reference and target states.
In order to derive bounds on the (geometrically-) local complexity and the multipartite binding complexity, we introduce some useful definitions:
\begin{itemize}
    \item The bipartite binding complexity $\mathcal{BC}^{\rm state}_{1, \rm hom}[\bar{A}]$, computed using the norm~\eqref{eq:norm_hom_after_limit} with the two subsystems given by the union of the subsystems $\lbrace 1, \dots , \bar{A} \rbrace$ in one part, and $ \lbrace \bar{A}+1, \dots , N \rbrace$ in the other part. This is the kind of binding complexity that we studied, \eg in section \ref{ssec:F1norm}.
    \item  The bipartite binding complexity $\mathcal{BC}^{\rm state}_{1, g}[\bar{A}]$, where all the non-local gates which have support on both sides of the partitioning, are not allowed, except from nearest-neighbour interactions of the form $T^{\bar A}_i T^{\bar A+1}_j$. This is similar to $\mathcal{BC}^{\rm state}_{1, \rm set}$, except that we allow several generators $T^{\bar A}_i T^{\bar A+1}_j$ with support on parties $\bar A$ and $\bar A+1$, instead of just one.
\end{itemize}

\begin{figure}[ht]
    \centering
\subfigure[Circuit contributing to $\mathcal{BC}^{\rm state}_{1, \rm hom}(P_3)$]{ \includegraphics[scale=0.2]{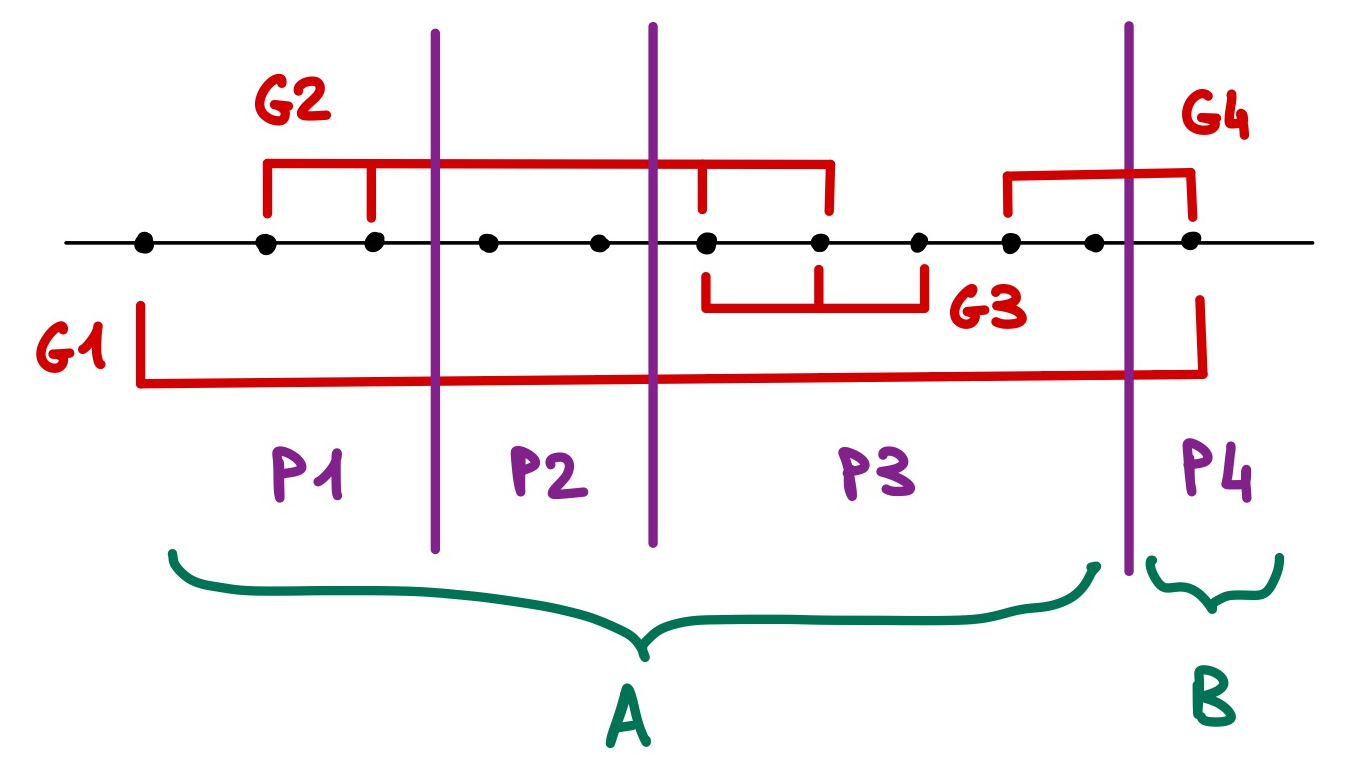} } \,
\subfigure[Circuit contributing to $\mathcal{BC}^{\rm state}_{1, g} (P_3)$]{ \includegraphics[scale=0.2]{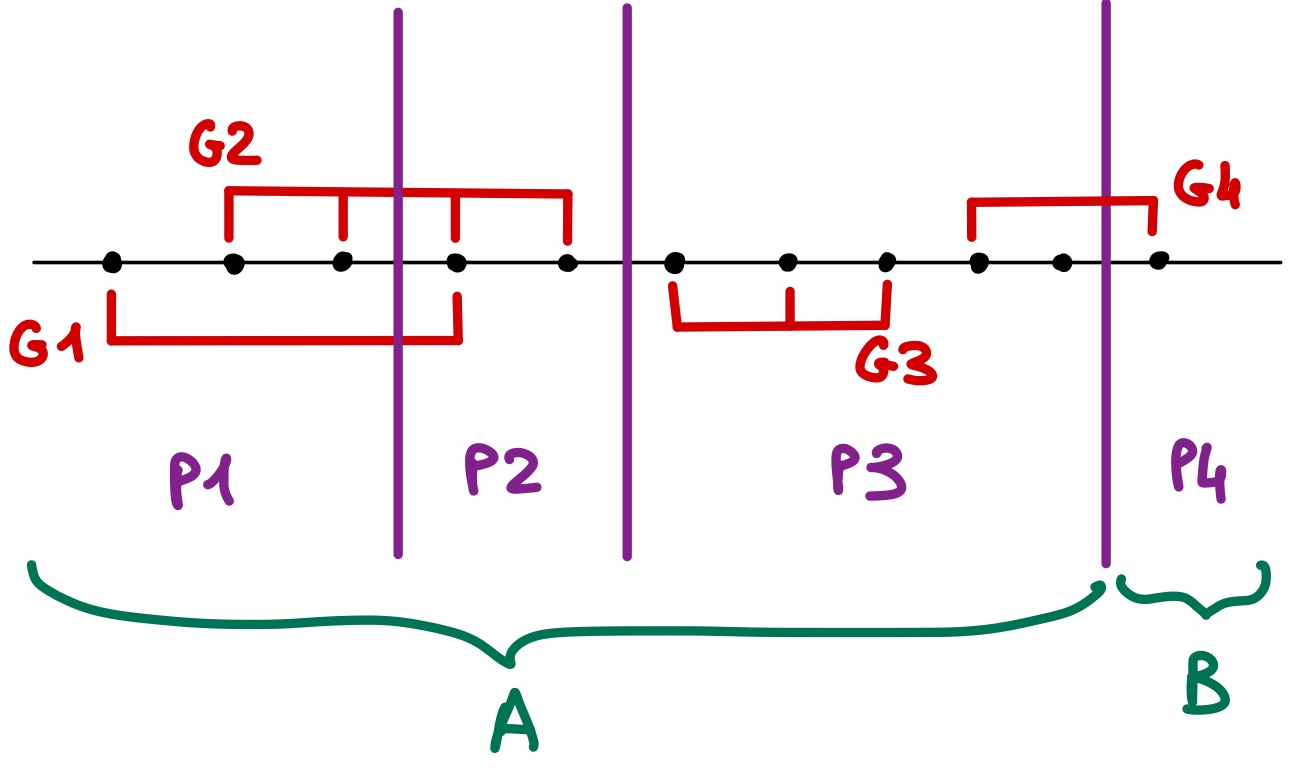} }
    \caption{ Pictorial representation of a spin chain where the vertical purple lines represent a separation between parts $P_1, \dots, P_4$. 
    (a) Circuit with gates acting on at most two parts of a system.
    (b) Circuit with gates acting only on nearest-neighbour parts or within each part. 
   Consider a bipartition of the system into $A=\lbrace P_1, P_2, P_3 \rbrace$ and $B=\lbrace P_4 \rbrace.$
Both circuits can be used to study the bipartite binding complexity $\mathcal{BC}^{\rm state}_{1, \rm hom}(P_3)$, but only case (b) is a circuit that can be used to study $\mathcal{BC}^{\rm state}_{1, g} (P_3)$ because $G_1$ in subfigure (a) does not act on nearest-neighbor-parts. The complexity is evaluated according to a norm only assigning non-vanishing cost to gates connecting the parts $A$ and $B$ among the allowed gates.}
    \label{fig:bind_geom_local_compl}
\end{figure}

The previous notions of bipartite binding complexity are pictorially represented in fig.~\ref{fig:bind_geom_local_compl}. 
We note that those quantities refer to the bipartite binding complexity and therefore, all generators that have support on just one side of the partitioning are free.
Of course, this implies that 
\be
\mathcal{BC}^{\rm state}_{1, g} [\bar{A}] \geq \mathcal{BC}^{\rm state}_{1, \rm hom}[\bar{A}] \, ,
\label{eq:inequality2_sec6}
\ee
because for the circuit that minimizes $\mathcal{BC}^{\rm state}_{1, g}$ the cost function is the same, but on the right-hand side we can build circuits using a larger set of generators, which allows to find in principle a better geodesic compared to the left-hand side. 

Consider now the following term involving nearest-neighbour parts in the expression for the geometrically-local complexity \eqref{eq:def_multipartite_complexity}
\be \label{eq:LikeBc1g}
\int_0^1 dt \, \sum^{N^2-1}_{i,j=1} |Y^{\bar{A},\bar{A}+1}_{i j}| \, ,
\ee
where the value of the control functions $Y$ in this expression are those determined by the minimization done in eq.~\eqref{eq:def_multipartite_complexity}.
Intuitively, when local generators (with respect to the bipartitioning) have no cost, the minimal circuit can be further optimized by using those gates.  In other words, $\mathcal{BC}^{\rm state}_{1, g} [\bar{A}] $ minimizes the cost of producing the target state with the norm of \eqref{eq:LikeBc1g}. If we consider a sub-optimal trajectory given by the minimization done in eq.~\eqref{eq:def_multipartite_complexity}, we will certainly obtain a larger value for this norm.
This formally corresponds to the inequality 
\be
\int_0^1 dt \, \sum^{N^2-1}_{i,j=1} |Y^{\bar{A},\bar{A}+1}_{i j}| \geq 
\mathcal{BC}^{\rm state}_{1, g} [\bar{A}] \, .
\ee
Summing over all parties gives on the left-hand side the definition \eqref{eq:def_multipartite_complexity}, and on the right-hand side a bound involving the bipartite notion of binding complexity:
\be
\mathcal{MBC}^{\rm state}_{1,g} \geq \sum_{A=1}^{n-1} \mathcal{BC}^{\rm state}_{1, g} [A] \, .
\label{eq:inequality3_sec6}
\ee
If we now combine the inequalities \eqref{eq:inequality1_sec6}, \eqref{eq:inequality2_sec6} and \eqref{eq:inequality3_sec6}, we obtain a lower bound on geometric complexity:
\be
\mathcal{C}^{\rm state}_{1,g, \rm hom} \geq \sum_{A=1}^{n-1} \mathcal{BC}^{\rm state}_{1, \rm hom} [A] \, .
\label{eq:lower_Cg}
\ee
Similar arguments (applied to the subset of terms in the expression \eqref{eq:def_multipartite_complexityA} corresponding to the binding complexity $\mathcal{BC}^{\rm state}_{1, \rm hom}[A]$), lead to a lower bound on local complexity:
\be
\mathcal{C}^{\rm state}_{1,l, \rm hom} \geq \max_{A}  \mathcal{BC}^{\rm state}_{1, \rm hom} [A] \, .
\ee
The bounds above are an improvement of the results obtained by \cite{Eisert:2021} using the notion of entanglement power; the results of \cite{Eisert:2021} are recovered by also using the inequality \eqref{Slwrbnds}. 

The bounds we derived in terms of bipartite binding complexity are tighter than the ones of \cite{Eisert:2021}. 
To illustrate this, consider a system consisting of a three-qubit chain, with reference and target states
\be
\ket{\psi_R} = \ket{000}, \quad 
\ket{\psi_T} = \cos \alpha \ket{000}+\sin \alpha \ket{111},
\ee
where we assume $0 \leq \alpha \leq \pi/4$.
To use \eqref{eq:lower_Cg}, when $A=1$ we partition the system between the first qubit and the rest, and when $A=2$, the partition is between the third qubit and the rest. In both partitions, the Schmidt decomposition is straightforward, and the coefficients are $\vec{\lambda} = (\cos \alpha, \sin \alpha)$. In section \ref{ssec:Pauli} we argued that the binding complexity for such partitions is given by the binding complexity between two qubits having the same Schmidt coefficients. 
Using eq.~\eqref{eq:binding_state_compl_2qubits}, we find that the binding complexity for each partition is given by $\alpha$, therefore the bound in \eqref{eq:lower_Cg} becomes 
\be
\mathcal{C}^{\rm state}_{1,g, \rm hom}[\ket{\psi}] \geq \sum_{A=1}^{n-1} \mathcal{BC}^{\rm state}_{1, \rm hom} [A] \, = 2 \alpha \geq -\cos^2 \alpha \,\log(\cos^2 \alpha) -\sin^2 \alpha \, \log(\sin^2 \alpha) \, .
\ee
The rightmost term in this inequality is the bound derived in reference \cite{Eisert:2021}, which we then proved to be less restrictive, in this case, than \eqref{eq:lower_Cg}.

\section{Discussion}
\label{sec:discussion}

\subsection{Summary of results}
\label{sec:ResultsSumm}

In this paper, we studied \emph{binding state complexity}, a notion of Nielsen's geometric complexity for a multipartite system \eqref{multiH} such that gates acting on each subsystem separately (we refer to such gates as local) have zero cost.
Here we focused, mainly, on the bipartite case \eqref{eq:splitting_Hilbert_spaces}.
The freedom to act with local generators implies that the norm $F^{\rm state}$ induced on the space of states can be reduced to a norm that only depends on the Schmidt coefficients $\vec{\lambda}$ of the state and on their derivatives $\dot{\vec{\lambda}}$ along a 
trajectory connecting the reference and target states:
\be
F^{\rm state}_p [\ket{\psi(t)}, |\dot{\psi}(t)\rangle] \rightarrow \mathcal{BF}_p [\vec{\lambda}, \dot{\vec{\lambda}}] \, .
\ee
To find this norm, we took an intermediate step in which we showed how the problem of finding the geodesics in SU($N_A N_B$) reduces to the manifold SO($N_A$), which is the rotation group of the Schmidt coefficients. 
The norm on SO($N_A$) is in terms of the parameters of $R$, the generator of rotations, and the equations of motion for the Schmidt coefficients read $\dot{\vec{\lambda}}(t) = R \vec{\lambda}(t)$.
An additional minimization over the stabilizer of $\vec{\lambda}$ inside SO$(N_A)$ is required to find the norm in terms of the Schmidt coefficients. Finally, the optimal trajectory whose length computes the binding state complexity is found by minimizing the length according to this last norm.

The main results of the work are the {\bf exact analytic expressions} for binding state complexity using the homogeneous norm defined in eq.~\eqref{eq:Fpenalties_flat}.
We summarize them in table~\ref{tab:results} and in the text below:
\begin{itemize}
    \item The $\mathcal{BF}_1$ norm depends on the choice of a basis of generators of the unitary group. In table~\ref{tab:results} we report the results obtained using the generalized Gell-Mann generators \eqref{eq:gell}.
    The optimal trajectory corresponds to a set of rotations along planes spanned by pairs of Cartesian axes to get from $\vec{\lambda}(0) = (1,0,...,0)$ to $\vec{\lambda}(1)$, see eq.~\eqref{eq:piecewise_trajectory_GellMann}.
     \item The $\mathcal{BF}_2$ norm is independent of the basis of generators in the unitary group, and turns out to be the standard round metric of the unit sphere.
    The binding state complexity is the arc length from the reference to the target state.
\end{itemize}

\begin{table}[ht]   
\begin{center}    
\begin{tabular}  {|c|c|c|} \hline  & $\mathcal{BF}_{p, \rm hom} $ \textbf{norm}  & \textbf{Binding state complexity}  \\ \hline
\rule{0pt}{4.9ex} $p=1$ \textbf{(Gell-Mann basis)}  & $ \sum^{N_A}_{m>1} \frac{ |\dot \lambda_m|}{\lambda_1} $  & Eq.~\eqref{eq:result_stateF1_hom_GellMann}  \\
\rule{0pt}{4.9ex} $p=2$ \textbf{(any basis)} & $\le \frac{2}{\mathcal{N}_A \mathcal{N}_B } \, \dot{\vec{\lambda}}(t)^2 \ri^{\frac{1}{2}} $  &  $  \sqrt{\frac{2}{\mathcal{N}_A \mathcal{N}_B }} \, \mathrm{arccos} \le e^{-\frac{1}{2}S_{\infty}(\ket{\psi_T})} \ri  $       \\[0.2cm]
\hline
\end{tabular}   
\caption{Results for the homogeneous norm \eqref{eq:Fpenalties_flat}.
\textbf{First row:} based on eqs.~\eqref{eq:distance_F1norm_GellMann}, \eqref{eq:result_stateF1_hom_GellMann}. $N_A$ represents the dimension of the smaller Hilbert space in the decomposition \eqref{eq:splitting_Hilbert_spaces}, while $\lambda_1$ is the maximal Schmidt coefficient.
\textbf{Second row:} based on eqs.~\eqref{eq:F2_hom_gennom}, \eqref{eq:bind_compl_GellMann}.
$\mathcal{N}_{A(B)}$ are overall factors related to the normalization of the generators, see eq.~\eqref{GenNormaliz}, and it is understood that $|\vec{\lambda}(t)|^2 =1.$
$S_{\infty}(\ket{\psi_T})$ is the minimal Rényi entropy of the reduced density matrix of the target state $\ket{\psi_T}$. } 
\label{tab:results}
\end{center}
\end{table}

For the $\mathcal{BF}_{1, \rm hom}$ with the Pauli basis of generators \eqref{eq:full_Pauli_basis}, we gave an exact result for a system of any number of qubits which is bipartitioned to a single qubit and the rest, and we found bounds in the general case.

We argued that
the binding complexity computed by any $F_1$ norm of the form \eqref{eq:penalty_intermediate} is the same, up to an overall normalization, as long as the generators forming the optimal trajectory all lie on the same adjoint orbit of the $\mathrm{SU}(N_A)\otimes \mathrm{SU}(N_B)$ action as the cheapest generator. The latter condition is satisfied for the Gell-Mann and Pauli bases.  As a consequence, the binding state complexity is rather insensitive to the penalty factors chosen for the non-local gates. 
This provides a huge class of systems that are accessible analytically.
The freedom to choose the penalty factors for non-local generators may help in implementing these circuits in an experimental setting, with the caveat that local gates can have a small but non-vanishing cost, as discussed at the end of section \ref{app:1=2}.

We have shown that the binding state complexity $\mathcal{BC}^{\rm state}_{1, \rm hom}$  
applied to a system composed of two qutrits is equivalent to the $F_1$ state complexity (\emph{not} binding) for a single qubit, for target states that belong to a certain region around the reference state. 
This extends previous studies in the literature that focused on the Riemannian $F_2$ norm for a single spin \cite{Brown:2019whu}.

Finally, we inferred lower bounds for the multipartite binding complexity and the geometrically local complexity \cite{Eisert:2021}, in terms of the bipartite binding complexity for different bipartitions of a given system.

\subsection{Future directions}

The approach presented in this work allowed us to pursue various directions that were previously hindered by the fact that the $F_1$ norm is not a Riemannian metric.
Several questions remain open, which we intend to study in the future:    
\begin{enumerate}
\item \textbf{Generalized Pauli basis.}
The generalized Pauli basis is natural when considering systems of spins on a lattice that entail a notion of spatial locality in which each spin resides at a different point, and therefore it is commonly used in the literature on Nielsen's complexity. 
We showed in section \ref{ssec:F2norm} that the binding complexity using the $\mathcal{BF}_2$ norm in the generalized Pauli-basis differs from the one obtained using the Gell-Mann basis just by a normalization factor. The $\mathcal{BF}_1$ norm case, however, is more involved.
We derived several bounds in section \ref{ssec:Pauli} for the complexity associated with the $\mathcal{BF}_1$ norm in the Pauli-basis, but it would be desirable to find the optimal trajectory and the exact complexity in the Pauli basis for an arbitrary number of spins.
A discrete version of this problem was considered in \cite{Zhang:2021xwx}.

\item \textbf{Extension of our results for other $\mathcal{BF}$ norms.}
We were able to calculate $\mathcal{BC}_{1, \rm hom}$ only in a region of the group manifold close to the reference state, see section \ref{ssec:F1norm}.
We would like to extend the analysis to the whole space.
Furthermore, here we focused on the $\mathcal{BF}_1$ and $\mathcal{BF}_2$ norms, finding that the binding complexity for the $p=2$ case is a function of the Rényi min-entropy. 
All the other $\mathcal{BF}_p$ norms will result in binding complexities that depend on the Schmidt coefficients. 
It is reasonable to assume that one can obtain all the Schmidt coefficients from the knowledge of binding complexities for all the $\mathcal{BF}_p$ norms, as one can obtain them from knowing all the Rényi entropies. 
We plan to obtain explicit results for these relations.
\item \textbf{Curvature properties of binding complexity geometry.}
The Gell-Mann basis for the unitary group $\mathrm{SU}(N)$ has the advantage of clearly distinguishing between elements of the stabilizer of the reference state, and elements of the coset space.\footnote{The Gell-Mann matrices introduced in eq.~\eqref{eq:gell} are proportional to the set of generators $E_{ij}, F_{ij}$ defined in eqs.~(I.6.60)-(I.6.61) on page 208 of reference \cite{Castellani:1991et}. The latter form a basis that distinguishes between the generators of the maximal subgroup $\mathrm{SU}(N-1)\times \mathrm{U}(1)$ and the coset space.}
In the framework of Nielsen's complexity, it was argued that geodesic deviation and the switchback effect, which are essential features of complexity geometry, require regions with negative sectional curvature \cite{Brown:2016wib,Brown:2017jil}.
In the Riemannian case, a negatively curved geometry can be obtained whenever the commutator structure of the algebra satisfies a relation of the kind [\textit{easy}, \textit{easy}] = \textit{hard}, where \textit{easy} and \textit{hard} refer to a smaller or bigger penalty factor inside the cost function, respectively.
This statement was argued to be a consequence of Pythagorean's theorem in curved space in reference \cite{Brown:2019whu}, and it was later confirmed for the state complexity of a single qutrit in \cite{Auzzi:2020idm}.
In the case of the $\mathcal{BF}_{p, \rm set}$ norm, we assign penalties 0 to local generators, 1 to a single non-local generator $T_{11} \equiv T^A_1 \otimes T^B_1$ and the formal cost $\infty$ to all the other non-local generators, which are not allowed.
As a consequence, the algebra naturally admits a subset of commutation relations of the form [\textit{local}, $T_{11}$] = \textit{non-local}, which satisfies the above-mentioned pattern [\textit{easy}, \textit{easy}] = \textit{hard}.  
We therefore expect that the sectional curvature is negative along the planes generated by any local generator and the single allowed gate $T_{11}$.
Since binding state complexity involves non-trivial limits on the penalty factors, it would be interesting to check whether these statements on the negative sectional curvature are still valid. Furthermore, it is interesting to understand the implications of the above structure of commutators in the context of the switchback effect.
Let us further remind the reader that in the context of a single qubit, we have shown in section \ref{ssec:example_2qutrits} that the circuit (as opposed to binding) complexity associated with the $F_1$ norm provides a natural way of suppressing hard gates which was achieved in previous works by the assignment of penalty factors. It is interesting to study under which conditions the $F_1$ norm can also provide an alternative mechanism for geodesic deviation.

\item \textbf{Characterizing multipartite entanglement.}
In reference \cite{Balasubramanian:2018hsu}, the authors suggested that the multipartite version of binding complexity can distinguish whether a multipartite state will be entangled after tracing over one of the parts. The authors use as an example the GHZ state which is separable after the tracing and has an estimated binding complexity of $\mathcal{O}(n)$, while the $W$ state which is not and has an estimated binding complexity of $\mathcal{O}(n^2)$, where $n$ is the number of spins. 
We plan to consider the multipartite version in spin systems and understand more generally the relation between multipartite binding complexity and the robustness of multipartite entanglement.
\item  \textbf{Generalized complexity measures.}
Binding state complexity can be interpreted as an example of a larger class of complexity measures in which one does not require to reach a specific target state, but only an equivalence class of states that satisfy certain properties. 
In this case, the required property is based on the entanglement structure encoded by the Schmidt coefficients.
Other examples could require that the equivalence class of target states has a fixed vacuum expectation value of certain operators, \ie $\langle \psi_T | O_i | \psi_T \rangle = Q_i$. 
This seems natural because usually we cannot distinguish states with absolute precision. 

\item \textbf{Holographic interpretation.}
It remains to be investigated if there is a holographic dual to the quantities we considered. 
In \cite{Balasubramanian:2018hsu}, it was conjectured that the binding complexity is dual to the volume of the interior of a multi-boundary wormhole (see \eg \cite{Skenderis:2009ju}), minimized over all the boundary times (see their discussion in sections 5 and 7). 
While in this work we mostly considered the bipartite case, which is conjectured to be related to the two-boundary wormhole, we should still be able to compare the binding complexity to the expected results from the gravity side in that specific case.
A holographic computation of the volume of a three-dimensional wormhole with three boundaries has been performed in \cite{Zolfi:2023bdp}. It would be interesting to compare this case to the multipartite binding complexity, partially considered in section \ref{sec:geometrically_local_complexity} and which we hope to investigate further in the near future.
Another interesting candidate for the holographic dual may be deduced as follows. Given the relation between the $\mathcal{BF}_2$ binding complexity and the minimal Rényi entropy, the holographic dual in that case may be obtained along the lines of \cite{Hung:2011nu,Dong:2016}. It would be interesting to compare the proposal obtained this way to the previous one.  

\item \textbf{Relation to kinematic space.}
For the vacuum state in a two-dimensional CFT, there is a conjectured relation between volume complexity and entanglement via the Crofton formula \cite{Abt:2017pmf,Abt:2018ywl}.
Furthermore, reference \cite{Orus:2005dpa} claims that the minimal Rényi entropy of a quantum chain at the critical point is related to the entanglement entropy. Combining these observations, one may argue that the volume complexity could be expressed as an integral over the binding complexity of an appropriate trajectory in kinematic space.
We intend to investigate this link more deeply.

\item \textbf{Subsystem complexity.}
Binding complexity is naturally defined when there is a partition of the Hilbert space into subsystems. In this case, the state of each subsystem separately is generically mixed.
Other interesting notions of complexity have also been defined in the context of mixed states, such as the purification or spectrum complexity proposed in 
reference \cite{Agon:2018zso}.\footnote{These definitions were originally suggested in the context of holographic complexity for subregions \cite{Alishahiha:2015rta,Carmi:2016wjl,Abt:2017pmf,Abt:2018ywl,Agon:2018zso,Alishahiha:2018lfv,Caceres:2018blh,Caceres:2019pgf,Auzzi:2019vyh,Auzzi:2019mah} and were further developed for Gaussian states, see \eg \cite{DiGiulio:2020hlz,Caceres:2019pgf,Jiang:2018nzg}.}  It would be interesting to compare these notions of complexity to the binding complexity of a state which is the purification of such mixed states (where the bipartition is between the system and its purification). 

\item  \textbf{Ancillary qubits.} We would like to understand the role of ancillary qubits in binding complexity.
In the Gell-Mann basis the result \eqref{eq:result_stateF1_hom_GellMann} is unaffected by enlarging the dimension of $\mathcal{H}_B$. Therefore there is no influence of ancillary qubits on the results in the Gell-Mann basis. Furthermore, in section \ref{ssec:Pauli}, we discussed an exact result for the complexity in the Pauli basis for two systems, one of which (system $A$) contains a single spin and the other (system $B$) an arbitrary number. In this case, we were able to prove that when the single spin in system $A$ was entangled with a single spin in system $B$, the extra spins in system $B$, which we referred to as ancillary, did not participate in the optimal trajectory and did not improve the result for the complexity. It would be interesting to understand if this statement is generally valid in the Pauli basis for the addition of spins which are unentangled both in the reference and the target state for arbitrary numbers of spins in the systems $A$ and $B$, and if a similar statement can be made for binding complexity with more than two parties. 
\end{enumerate}

\section*{Acknowledgements}
We are happy to thank Vijay Balasubramanian, Ramy Brustein, Hugo Camargo, Nat Levine, Christian Northe, Yaron Oz, Eran Palti, Daniel Rohrlich, Michael Walter and Erez Zohar for valuable discussions.  We are particularly grateful to Adam Chapman for providing us with the proof of the inequalities in appendix \ref{ssec:bounds_norms}.
The work of SB, SC and TS is supported by the Israel Science Foundation (grant No. 1417/21), by the German Research Foundation through a German-Israeli Project Cooperation (DIP) grant “Holography and the Swampland”, by Carole and Marcus Weinstein through the BGU Presidential Faculty Recruitment Fund and by the ISF Center of Excellence for theoretical high energy physics. 
SB is supported by an Azrieli fellowship funded by the Azrieli foundation. SB gratefully acknowledges  the Perimeter Institute for Theoretical Physics, Brandeis University and the Simons Center for Geometry and Physics  at Stony Brook University, where part of the research in this paper was carried out.
Research at Perimeter Institute is supported by the Government of Canada through the Department of Innovation, Science and Economic Development and by the Province of Ontario through the Ministry of Colleges and Universities.
SB and GP thank the organizers and participants of the workshop YITP-T-23-01 “Quantum information, quantum matter and quantum gravity” in Kyoto for hospitality and for interesting discussions.

\appendix

\section{An inequality for the $F_1$ cost function }
\label{ssec:bounds_norms}

This appendix is devoted to the proof of the inequality \eqref{eq:chain_relations_F1normT}.
We derive the following chain of relations:
\be
\begin{aligned}
 & \min_{ \mathrm{Ad}_U} \sum_{b,j} \left|\tilde{Y}^{ai} (\text{Ad}_{U^A})^{\,\,b}_{a}(\text{Ad}_{U^B})^{\,\,j}_{i}  \right| \geq \min_{ O} \sum_{b,j} |\tilde{Y}^{ai} (O^A)^{\,\,b}_{a} (O^B)^{\,\,j}_{ i} | = \\
 & =  \sum_a \sigma_a (\tilde{Y})  \geq  \sum_a |\tilde{Y}^{aa}|  \, .
 \end{aligned}
 \label{eq:chain_relations_F1norm}
\ee
The first minimum in the first line is over all possible adjoint matrices. 
$O^{A(B)}$ denote general orthogonal matrices and the second minimum in the first line is with respect to these matrices.  
$\sigma_a (\tilde{Y})$ are the singular values of the matrix $\tilde{Y}_{ai}$ and the first quantity in the second line is the so-called nuclear or Schatten 1-norm \cite{bhatia2013matrix}. 
Let us go through the derivation of the previous steps:
\begin{itemize}
    \item The inequality in the first line follows from the fact that the adjoints $\mathrm{Ad}_{U}$ are specific orthogonal matrices, thus the complete set of orthogonal matrices is larger than those in the adjoint. Note that here we are comparing the adjoint of $\mathrm{SU}(N_A)$ to $\mathrm{SO}(N_A^2-1)$ (not to $\mathrm{SO}(N_A)$).
    \item In going from the first to the second line, we used the inequality\footnote{In turn, this inequality can be proven using the following property of the singular values $\sum_a \sigma_a (M+N)\leq \sum_a \sigma_a (M) + \sigma_a (N)$ \cite{horn_johnson_1991} if we decompose the matrix $T$ into matrix units of the form $t_{a,b}\delta_{a,b}$ and apply this inequality to the sum. In this case $|t_{a,b}|$ is the only non-zero singular value of each matrix unit.}  $\sum_a \sigma_a ( T)\leq \sum_{a,b}  | T_{a,b}|$ with $T=(O^A)^T \tilde Y O^B$, together with the observation that the nuclear norm is invariant under orthogonal transformations. 
 The inequality is saturated when performing the minimization over the orthogonal matrices, once we choose $O^A,O^B$ according to the singular value decomposition of $\tilde Y^{ai}$.
    \item The last inequality (see \eg \cite{CHEN1992109}) can be proven by applying von Neumann's trace inequality $|\Tr (\tilde YB)|\leq \sum_i \sigma_i(\tilde Y)\sigma_i(B)$. This is done by choosing $B$ as a diagonal matrix with elements $\pm 1$, where the sign is chosen such that $|\text{Tr}(\tilde YB)| =\sum_a |\tilde Y^{aa}|$ recovers the last inequality in \eqref{eq:chain_relations_F1norm}. The singular values of the matrix $B$ are of course $\sigma_i(B)=1$.
    Notice that this argument can also be applied to rectangular $\tilde{Y}$ matrices. In fact, one can pad $\tilde{Y}$ with zeroes to define a square matrix $\hat{Y}$, and then perform the same steps outlined above to obtain the last inequality for $\hat{Y}$. The non-zero singular values and the non-zero diagonal elements of $\hat Y$ are equal to the ones of $\tilde Y$.  
\end{itemize}

\noindent
Once we recognize that the first term in \eqref{eq:chain_relations_F1norm} is the definition of the homogeneous $F$-norm in eq.~\eqref{eq:Fp_flat_Ytilde} with $p=1$, and upon an additional minimization over $\tilde Y$ subject to the constraint \eqref{eq:Rcnstrain}, the chain of relations \eqref{eq:chain_relations_F1norm} can be written as \eqref{eq:ineq2}.

\section{Complexity with a large number of Schmidt coefficients}\label{app:ContSchmi}

In this appendix, we focus on a quantum-mechanical system with a large number of Schmidt coefficients. In this case, we may recast the binding complexity \eqref{eq:result_stateF1_hom_GellMann} in the form of an integral. The integral is generally easier to compute than the original discrete sum. Furthermore, as mentioned at the end of section \ref{ssec:F1norm}, this can be viewed as a first step towards the study of binding complexity in QFT.

First, let us re-write eq.~\eqref{eq:result_stateF1_hom_GellMann} in reverse order of summation,\footnote{Note that we are using the form of eq.~\eqref{eq:result_stateF1_hom_GellMann} before performing the integration. The reason is that we will be expanding under the assumption that successive Schmidt coefficients are close to each other and therefore the integrands themselves will appear naturally in our expressions.}
\be
\begin{aligned} \label{BCreverse}
	& \mathcal{BC}^{\rm state}_{1, \rm hom}  = - \sum_{k=1}^{N_A-1}  \int_{\bar \lambda_{k+1}}^{\bar \lambda_{k+2}} \frac{k d\lambda}{\sqrt{1-\sum_{j=k+2}^{N_A} \bar \lambda_{j}^2 -k \lambda^2}} \, .
	\end{aligned}
 \ee
 The index $k$ counts the number of Schmidt coefficients larger than $\bar \lambda_{k+1}$. In the limit of an infinite number of Schmidt coefficients $N_A\rightarrow \infty$, we obtain a distribution $\tilde \lambda(k)$. We will treat $\bar \lambda_1$ as a normalization that does not have to be a part of the continuous distribution. This means that the distribution represents all the Schmidt coefficients apart from $\bar \lambda_1$ and that  $\bar{\lambda}_1^2+\int_{0}^{\infty} \tilde \lambda(k)^2 dk = \bar{\lambda}_1^2+\int_{0}^{\tilde \lambda(0)} \rho(\tilde \lambda)\tilde \lambda^2 d\tilde \lambda=1$, where $\rho(\tilde\lambda)$ is the density of Schmidt coefficients. This type of separation can be seen for the entanglement spectrum derived in \cite{calabrese2008entanglement} for a one-dimensional system near the scale-invariant regime. 
 
 In the continuum, the index $k$ becomes a continuous function of $\tilde{\lambda}$, such that $k(\tilde{\lambda})= \int_{\tilde{\lambda}}^{\tilde \lambda(0)} \rho(\lambda) d\lambda$, and the sum $\sum_{j=k+2}^\infty \bar \lambda_j^2$ turns to $\int_0^{\tilde{\lambda}(k)} \rho(\lambda) \lambda^2 d\lambda$.  With these replacements the complexity reads
 \be\label{bestcontin2}
\mathcal{BC}^{\rm state}_{1, \rm hom}=\int_0^{\tilde\lambda(0)} d\lambda \, \frac{ k(\lambda)}{\sqrt{
\bar \lambda_1^2+\int_{\lambda}^{ \tilde\lambda(0)} \rho(x) x^2 dx
- \lambda^2 k(\lambda)}} =  \int_0^{\tilde\lambda(0)} d\lambda \, \frac{ k(\lambda)}{\sqrt{
\bar \lambda_1^2+2 \int_\lambda^{ \tilde\lambda(0)} x k(x) dx}  }  \, .
\ee
With an additional change of variables, the above formula can be rewritten as
 \be
\begin{aligned} \label{bestcontin}
	& \mathcal{BC}^{\rm state}_{1, \rm hom}  = - \int_{0}^{\infty} dk \frac{k \tilde \lambda'(k)}{\sqrt{1-\int_{k}^{\infty} \tilde \lambda(x)^2 dx -k \tilde \lambda(k)^2}}= - \int_{0}^{\infty} dk \frac{k \tilde \lambda'(k)}{\sqrt{\bar \lambda_1^2 + \int_{0}^{k} \tilde \lambda(x)^2 dx -k \tilde \lambda(k)^2}} \, .
\end{aligned}
 \ee
We note that this result can be obtained by considering a continuous version of the norm of \eqref{eq:distance_F1norm_GellMann} of the form $\mathcal{BF}_{1,\rm hom} dt = \int_0^{\infty} dk \frac{| d \tilde{\lambda} (k,t)|}{\lambda_1(t)}$, and by then evaluating it on a trajectory that is analogous to the discrete one.
A caveat should be made regarding the range of validity of \eqref{bestcontin2} and \eqref{bestcontin}. We remind the reader that the result in eq.~\eqref{BCreverse} was derived in the region $\bar \lambda_1 \geq \bar \lambda_2+\bar \lambda_3$, where we obtained the exact form of the norm $\mathcal{BF}_{1,\rm hom}$. 
A natural guess for the continuum version of the previous constraint is $\lambda_1\geq 2\tilde\lambda(0)$. When the spectrum distribution does not have a separation between the largest Schmidt coefficient and the continuous part, \ie $\int_{0}^{\infty} \tilde \lambda(k)^2 dk = 1$, the expected constraint reads  $\tilde\lambda(0)\geq2\tilde\lambda(0)$, and is clearly violated. Nevertheless, we suspect that the result \eqref{eq:distance_F1norm_GellMann} may in fact be valid even outside  the range dictated by the inequality $\bar \lambda_1 \geq \bar \lambda_2+\bar \lambda_3$, even for the discrete case, and in that case the continuum result could perhaps be used for distributions without a spectrum separation.

The expressions above, besides being suitable for a continuous distribution, can help in approximating the discrete $\mathcal{BC}^{\rm state}_{1, \rm hom}$ by a simpler integral expression. In this case, we have to explain how to turn our discrete distribution into a continuous function over which we can integrate. We do this as follows. Let $\bar \lambda_{n}$, where $n$ is an integer that goes from $1$ to $\infty$, be a set of ordered Schmidt coefficients. Define a smooth monotonic function $\tilde \lambda(x)$ that interpolates the set starting from $n \geq 2$, at points that differ by a unit interval, \ie 
\be
\tilde \lambda(x_0 + m) = \bar \lambda_{m+2} \, , \qquad
m \geq 0 \, .
\label{eq:def_x0}
\ee  
In addition, set $x_0$ such that $\int_0^\infty \tilde \lambda(x)^2 dx = \sum_{n=2}^\infty \bar \lambda_n^2=1 - \bar \lambda_1^2$. Approximating the integrals of \eqref{bestcontin} (or \eqref{bestcontin2}) by a sum using the values of the integrands at integer values gives an expression that is close to the one obtained by approximating the integrals of the exact result in \eqref{BCreverse}. 

The error in approximating \eqref{BCreverse} by  \eqref{bestcontin} (or \eqref{bestcontin2}) can be determined by standard numerical techniques, and will depend on the difference between adjacent Schmidt coefficients and the exact choice of the interpolating function $\tilde \lambda(x)$.
We note that if $ \bar \lambda_{k}-\bar \lambda_{k+1}$ is not sufficiently small for some specific value of $k$, one could combine the continuous approximation with the exact calculation. 
A particular case of interest would be if there is a finite number $N_A$ Schmidt coefficients, where $\bar \lambda_{N_A}$ is of order $\frac{1}{\sqrt{N_A}}$. In this case, the boundaries of integration in \eqref{bestcontin} could be taken from $0$ to $N_A-2$ and an additional term of value $\sqrt{N_A-1} \, \mathrm{arcsin} \, [ \bar \lambda_{N_A}\sqrt{N_A-1}]$ coming from the $N_A-1$ term of \eqref{BCreverse}, should be added.

Let us exemplify the above-mentioned continuous approximation
in the case of the thermofield double state of two harmonic oscillators.
We start from a discrete system whose Hilbert space decomposes according to eq.~\eqref{eq:splitting_Hilbert_spaces}, with Schmidt coefficients\footnote{Notice that the Schmidt coefficients do not satisfy $\bar \lambda_1 \geq \bar \lambda_2+\bar{\lambda}_3$ when $\beta$ is sufficiently small, and therefore we are uncertain whether the binding complexity should indeed be computed using \eqref{BCreverse}. Nevertheless, this serves as a good example for the continuous approximation \eqref{bestcontin}.
Our methods can be applied to systems that have an entanglement gap in the thermodynamic limit, for instance, those discussed in \cite{Wang:2020}. 
}
\be
\bar\lambda_n = \sqrt{e^{\beta} -1} \, e^{- \frac{\beta n}{2}}\qquad \, (n\geq 1), \qquad
\mathrm{such \,\, that \quad}
\sum_{n=1}^{\infty} \bar\lambda^2_n = 1 \, .
\label{eq:discrete_Schmidt}
\ee
In particular, $\bar{\lambda}_1$ is the largest among the previous Schmidt coefficients.
According to the discussion around eq.~\eqref{eq:def_x0}, in the continuum we define a new distribution related to the discrete one by
\be
\tilde{\lambda} (x) = \sqrt{e^{\beta} -1} \, e^{- \frac{\beta (x-x_0)}{2}} \, , \qquad
\text{with $x_0$ such  that }\quad
\bar{\lambda}_1^2 + \int_0^{\infty} dx \, \tilde{\lambda}(x)^2 = 1 \, .
\label{eq:continuous_Schmidt}
\ee
We can now compare the exact result for the binding complexity in the discrete case \eqref{eq:result_stateF1_hom_GellMann} using the Schmidt coefficients \eqref{eq:discrete_Schmidt} with the continuous approximation \eqref{bestcontin} using the distribution \eqref{eq:continuous_Schmidt}. 
In the regime $\beta \rightarrow 0$, the binding complexity is of order $1/\sqrt{\beta}$, while the \emph{relative} error found by this procedure can be verified numerically to be smaller than $\beta\sqrt{\beta}$.

\section{Gell-Mann trajectory length in the Pauli basis}\label{app:Pauli}

In this appendix, we prove the claim of section \ref{ssec:Pauli} that if we use the trajectory optimizing the $\mathcal{BF}_{1,\rm hom, Gell-Mann}$ norm and evaluate its length according to the $\mathcal{BF}_{1,\rm hom, Pauli}$ norm, we recover the same length. More explicitly, we claim that the $\mathcal{BF}_{1,\rm hom, Pauli}$ norm evaluated on the Hamiltonian \eqref{eq:optimal_Ham_F1GellMann} 
\begin{equation}
    H = \sum_{m>1}^{N_A=2^{n_A}} w_m  (t) T^A_{c(m,1)} T^B_{c(1,m)} 
    \, , \qquad
w_m  = \frac{\dot{\lambda}^g_m(t)}{\lambda^g_1(t)} \, ,
\end{equation}
reads
\begin{equation}\label{eq:appPauliresfin}
    \mathcal{BF}_{1, \rm hom,Pauli}=  
    \sum^{N_A}_{m>1} | w_m(t)| \, ,
\end{equation}
which is the same as the result for the Gell-Mann basis
\eqref{eq:distance_F1norm_GellMann}.

To demonstrate this, let us use the definition \eqref{eq:gell} for the Gell-Mann basis, in particular
\begin{equation}
    T_{c(m,1)} = i (\ket{m}\bra{1}-\ket{1}\bra{m}),\qquad
    T_{c(1,m)} =  \ket{m}\bra{1}+\ket{1}\bra{m} \, .
\end{equation}
Here it is understood that the basis vectors used in this definition are 
\be \label{pau:m1m2m3}
\ket{m}^A \equiv \ket{\vec m}^A = \ket{m_1 m_2 \dots m_{n_A}  } \, , \qquad
\ket{ m}^B = \ket{\vec m}^B = |m_1 m_2 \dots m_{n_A} \underbrace{\uparrow \dots \uparrow}_{n_B-n_A} \, \rangle \, ,
\ee
where $m_i\in \{\ket{\uparrow},\ket{\downarrow} \}$ represent spins up or down, and the basis vectors for the system $B$ have been padded with $\ket{\uparrow}$ to reach the correct length. The different generators in the Pauli basis can be enumerated by vectors $\vec\mu$, $\vec \nu$ as in the right-hand side of eq.~\eqref{eq:full_Pauli_basis}
\be
T_{\vec\mu \vec \nu}  = \sigma_{\mu_{1}} \otimes \ldots \sigma_{\mu_{n_A}} \otimes \sigma_{\nu_{1}} \otimes \ldots \otimes \sigma_{\nu_{n_B}} \, .
\ee 
To find the control functions, we should take the normalized trace of each generator against the Hamiltonian for our trajectory, that is
\begin{equation}
    Y_{\vec\mu \vec \nu} = \frac{1}{2^{n_A+n_B}} \, \trace[H\cdot T_{\vec \mu \vec \nu}] \, .
\end{equation}
Using the following properties of the trace and tensor products $\tr(A\otimes B)=\tr(A)\cdot \tr(B)$ and $(a_1\otimes b_1)\cdot(a_2\otimes b_2)=(a_1\cdot a_2)\otimes (b_1\cdot b_2)$, as well as the cyclic property of the trace, we can rewrite the above control functions as 
\begin{equation}
    Y_{\vec\mu \vec \nu} =  \frac{i}{2^{n_A+n_B}} \, \sum_{\vec m\neq \vec 1} \tilde v_{\vec m,1} (\langle \vec 1 | \sigma_{\vec\mu}\ket{\vec m}-\mathrm{c.c.})_{A} \, (\langle \vec 1 | \sigma_{\vec\nu}\ket{\vec m}+ \mathrm{c.c.})_B \, ,
\end{equation}
where we used $\vec 1$ to denote the state with all spins up, “c.c.”
 stands for complex conjugation and $\sigma_{\vec{\mu}}, \sigma_{\vec{\nu}}$ collectively denote the generalized Pauli matrices in the two subsystems $A,B$ separately.
By splitting these generalized Pauli generators into tensor products of standard Pauli matrices, we compute the matrix elements as follows: 
\begin{equation}\label{pauliexplicit}
    \langle \vec 1 | \sigma_{\vec\mu}\ket{\vec m} = \prod_{k=1}^{n_A} (\delta_{m_k,\uparrow} (\delta_{{\mu_k},0} +\delta_{{\mu_k},3})+ \delta_{m_k,\downarrow} (\delta_{{\mu_k},1} -i\delta_{{\mu_k},2})).
\end{equation}
We now use the above machinery to evaluate 
\begin{equation}
\mathcal{BF}_{1, \rm hom,Pauli}
=  \sum_{\vec\mu\vec\nu}|Y_{\vec\mu\vec\nu}| =  \frac{1}{2^{n_A+n_B}} \, \sum_{\vec m\neq \vec 1} | w_m | \, \sum_{\vec\mu\vec\nu}   |\langle \vec 1 | \sigma_{\vec\mu}\ket{\vec m}- \mathrm{c.c}| \times |\langle \vec 1 | \sigma_{\vec\nu}\ket{\vec m}+ \mathrm{c.c}| \, .
\end{equation}
We see that for any given $\vec m$, we are left with a combinatorial exercise where we have to sum products similar to those in eq.~\eqref{pauliexplicit} over all values of $\vec \mu$ and $\vec \nu$. For the term in the first absolute value not to vanish when subtracting the complex conjugate, we need an odd number of $\mu_k=2$ components, in this case the “c.c.” simply doubles the contribution. Similarly, for the second absolute value not to vanish, we need an even number of $\nu_k=2$ components. In this case, the contribution of “c.c.” is again a factor of 2. 
For specific values of $\vec \mu, \vec \nu, \vec m$, we have at most one out of all the products of delta functions which survives. Therefore, every single contribution to this sum is equal to $4$ and we just need to understand how many of those contributions do not vanish for a given value of $\vec m$. 

What is the number of non-vanishing contributions? 
Assume to have a certain vector $\vec m$ with $m_\downarrow$ number of spins down. Notice that the number of spins down is the same for $\ket{m}^A$ and $\ket{m}^B$. For every spin up, we have two options: either $\delta_{\mu_k,0}$ or $\delta_{\mu_k,3}$ will contribute for this value of $k$, so in total this yields a factor of $2^{n_A-m_\downarrow}\times 2^{n_B-m_\downarrow}$ possible contributions. Then we have to choose the location of an odd (even) number of $\delta_{\mu_k,2}$ in order for the first (second) term not to vanish after subtracting (adding) the complex conjugation. These are chosen out of the down arrows, resulting in a multiplicative factor $\sum_{k(\mathrm{odd/even})}^{m_\downarrow}\left( m_\downarrow \atop k\right) = 2^ { m_\downarrow-1}$ for each system. Multiplying everything together we obtain $4\cdot 2^{n_A-m_\downarrow}\cdot 2^{n_B-m_\downarrow}\cdot 2^ { m_\downarrow-1}\cdot 2^ { m_\downarrow-1}=2^{n_A+n_B}$, which precisely cancels the factor of $2^{n_A+n_B}$ in the denominator, and we get the same result \eqref{eq:appPauliresfin} as advertised in the main text.

 \section{$\mathcal{BF}_2$ homogeneous norm from the Euler-Arnold equations} 
 \label{app:ssec:Pauli_result}

In this appendix, we compute binding state complexity for a system of two qudits, whose Hilbert space dimensions are $N_A$ and $N_B$, see eq.~\eqref{eq:splitting_Hilbert_spaces}. For convenience, we assume that $N_A\leq N_B$. The complexity is computed according to the $F_{2, \rm hom}$ norm defined in eq.~\eqref{eq:Fpenalties_flat} using an arbitrary basis of the unitary algebra. 
The following calculation is based on the Euler-Arnold equations and provides a cross-check of the result \eqref{eq:bind_compl_GellMann}.  
\subsection{Geodesics from the Euler-Arnold equations}
\label{app:geodesics_EA}

To compare more easily with the literature, we use the notation of \cite{Nielsen3}. 
Given the space of unitary operators $\mathrm{SU}(N\equiv N_A N_B)$ acting on the qudits, we define the length $\ell$ of a trajectory $U(t)$ \eqref{eq:generic_path_unitaries} generated by a Hamiltonian $H(t)$ as
\be
\ell \equiv \int_0^1 dt \sqrt{\langle H(t), H(t) \rangle} \, ,
\label{eq:Riem_length}
\ee
where $\langle \cdot, \cdot \rangle$ is a suitable Riemannian metric independent of the group point $U$.
Explicitly, we define the norm at the origin as
\be
\langle H, K \rangle \equiv \frac{1}{\mathcal{N}} \Tr \le H \mathcal{G}(K) \ri \, . 
\label{eq:inner_product}
\ee
Here $H$ and $K$ are traceless Hermitian Hamiltonians,    $\mathcal{G}$ is a positive-definite linear operator acting on  the algebra $\mathfrak{su}(N)$ as $\mathcal{G}(T_I)= q_I T_I$, where $T_I$ are the generators of the algebra and $q_I$ the corresponding penalty factors, and $\mathcal{N}$ is a normalization constant chosen such that $\Tr ( T^I T^J) = \mathcal{N}\delta^{IJ}$. 
Eq.~\eqref{eq:Riem_length} is equivalent to the definition of the $F_2$ norm according to eq.~\eqref{eq:cost_function_Finsler}.
Unitary complexity is given by the length \eqref{eq:Riem_length} of the geodesic $U(t)$ with boundary conditions $U(0)=\mathbf{1}$ and $U(t)=U_{T},$ the latter being a target unitary.
This corresponds to the definition \eqref{eq:unitary_complexity}.

We now consider a bipartition of the two-qudit system and decompose the Hamiltonian as in \eqref{eq:decomposition_Hamiltonian}. 
We define an $F_2$ norm corresponding to the following choice of $\mathcal{G}$
\be 
\mathcal{G}_{2}(H) =  \varepsilon S + Q \, , 
\label{eq:G2_appD}
\ee  
where $S$ and $Q$ represent respectively the local and non-local parts of $H$:
\be
S \equiv Y^a_A T^A_a \otimes \mathbf{1}^B +Y^i_B \mathbf{1}^A \otimes T^B_i  \, , \quad
Q \equiv Y^{ai} T^A_a \otimes T^B_i \, , \quad
H(t) = S(t) + Q(t) \, .
\label{eq:distinction_local_nl}
\ee
The geodesic equations in the space of unitaries with inner product \eqref{eq:inner_product} are the Euler-Arnold equations \cite{AIF_1966__16_1_319_0}\footnote{Euler-Arnold equations have been recently used to study complexity in the 
space of unitaries, see \eg \cite{Balasubramanian:2019wgd,Flory:2020eot,Flory:2020dja,Balasubramanian:2021mxo,Basteiro:2021ene}.}
 \be
\dot{H} = - i \mathcal{G}^{-1}([H,\mathcal{G}(H)]) \, .
\label{eq:Euler_Arnold}
 \ee
When working with the operator $\mathcal{G}_{2}$, we can use the
commutators $[S,S]=[Q,Q]=0$ and the fact that $[S,Q] $ is non-local to show that the 
Euler-Arnold equations take the form 
\be
\dot{S} = 0 \, , \qquad 		 
\dot{Q} = -i(1-\varepsilon)[S,Q] \, .
\ee
The $F_{2, \rm hom}$ norm in eq.~\eqref{eq:Fpenalties_flat} corresponds to~\eqref{eq:inner_product} with the operator $\mathcal{G}_{2, \rm hom} \equiv \lim_{\varepsilon \rightarrow 0} \mathcal{G}_2 $.
The solution for the Hamiltonian in this case reads
	\be
		H(t) = \lim_{\varepsilon \rightarrow 0} \le  S_0 +e^{-i (1-\varepsilon) S_0 t} Q_0 e^{i(1-\varepsilon)S_0 t} \ri 
		= S_0 +e^{-i S_0 t} Q_0 e^{i S_0 t} \, ,
		\label{eq:solution_Ham}
	\ee
where $S_0$ and $Q_0$ are the values of the local and non-local parts of the Hamiltonian at $t=0$, respectively.
By solving the Schr\"{o}dinger equation $\dot{U} = -i H U$ for the unitary along the curve, we get
\begin{align}
U(t) = \lim_{\varepsilon \rightarrow 0} e^{-i(1-\varepsilon) S_0 t}e^{-i (1-(1-\varepsilon)) S_0 t-i Q_0t} =  e^{-i  S_0 t}e^{-i Q_0t} \, .
\label{eq:app:unitary_path}
\end{align}
The length of this geodesic in the space of unitaries is
\be
\ell_{2, \rm hom} = \int_0^1 dt \, \lim_{\varepsilon \rightarrow 0} \sqrt{ \frac{\varepsilon}{\mathcal{N}} \Tr (S(t) S(t) ) + \frac{1}{\mathcal{N}} \Tr (Q(t) Q(t))} = 
\sqrt{\Tr \le \frac{Q_0 Q_0}{\mathcal{N}} \ri} \, ,
\label{eq:length_geodesic_app}
\ee
where we used the result \eqref{eq:solution_Ham} and the time-independence of $Q_0$ to directly perform the trivial integration.
Binding unitary complexity $\mathcal{BC}_{2, \rm hom}$ is defined as the minimal length \eqref{eq:length_geodesic_app} of the global geodesic connecting the identity with the target unitary $U_T$. 
We immediately observe that this corresponds to minimizing $\Tr(Q_0 Q_0)$, independently of the initial value $S_0$  of the local part of the Hamiltonian.

\subsection{Computation of binding state complexity}
\label{app:binding_compl_appendix}
Let us now move to the computation of binding state complexity.
First of all, we specify the reference and target states in the Hilbert space \eqref{eq:splitting_Hilbert_spaces} of the two qudits 
\begin{align}
\begin{split}
\ket{\psi_{R}} &= \ket{1 \tilde{1}} \, , 
\\
\ket{\psi_{T}} &= 
\sum_{m=1}^{N_A} \bar{\lambda}_m \ket{\bar{\phi}_m \bar{\chi}_m} = U_A \otimes U_B
\sum_{m=1}^{ N_A } \bar{\lambda}_m \ket{m \tilde{m} } \, ,
\quad
\bar{\lambda}_1 \geq \bar{\lambda}_2 \geq \dots \geq \bar{\lambda}_{N_A} \, ,
\label{eq:app:ref_target_states}
\end{split}
\end{align}
where $\ket{m \tilde{m}}$ is a fixed basis in the full Hilbert space. 
The state at a generic time along a geodesic trajectory is given by
\begin{align} \label{eq:app:geodsic}
\ket{\psi(t)} = e^{-i  S_0 t}e^{-i Q_0t} \ket{1\tilde{1}} \, .
\end{align}
The globally minimal geodesic is obtained by minimizing eq.~\eqref{eq:length_geodesic_app} over all the $Q_0$ that bring the reference state to a state with the same Schmidt coefficients $\bar{\lambda}_m$ as specified in eq.~\eqref{eq:app:ref_target_states}. 
Since $S_0$ can be chosen without affecting the length of the trajectory, we are free to act with LU operators (which modify the basis of the Schmidt decomposition) until we precisely reach the target state.

Let us further comment that there are Hamiltonians relating the reference and the target state continuously, \ie without a final LU transformation. If the target state $\ket{\psi_T}$ and the final state of the geodesic trajectory $| \tilde{\psi}_T \rangle = e^{-i H t} \ket{\psi_R}|_{t=1} $ are related by $ \ket{\psi_T} = e^{-i S_0} | \tilde{\psi}_T \rangle$ where $S_0$ is a time-independent generator of LU transformations, then the unitary $U(t)$ such that
\be
U(t) = e^{-i S_0 t}e^{-i Q_0 t}  \, ,
\ee
will have the same binding complexity as $e^{- i Q_0 t}$. 
In summary, we have shown by a different method than the one used in section \ref{ssec:reduction_set} 
that binding state complexity is defined in terms of equivalence classes of LUE reference and target states, see eq.~\eqref{eq:def_LUE}.

We are now going to show that the length of the geodesic in the space of states 
can be expressed in terms of the Fubini-Study distance on the projective space $\mathbb{CP}^{ N -1}$.
The Fubini-Study distance is proportional to the state complexity $\mathcal{C}^{\rm state}_{\rm IP}$, defined according to eq.~\eqref{eq:def_compl_state}, and associated with the standard inner-product norm on the unitary group $G = \mathrm{U}(N)$, which assigns penalties 1 to all the generators (either local or non-local), \ie
\begin{equation}
\begin{aligned}\label{app:eq:normalization}
    & \,  \mathcal{C}^{\rm state}_{\rm IP} (\ket{\psi_{T}}, \ket{\psi_{R}})  \equiv  
\min_{ \substack{ \lbrace U \in \mathrm{SU}(N) : \,
  \ket{\psi_T}
= U 
  \ket{\psi_R}
\rbrace } } \mathcal{C}_{F_{\rm IP}} [U], & \\
&  
F_{\text{IP}}[\vec{Y}] = \sqrt{\frac{\text{Tr}(H^2)}{\mathcal{N}}} =\sqrt{ \sum_{a}  |Y^a_A|^2+\sum_{i}  |Y^i_B|^2 +\sum_{a,i}|Y^{ai}|^2 } \,  ,  &
\end{aligned}
\end{equation}
where IP stands for inner product.
The norm on the projective space $\mathbb{CP}^{ N-1}$  obtained after the minimization over the stabilizer as in eq.~\eqref{eq:def_Fstate} is (proportional to) the Fubini-Study metric \cite{Fubini,Study}\footnote{The minimization can be done by expressing the elements of $H$ in a Hilbert space basis for which the first and second basis vectors are $\ket {\psi_1} \equiv \ket{\psi(t)}$ and $\ket {\psi_2} \equiv \left[
|\dot \psi(t) \rangle-\langle \psi(t) |\dot \psi(t) \rangle \ket{\psi(t)}\right]/\sqrt{ds^2_{\rm FS}}$. In this basis, $H_{1n}$ and $H_{n1}$ for any $n$ are fixed by the Schr\"{o}dinger equation, and the rest of the elements belong to the stabilizer over which we minimize. A further minimization over the time derivative of the global phase of the state $\ket{\psi(t)}$ will result in \eqref{FubiniFIP}.}
\be \label{FubiniFIP}
F^{\rm state}_{\text{IP}} dt = \sqrt{\frac{2}{\mathcal{N}} \, ds^2_{\rm FS}}
\ee
where 
\begin{equation}
\label{eq:Fubini_Study_metric}
 ds^2_{\rm FS} = 
 \langle d \psi \vert d \psi \rangle
- 
\langle d \psi \vert \psi \rangle \; 
\langle \psi \vert d \psi \rangle \, ,
\end{equation}  
and the associated distance reads
\begin{equation}
d_{\rm{FS}} (\ket{\psi_{T}}, \ket{\psi_{R}})=
\mathrm{arccos} \, |\langle\psi_T| \psi_R \rangle| \, .
\end{equation}
From this, we conclude that
\be
\mathcal{C}^{\rm state}_{\rm IP} = \sqrt{\frac{2}{\mathcal{N}}} \, d_{\rm{FS}}.
\label{eq:finite_IP_norm}
\ee
Now the main idea of this section is to define a metric over equivalence classes of states by minimizing the above metric over LUE reference and target states: 
\be
\begin{aligned}
& \, \mathcal{C}^{\rm state}_{2, \rm LUE } ( \{\ket{\psi_{T}} \},\{ \ket{\psi_{R} } \}) \equiv  
\min_{ \rm LUE } \mathcal{C}^{\rm state}_{\rm IP}( \ket{\psi_{T}} ,\ket{\psi_{R} } ) 
\end{aligned}\label{eq:app:compl_state_FS_LUE}
\ee
where the curly brackets denote equivalence classes of states.
Let us now explain how all this relates to the binding complexity defined in the main text, where zero cost is associated with local generators. A trivial observation is that cost functions with bigger penalty factors imply a bigger complexity, leading to $\mathcal{BC}^{\rm state}_{2, \rm hom} \leq \mathcal{C}^{\rm state}_{\rm IP} $. Next, by minimizing this inequality over LUE states, we find that $\mathcal{BC}^{\rm state}_{2, \rm hom} \leq
\mathcal{C}^{\rm state}_{2, \rm LUE}$, because the value of $\mathcal{BC}^{\rm state}_{2, \rm hom}$ does not change within the LUE class of states. On the other hand, taking the Hamiltonian of the optimal circuit for the $\mathcal{BC}^{\rm state}_{2, \rm hom}$ and setting $S_0=0$ will result in a different circuit connecting the reference state to a target state in the same LUE class. This circuit has the same cost according to either the inner-product metric, or the distance defined by eq.~\eqref{eq:length_geodesic_app}, therefore 
\be
\mathcal{BC}^{\rm state}_{2, \rm hom} =
\mathcal{C}^{\rm state}_{2, \rm LUE} \, .
\label{eq:app:chain_equalities}
\ee
Let us now exploit this result to find the binding complexity by actually computing the geodesic length using the norm \eqref{eq:app:compl_state_FS_LUE}.
The state complexity $\mathcal{C}^{\rm state}_{2, \rm LUE} $ is nothing but the length of a global geodesic in the geometry \eqref{FubiniFIP}, minimized over all LUE target states.
We then find 
\be
\mathcal{C}^{\rm state}_{2, \rm LUE} =  \sqrt{\frac{2}{\mathcal{N}}} \min_{ \{\ket{\psi_{T}} \}, \{\ket{\psi_{R}}  \} }  d_{\rm FS} (\ket{\psi_{T}}, \ket{\psi_{R}}) \, ,
\label{eq:relation_FubiniStudy}
\ee
where $d_{\rm FS}$ denotes the Fubini-Study distance between a given representative of the equivalence classes of the reference and target states. 
Recall that the Fubini-Study metric is left-invariant, and that we are minimizing modulo LUE states.
For these reasons, we can perform arbitrary local unitary transformations on the reference state 
\be
U_A \otimes U_B \ket{1 \tilde{1}} = \ket{\phi \chi } \, ,
\ee
and equivalently express the complexity as
\be
\mathcal{C}^{\rm state}_{2, \rm LUE} =\sqrt{\frac{2}{\mathcal{N}}}\min_{\phi,\chi} \, \mathrm{arccos} \, \le \sum_{m=1}^{N_A} \bra{\phi \chi} \bar{\lambda}_m \ket{m \tilde{m}}  \ri \, ,
\ee
where the definition \eqref{eq:app:ref_target_states} has been used.
By performing a change of basis $\ket{ \chi } = \sum_m a^*_m \ket{m} $, such that $\sum_m |a_m|^2 =1$, we find the chain of relations 
\be
\left| \bra{\phi \chi} \bar{\lambda}_m \ket{m \tilde{m}}  \right|
= \left| \bra{\phi} a_m \bar{\lambda}_m \ket{m }   \right| \leq 
| \bra{\phi} \ket{\phi} | \, | \bra{n} a_m a_n^* \bar{\lambda}_m \bar{\lambda}_n \ket{m} | =
\sqrt{ | a_m \bar{\lambda}_m|^2 } \leq \bar{\lambda}_1 \, .
\ee
In the second step we used the Cauchy-Schwarz inequality, while in the last step we used the fact that $\bar{\lambda}_1$ is the largest Schmidt coefficient.
Since the $\mathrm{arccos}$ is a monotonically decreasing function of its argument, we conclude that
\be
 \mathrm{arccos} \, \le \sum_{m=1}^{N_A} \bra{\phi \chi} \bar{\lambda}_m \ket{m \tilde{m}}  \ri \geq 
 \mathrm{arccos} (\bar{\lambda}_1) \, .
\ee
The minimum is actually achieved by constructing a unitary \eqref{eq:generic_path_unitaries} with a Hamiltonian which has the following non-local part at $t=0$
	\begin{align}
		Q_0 = - i \, \frac{ \mathrm{arccos} \, (\bar{\lambda}_1)}{\sqrt{1-\bar{\lambda}_1^2}} 
			\le \sum_{m=1}^{N_A}  
			\bar{\lambda}_m ( \ket{m\tilde m} \bra{1\tilde 1} - \ket{1\tilde 1} \bra{m\tilde m} \ri
	 \, .
  \label{eq:nonlocal_Ham_app}
	\end{align}
Plugging the explicit form of $Q_0$ inside eq.~\eqref{eq:length_geodesic_app}, we obtain that the length of this geodesic is indeed proportional to the Fubini-Study distance, with the proportionality factor determined by eq.~\eqref{eq:finite_IP_norm}. 
Using $\mathcal{N}= \mathcal{N}_A \mathcal{N}_B$,\footnote{Here, $\mathcal{N}_A$ and $ \mathcal{N}_B$ are the normalization constants of the traces of pairs of generators in $\mathrm{SU}(N_A)$ and $\mathrm{SU}(N_B)$ describing each of the qudits, respectively, as defined in the text below eq.~\eqref{eq:inner_product}.} this implies that the binding state complexity reads
\be
\mathcal{BC}^{\rm state}_{2, \rm hom} =
\mathcal{C}^{\rm state}_{2, \rm IP} =  \sqrt{\frac{2}{\mathcal{N}_A \mathcal{N}_B }}   \, \mathrm{arccos} (\bar{\lambda}_1) \, ,
\label{eq:result_compl_appD}
\ee
which matches the result \eqref{eq:bind_compl_GellMann}.

To conclude, we compare this procedure, which uses the Euler-Arnold equations, to the SO$(N_A)$ method presented in section \ref{sec:SON_picture}. 

\textbf{Advantages:}
\begin{itemize}
\item The use of Euler-Arnold equations may be more familiar to the reader.
    \item The trajectory in the unitary space is simple and smooth, see eq.~\eqref{eq:app:geodsic}.
\end{itemize}

\textbf{Disadvantages:}
\begin{itemize}
    \item The reduction of the problem to the space of Schmidt coefficients is not manifest until the end of the computation.
    \item This method only applies to the $F_2$ norm.
\end{itemize}

\bibliographystyle{unsrtnat}
\bibliography{bibliography}

\begin{thebibliography}{124}
\providecommand{\natexlab}[1]{#1}
\providecommand{\url}[1]{\texttt{#1}}
\expandafter\ifx\csname urlstyle\endcsname\relax
  \providecommand{\doi}[1]{doi: #1}\else
  \providecommand{\doi}{doi: \begingroup \urlstyle{rm}\Url}\fi

\bibitem[Watrous(2008)]{watrous2008quantum}
John Watrous.
\newblock {Quantum Computational Complexity}.
\newblock 4 2008.
\newblock \doi{10.1007/978-0-387-30440-3_428}.

\bibitem[Aaronson(2016)]{Aaronson:2016vto}
Scott Aaronson.
\newblock {The Complexity of Quantum States and Transformations: From Quantum Money to Black Holes}.
\newblock 7 2016.
\newblock \doi{10.48550/arXiv.1607.05256}.

\bibitem[Balasubramanian et~al.(2019)Balasubramanian, DeCross, Kar, and Parrikar]{Balasubramanian:2018hsu}
Vijay Balasubramanian, Matthew DeCross, Arjun Kar, and Onkar Parrikar.
\newblock {Binding Complexity and Multiparty Entanglement}.
\newblock \emph{JHEP}, 02:\penalty0 069, 2019.
\newblock \doi{10.1007/JHEP02(2019)069}.

\bibitem[Nielsen and Chuang(2010)]{nielsen_chuang_2010}
Michael~A. Nielsen and Isaac~L. Chuang.
\newblock \emph{Quantum Computation and Quantum Information: 10th Anniversary Edition}.
\newblock Cambridge University Press, 2010.
\newblock \doi{10.1017/CBO9780511976667}.

\bibitem[Nielsen(2006)]{Nielsen1}
Michael~A. Nielsen.
\newblock {A geometric approach to quantum circuit lower bounds}.
\newblock \emph{Quant. Inf. Comput.}, 6\penalty0 (3):\penalty0 213--262, 2006.
\newblock \doi{10.26421/QIC6.3-2}.

\bibitem[Dowling and Nielsen(2008)]{Nielsen3}
Mark~R. Dowling and Michael~A. Nielsen.
\newblock {The geometry of quantum computation}.
\newblock \emph{Quant. Inf. Comput.}, 8\penalty0 (10):\penalty0 0861--0899, 2008.
\newblock \doi{10.26421/QIC8.10-1}.

\bibitem[Nielsen et~al.(2006)Nielsen, Dowling, Gu, and Doherty]{Nielsen_2006}
Michael~A. Nielsen, Mark~R. Dowling, Mile Gu, and Andrew~C. Doherty.
\newblock Quantum computation as geometry.
\newblock \emph{Science}, 311\penalty0 (5764):\penalty0 1133--1135, feb 2006.
\newblock \doi{10.1126/science.1121541}.
\newblock URL \url{https://doi.org/10.1126%2Fscience.1121541}.

\bibitem[Carlini et~al.(2006)Carlini, Hosoya, Koike, and Okudaira]{carlini:2006}
Alberto Carlini, Akio Hosoya, Tatsuhiko Koike, and Yosuke Okudaira.
\newblock Time-optimal quantum evolution.
\newblock \emph{Phys. Rev. Lett.}, 96:\penalty0 060503, Feb 2006.
\newblock \doi{10.1103/PhysRevLett.96.060503}.
\newblock URL \url{https://link.aps.org/doi/10.1103/PhysRevLett.96.060503}.

\bibitem[Werschnik and Gross(2007)]{werschnik:2007}
J~Werschnik and EKU Gross.
\newblock Quantum optimal control theory.
\newblock \emph{Journal of Physics B: Atomic, Molecular and Optical Physics}, 40\penalty0 (18):\penalty0 R175, 2007.
\newblock \doi{10.1088/0953-4075/40/18/R01}.
\newblock URL \url{https://dx.doi.org/10.1088/0953-4075/40/18/R01}.

\bibitem[Koch et~al.(2022)]{koch:2022gll}
Christiane~P. Koch et~al.
\newblock {Quantum optimal control in quantum technologies. Strategic report on current status, visions and goals for research in Europe}.
\newblock \emph{EPJ Quant. Technol.}, 9\penalty0 (1):\penalty0 19, 2022.
\newblock \doi{10.1140/epjqt/s40507-022-00138-x}.

\bibitem[Hartman and Maldacena(2013)]{Hartman:2013qma}
Thomas Hartman and Juan Maldacena.
\newblock {Time Evolution of Entanglement Entropy from Black Hole Interiors}.
\newblock \emph{JHEP}, 05:\penalty0 014, 2013.
\newblock \doi{10.1007/JHEP05(2013)014}.

\bibitem[Susskind(2016{\natexlab{a}})]{Susskind:2014moa}
Leonard Susskind.
\newblock {Entanglement is not enough}.
\newblock \emph{Fortsch. Phys.}, 64:\penalty0 49--71, 2016{\natexlab{a}}.
\newblock \doi{10.1002/prop.201500095}.

\bibitem[Freivogel et~al.(2015)Freivogel, Jefferson, Kabir, Mosk, and Yang]{Freivogel:2015}
Ben Freivogel, Ro~Jefferson, Laurens Kabir, Benjamin Mosk, and I-Sheng Yang.
\newblock {Casting Shadows on Holographic Reconstruction}.
\newblock \emph{Phys. Rev. D}, 91\penalty0 (8):\penalty0 086013, 2015.
\newblock \doi{10.1103/PhysRevD.91.086013}.

\bibitem[Balasubramanian et~al.(2015)Balasubramanian, Chowdhury, Czech, and de~Boer]{Balasubramanian:2014sra}
Vijay Balasubramanian, Borun~D. Chowdhury, Bartlomiej Czech, and Jan de~Boer.
\newblock {Entwinement and the emergence of spacetime}.
\newblock \emph{JHEP}, 01:\penalty0 048, 2015.
\newblock \doi{10.1007/JHEP01(2015)048}.

\bibitem[Susskind(2016{\natexlab{b}})]{Susskind:2014rva}
Leonard Susskind.
\newblock {Computational Complexity and Black Hole Horizons}.
\newblock \emph{Fortsch. Phys.}, 64:\penalty0 24--43, 2016{\natexlab{b}}.
\newblock \doi{10.1002/prop.201500092}.
\newblock [Addendum: Fortsch.Phys. 64, 44--48 (2016), \href{https://onlinelibrary.wiley.com/doi/10.1002/prop.201500093}{DOI: 10.1002/prop.201500093}]".

\bibitem[Stanford and Susskind(2014)]{Stanford:2014jda}
Douglas Stanford and Leonard Susskind.
\newblock {Complexity and Shock Wave Geometries}.
\newblock \emph{Phys. Rev. D}, 90\penalty0 (12):\penalty0 126007, 2014.
\newblock \doi{10.1103/PhysRevD.90.126007}.

\bibitem[Belin et~al.(2019)Belin, Lewkowycz, and S\'arosi]{Belin:2018bpg}
Alexandre Belin, Aitor Lewkowycz, and G\'abor S\'arosi.
\newblock {Complexity and the bulk volume, a new York time story}.
\newblock \emph{JHEP}, 03:\penalty0 044, 2019.
\newblock \doi{10.1007/JHEP03(2019)044}.

\bibitem[Rabinovici et~al.(2023)Rabinovici, S\'anchez-Garrido, Shir, and Sonner]{Rabinovici:2023yex}
E.~Rabinovici, A.~S\'anchez-Garrido, R.~Shir, and J.~Sonner.
\newblock {A bulk manifestation of Krylov complexity}.
\newblock \emph{JHEP}, 08:\penalty0 213, 2023.
\newblock \doi{10.1007/JHEP08(2023)213}.

\bibitem[Brown et~al.(2016{\natexlab{a}})Brown, Roberts, Susskind, Swingle, and Zhao]{Brown:2015lvg}
Adam~R. Brown, Daniel~A. Roberts, Leonard Susskind, Brian Swingle, and Ying Zhao.
\newblock {Complexity, action, and black holes}.
\newblock \emph{Phys. Rev. D}, 93\penalty0 (8):\penalty0 086006, 2016{\natexlab{a}}.
\newblock \doi{10.1103/PhysRevD.93.086006}.

\bibitem[Brown et~al.(2016{\natexlab{b}})Brown, Roberts, Susskind, Swingle, and Zhao]{Brown:2015bva}
Adam~R. Brown, Daniel~A. Roberts, Leonard Susskind, Brian Swingle, and Ying Zhao.
\newblock {Holographic Complexity Equals Bulk Action?}
\newblock \emph{Phys. Rev. Lett.}, 116\penalty0 (19):\penalty0 191301, 2016{\natexlab{b}}.
\newblock \doi{10.1103/PhysRevLett.116.191301}.

\bibitem[Couch et~al.(2017)Couch, Fischler, and Nguyen]{Couch:2016exn}
Josiah Couch, Willy Fischler, and Phuc~H. Nguyen.
\newblock {Noether charge, black hole volume, and complexity}.
\newblock \emph{JHEP}, 03:\penalty0 119, 2017.
\newblock \doi{10.1007/JHEP03(2017)119}.

\bibitem[Belin et~al.(2022)Belin, Myers, Ruan, S\'arosi, and Speranza]{Belin:2021bga}
Alexandre Belin, Robert~C. Myers, Shan-Ming Ruan, G\'abor S\'arosi, and Antony~J. Speranza.
\newblock {Does Complexity Equal Anything?}
\newblock \emph{Phys. Rev. Lett.}, 128\penalty0 (8):\penalty0 081602, 2022.
\newblock \doi{10.1103/PhysRevLett.128.081602}.

\bibitem[Belin et~al.(2023)Belin, Myers, Ruan, S\'arosi, and Speranza]{Belin:2022xmt}
Alexandre Belin, Robert~C. Myers, Shan-Ming Ruan, G\'abor S\'arosi, and Antony~J. Speranza.
\newblock {Complexity equals anything II}.
\newblock \emph{JHEP}, 01:\penalty0 154, 2023.
\newblock \doi{10.1007/JHEP01(2023)154}.

\bibitem[J\o{}rstad et~al.(2023)J\o{}rstad, Myers, and Ruan]{Jorstad:2023kmq}
Eivind J\o{}rstad, Robert~C. Myers, and Shan-Ming Ruan.
\newblock {Complexity=anything: singularity probes}.
\newblock \emph{JHEP}, 07:\penalty0 223, 2023.
\newblock \doi{10.1007/JHEP07(2023)223}.

\bibitem[Erdmenger et~al.(2022)Erdmenger, Flory, Gerbershagen, Heller, and Weigel]{Erdmenger:2021wzc}
Johanna Erdmenger, Mario Flory, Marius Gerbershagen, Michal~P. Heller, and Anna-Lena Weigel.
\newblock {Exact Gravity Duals for Simple Quantum Circuits}.
\newblock \emph{SciPost Phys.}, 13\penalty0 (3):\penalty0 061, 2022.
\newblock \doi{10.21468/SciPostPhys.13.3.061}.

\bibitem[Erdmenger et~al.(2023)Erdmenger, Weigel, Gerbershagen, and Heller]{Erdmenger:2022lov}
Johanna Erdmenger, Anna-Lena Weigel, Marius Gerbershagen, and Michal~P. Heller.
\newblock {From complexity geometry to holographic spacetime}.
\newblock \emph{Phys. Rev. D}, 108\penalty0 (10):\penalty0 106020, 2023.
\newblock \doi{10.1103/PhysRevD.108.106020}.

\bibitem[Chandra et~al.(2023)Chandra, de~Boer, Flory, Heller, H\"ortner, and Rolph]{Chandra:2022pgl}
A.~Ramesh Chandra, Jan de~Boer, Mario Flory, Michal~P. Heller, Sergio H\"ortner, and Andrew Rolph.
\newblock {Cost of holographic path integrals}.
\newblock \emph{SciPost Phys.}, 14\penalty0 (4):\penalty0 061, 2023.
\newblock \doi{10.21468/SciPostPhys.14.4.061}.

\bibitem[Lehner et~al.(2016)Lehner, Myers, Poisson, and Sorkin]{Lehner:2016vdi}
Luis Lehner, Robert~C. Myers, Eric Poisson, and Rafael~D. Sorkin.
\newblock {Gravitational action with null boundaries}.
\newblock \emph{Phys. Rev. D}, 94\penalty0 (8):\penalty0 084046, 2016.
\newblock \doi{10.1103/PhysRevD.94.084046}.

\bibitem[Barbon and Rabinovici(2016)]{Barbon:2015ria}
Jose L.~F. Barbon and Eliezer Rabinovici.
\newblock {Holographic complexity and spacetime singularities}.
\newblock \emph{JHEP}, 01:\penalty0 084, 2016.
\newblock \doi{10.1007/JHEP01(2016)084}.

\bibitem[Chapman et~al.(2017)Chapman, Marrochio, and Myers]{Chapman:2016hwi}
Shira Chapman, Hugo Marrochio, and Robert~C. Myers.
\newblock {Complexity of Formation in Holography}.
\newblock \emph{JHEP}, 01:\penalty0 062, 2017.
\newblock \doi{10.1007/JHEP01(2017)062}.

\bibitem[Cai et~al.(2016)Cai, Ruan, Wang, Yang, and Peng]{Cai:2016xho}
Rong-Gen Cai, Shan-Ming Ruan, Shao-Jiang Wang, Run-Qiu Yang, and Rong-Hui Peng.
\newblock {Action growth for AdS black holes}.
\newblock \emph{JHEP}, 09:\penalty0 161, 2016.
\newblock \doi{10.1007/JHEP09(2016)161}.

\bibitem[Carmi et~al.(2017{\natexlab{a}})Carmi, Chapman, Marrochio, Myers, and Sugishita]{Carmi:2017jqz}
Dean Carmi, Shira Chapman, Hugo Marrochio, Robert~C. Myers, and Sotaro Sugishita.
\newblock {On the Time Dependence of Holographic Complexity}.
\newblock \emph{JHEP}, 11:\penalty0 188, 2017{\natexlab{a}}.
\newblock \doi{10.1007/JHEP11(2017)188}.

\bibitem[Chapman et~al.(2018{\natexlab{a}})Chapman, Marrochio, and Myers]{Chapman:2018dem}
Shira Chapman, Hugo Marrochio, and Robert~C. Myers.
\newblock {Holographic complexity in Vaidya spacetimes. Part I}.
\newblock \emph{JHEP}, 06:\penalty0 046, 2018{\natexlab{a}}.
\newblock \doi{10.1007/JHEP06(2018)046}.

\bibitem[Chapman et~al.(2018{\natexlab{b}})Chapman, Marrochio, and Myers]{Chapman:2018lsv}
Shira Chapman, Hugo Marrochio, and Robert~C. Myers.
\newblock {Holographic complexity in Vaidya spacetimes. Part II}.
\newblock \emph{JHEP}, 06:\penalty0 114, 2018{\natexlab{b}}.
\newblock \doi{10.1007/JHEP06(2018)114}.

\bibitem[Chapman et~al.(2019{\natexlab{a}})Chapman, Ge, and Policastro]{Chapman:2018bqj}
Shira Chapman, Dongsheng Ge, and Giuseppe Policastro.
\newblock {Holographic Complexity for Defects Distinguishes Action from Volume}.
\newblock \emph{JHEP}, 05:\penalty0 049, 2019{\natexlab{a}}.
\newblock \doi{10.1007/JHEP05(2019)049}.

\bibitem[Braccia et~al.(2020)Braccia, Cotrone, and Tonni]{Braccia:2019xxi}
Paolo Braccia, Aldo~L. Cotrone, and Erik Tonni.
\newblock {Complexity in the presence of a boundary}.
\newblock \emph{JHEP}, 02:\penalty0 051, 2020.
\newblock \doi{10.1007/JHEP02(2020)051}.

\bibitem[Sato and Watanabe(2019)]{Sato:2019kik}
Yoshiki Sato and Kento Watanabe.
\newblock {Does Boundary Distinguish Complexities?}
\newblock \emph{JHEP}, 11:\penalty0 132, 2019.
\newblock \doi{10.1007/JHEP11(2019)132}.

\bibitem[Bernamonti et~al.(2019)Bernamonti, Galli, Hernandez, Myers, Ruan, and Sim\'on]{Bernamonti:2019zyy}
Alice Bernamonti, Federico Galli, Juan Hernandez, Robert~C. Myers, Shan-Ming Ruan, and Joan Sim\'on.
\newblock {First Law of Holographic Complexity}.
\newblock \emph{Phys. Rev. Lett.}, 123\penalty0 (8):\penalty0 081601, 2019.
\newblock \doi{10.1103/PhysRevLett.123.081601}.

\bibitem[Bernamonti et~al.(2020)Bernamonti, Galli, Hernandez, Myers, Ruan, and Sim\'on]{Bernamonti:2020bcf}
Alice Bernamonti, Federico Galli, Juan Hernandez, Robert~C. Myers, Shan-Ming Ruan, and Joan Sim\'on.
\newblock {Aspects of The First Law of Complexity}.
\newblock \emph{J. Phys. A}, 53:\penalty0 29, 2020.
\newblock \doi{10.1088/1751-8121/ab8e66}.

\bibitem[Chapman et~al.(2022)Chapman, Galante, and Kramer]{Chapman:2021eyy}
Shira Chapman, Dami\'an~A. Galante, and Eric~David Kramer.
\newblock {Holographic complexity and de Sitter space}.
\newblock \emph{JHEP}, 02:\penalty0 198, 2022.
\newblock \doi{10.1007/JHEP02(2022)198}.

\bibitem[Auzzi et~al.(2021{\natexlab{a}})Auzzi, Baiguera, Bonansea, Nardelli, and Toccacelo]{Auzzi:2021nrj}
Roberto Auzzi, Stefano Baiguera, Sara Bonansea, Giuseppe Nardelli, and Kristian Toccacelo.
\newblock {Volume complexity for Janus AdS$_{3}$ geometries}.
\newblock \emph{JHEP}, 08:\penalty0 045, 2021{\natexlab{a}}.
\newblock \doi{10.1007/JHEP08(2021)045}.

\bibitem[Baiguera et~al.(2021)Baiguera, Bonansea, and Toccacelo]{Baiguera:2021cba}
Stefano Baiguera, Sara Bonansea, and Kristian Toccacelo.
\newblock {Volume complexity for the nonsupersymmetric Janus AdS5 geometry}.
\newblock \emph{Phys. Rev. D}, 104\penalty0 (8):\penalty0 086030, 2021.
\newblock \doi{10.1103/PhysRevD.104.086030}.

\bibitem[Auzzi et~al.(2022)Auzzi, Baiguera, Bonansea, and Nardelli]{Auzzi:2021ozb}
Roberto Auzzi, Stefano Baiguera, Sara Bonansea, and Giuseppe Nardelli.
\newblock {Action complexity in the presence of defects and boundaries}.
\newblock \emph{JHEP}, 02:\penalty0 118, 2022.
\newblock \doi{10.1007/JHEP02(2022)118}.

\bibitem[Emparan et~al.(2022)Emparan, Frassino, Sasieta, and Toma\v{s}evi\'c]{Emparan:2021hyr}
Roberto Emparan, Antonia~Micol Frassino, Martin Sasieta, and Marija Toma\v{s}evi\'c.
\newblock {Holographic complexity of quantum black holes}.
\newblock \emph{JHEP}, 02:\penalty0 204, 2022.
\newblock \doi{10.1007/JHEP02(2022)204}.

\bibitem[J\o{}rstad et~al.(2022)J\o{}rstad, Myers, and Ruan]{Jorstad:2022mls}
Eivind J\o{}rstad, Robert~C. Myers, and Shan-Ming Ruan.
\newblock {Holographic complexity in dS$_{d+1}$}.
\newblock \emph{JHEP}, 05:\penalty0 119, 2022.
\newblock \doi{10.1007/JHEP05(2022)119}.

\bibitem[Auzzi et~al.(2023)Auzzi, Nardelli, Ungureanu, and Zenoni]{Auzzi:2023qbm}
Roberto Auzzi, Giuseppe Nardelli, Gabriel~Pedde Ungureanu, and Nicolo Zenoni.
\newblock {Volume complexity of dS bubbles}.
\newblock \emph{Phys. Rev. D}, 108\penalty0 (2):\penalty0 026006, 2023.
\newblock \doi{10.1103/PhysRevD.108.026006}.

\bibitem[Anegawa et~al.(2023)Anegawa, Iizuka, Sake, and Zenoni]{Anegawa:2023wrk}
Takanori Anegawa, Norihiro Iizuka, Sunil~Kumar Sake, and Nicol\`o Zenoni.
\newblock {Is action complexity better for de Sitter space in Jackiw-Teitelboim gravity?}
\newblock \emph{JHEP}, 06:\penalty0 213, 2023.
\newblock \doi{10.1007/JHEP06(2023)213}.

\bibitem[Anegawa and Iizuka(2023)]{Anegawa:2023dad}
Takanori Anegawa and Norihiro Iizuka.
\newblock {Shock waves and delay of hyperfast growth in de Sitter complexity}.
\newblock \emph{JHEP}, 08:\penalty0 115, 2023.
\newblock \doi{10.1007/JHEP08(2023)115}.

\bibitem[Baiguera et~al.(2023)Baiguera, Berman, Chapman, and Myers]{Baiguera:2023tpt}
Stefano Baiguera, Rotem Berman, Shira Chapman, and Robert~C. Myers.
\newblock {The cosmological switchback effect}.
\newblock \emph{JHEP}, 07:\penalty0 162, 2023.
\newblock \doi{10.1007/JHEP07(2023)162}.

\bibitem[Aguilar-Gutierrez et~al.(2023)Aguilar-Gutierrez, Heller, and Van~der Schueren]{Aguilar-Gutierrez:2023zqm}
Sergio~E. Aguilar-Gutierrez, Michal~P. Heller, and Silke Van~der Schueren.
\newblock {Complexity = Anything Can Grow Forever in de Sitter}.
\newblock 5 2023.
\newblock \doi{10.48550/arXiv.2305.11280}.

\bibitem[Aguilar-Gutierrez(2024)]{Aguilar-Gutierrez:2023pnn}
Sergio~E. Aguilar-Gutierrez.
\newblock {C=Anything and the switchback effect in Schwarzschild-de Sitter space}.
\newblock \emph{JHEP}, 03:\penalty0 062, 2024.
\newblock \doi{10.1007/JHEP03(2024)062}.

\bibitem[Akhavan and Omidi(2019)]{Akhavan:2019zax}
Amin Akhavan and Farzad Omidi.
\newblock {On the Role of Counterterms in Holographic Complexity}.
\newblock \emph{JHEP}, 11:\penalty0 054, 2019.
\newblock \doi{10.1007/JHEP11(2019)054}.

\bibitem[Omidi(2020)]{Omidi:2020oit}
Farzad Omidi.
\newblock {Regularizations of Action-Complexity for a Pure BTZ Black Hole Microstate}.
\newblock \emph{JHEP}, 07:\penalty0 020, 2020.
\newblock \doi{10.1007/JHEP07(2020)020}.

\bibitem[Caputa et~al.(2017)Caputa, Kundu, Miyaji, Takayanagi, and Watanabe]{Caputa:2017yrh}
Pawel Caputa, Nilay Kundu, Masamichi Miyaji, Tadashi Takayanagi, and Kento Watanabe.
\newblock {Liouville Action as Path-Integral Complexity: From Continuous Tensor Networks to AdS/CFT}.
\newblock \emph{JHEP}, 11:\penalty0 097, 2017.
\newblock \doi{10.1007/JHEP11(2017)097}.

\bibitem[Jefferson and Myers(2017)]{Jefferson:2017sdb}
Ro~Jefferson and Robert~C. Myers.
\newblock {Circuit complexity in quantum field theory}.
\newblock \emph{JHEP}, 10:\penalty0 107, 2017.
\newblock \doi{10.1007/JHEP10(2017)107}.

\bibitem[Chapman et~al.(2018{\natexlab{c}})Chapman, Heller, Marrochio, and Pastawski]{Chapman:2017rqy}
Shira Chapman, Michal~P. Heller, Hugo Marrochio, and Fernando Pastawski.
\newblock {Toward a Definition of Complexity for Quantum Field Theory States}.
\newblock \emph{Phys. Rev. Lett.}, 120\penalty0 (12):\penalty0 121602, 2018{\natexlab{c}}.
\newblock \doi{10.1103/PhysRevLett.120.121602}.

\bibitem[Khan et~al.(2018)Khan, Krishnan, and Sharma]{Khan:2018rzm}
Rifath Khan, Chethan Krishnan, and Sanchita Sharma.
\newblock {Circuit Complexity in Fermionic Field Theory}.
\newblock \emph{Phys. Rev. D}, 98\penalty0 (12):\penalty0 126001, 2018.
\newblock \doi{10.1103/PhysRevD.98.126001}.

\bibitem[Bhattacharyya et~al.(2018)Bhattacharyya, Caputa, Das, Kundu, Miyaji, and Takayanagi]{Bhattacharyya:2018wym}
Arpan Bhattacharyya, Pawel Caputa, Sumit~R. Das, Nilay Kundu, Masamichi Miyaji, and Tadashi Takayanagi.
\newblock {Path-Integral Complexity for Perturbed CFTs}.
\newblock \emph{JHEP}, 07:\penalty0 086, 2018.
\newblock \doi{10.1007/JHEP07(2018)086}.

\bibitem[Chapman et~al.(2019{\natexlab{b}})Chapman, Eisert, Hackl, Heller, Jefferson, Marrochio, and Myers]{Chapman:2018hou}
Shira Chapman, Jens Eisert, Lucas Hackl, Michal~P. Heller, Ro~Jefferson, Hugo Marrochio, and Robert~C. Myers.
\newblock {Complexity and entanglement for thermofield double states}.
\newblock \emph{SciPost Phys.}, 6\penalty0 (3):\penalty0 034, 2019{\natexlab{b}}.
\newblock \doi{10.21468/SciPostPhys.6.3.034}.

\bibitem[Camargo et~al.(2019)Camargo, Caputa, Das, Heller, and Jefferson]{Camargo:2018eof}
Hugo~A. Camargo, Pawel Caputa, Diptarka Das, Michal~P. Heller, and Ro~Jefferson.
\newblock {Complexity as a novel probe of quantum quenches: universal scalings and purifications}.
\newblock \emph{Phys. Rev. Lett.}, 122\penalty0 (8):\penalty0 081601, 2019.
\newblock \doi{10.1103/PhysRevLett.122.081601}.

\bibitem[Ge and Policastro(2019)]{Ge:2019mjt}
Dongsheng Ge and Giuseppe Policastro.
\newblock {Circuit Complexity and 2D Bosonisation}.
\newblock \emph{JHEP}, 10:\penalty0 276, 2019.
\newblock \doi{10.1007/JHEP10(2019)276}.

\bibitem[Brown and Susskind(2019)]{Brown:2019whu}
Adam~R. Brown and Leonard Susskind.
\newblock {Complexity geometry of a single qubit}.
\newblock \emph{Phys. Rev. D}, 100\penalty0 (4):\penalty0 046020, 2019.
\newblock \doi{10.1103/PhysRevD.100.046020}.

\bibitem[Balasubramanian et~al.(2020)Balasubramanian, Decross, Kar, and Parrikar]{Balasubramanian:2019wgd}
Vijay Balasubramanian, Matthew Decross, Arjun Kar, and Onkar Parrikar.
\newblock {Quantum Complexity of Time Evolution with Chaotic Hamiltonians}.
\newblock \emph{JHEP}, 01:\penalty0 134, 2020.
\newblock \doi{10.1007/JHEP01(2020)134}.

\bibitem[Chapman and Chen(2021)]{Chapman:2019clq}
Shira Chapman and Hong~Zhe Chen.
\newblock {Charged Complexity and the Thermofield Double State}.
\newblock \emph{JHEP}, 02:\penalty0 187, 2021.
\newblock \doi{10.1007/JHEP02(2021)187}.

\bibitem[Caputa and Magan(2019)]{Caputa:2018kdj}
Pawel Caputa and Javier~M. Magan.
\newblock {Quantum Computation as Gravity}.
\newblock \emph{Phys. Rev. Lett.}, 122\penalty0 (23):\penalty0 231302, 2019.
\newblock \doi{10.1103/PhysRevLett.122.231302}.

\bibitem[Auzzi et~al.(2021{\natexlab{b}})Auzzi, Baiguera, De~Luca, Legramandi, Nardelli, and Zenoni]{Auzzi:2020idm}
Roberto Auzzi, Stefano Baiguera, G.~Bruno De~Luca, Andrea Legramandi, Giuseppe Nardelli, and Nicol\`o Zenoni.
\newblock {Geometry of quantum complexity}.
\newblock \emph{Phys. Rev. D}, 103\penalty0 (10):\penalty0 106021, 2021{\natexlab{b}}.
\newblock \doi{10.1103/PhysRevD.103.106021}.

\bibitem[Caginalp and Leutheusser(2020)]{Caginalp:2020tzw}
Reginald~J. Caginalp and Samuel Leutheusser.
\newblock {Complexity in One- and Two-Qubit Systems}.
\newblock 10 2020.
\newblock \doi{10.48550/arXiv.2010.15099}.

\bibitem[Flory and Heller(2020{\natexlab{a}})]{Flory:2020eot}
Mario Flory and Michal~P. Heller.
\newblock {Geometry of Complexity in Conformal Field Theory}.
\newblock \emph{Phys. Rev. Res.}, 2\penalty0 (4):\penalty0 043438, 2020{\natexlab{a}}.
\newblock \doi{10.1103/PhysRevResearch.2.043438}.

\bibitem[Flory and Heller(2020{\natexlab{b}})]{Flory:2020dja}
Mario Flory and Michal~P. Heller.
\newblock {Conformal field theory complexity from Euler-Arnold equations}.
\newblock \emph{JHEP}, 12:\penalty0 091, 2020{\natexlab{b}}.
\newblock \doi{10.1007/JHEP12(2020)091}.

\bibitem[Chagnet et~al.(2022)Chagnet, Chapman, de~Boer, and Zukowski]{Chagnet:2021uvi}
Nicolas Chagnet, Shira Chapman, Jan de~Boer, and Claire Zukowski.
\newblock {Complexity for Conformal Field Theories in General Dimensions}.
\newblock \emph{Phys. Rev. Lett.}, 128\penalty0 (5):\penalty0 051601, 2022.
\newblock \doi{10.1103/PhysRevLett.128.051601}.

\bibitem[Basteiro et~al.(2022)Basteiro, Erdmenger, Fries, Goth, Matthaiakakis, and Meyer]{Basteiro:2021ene}
Pablo Basteiro, Johanna Erdmenger, Pascal Fries, Florian Goth, Ioannis Matthaiakakis, and Ren\'e Meyer.
\newblock {Quantum complexity as hydrodynamics}.
\newblock \emph{Phys. Rev. D}, 106\penalty0 (6):\penalty0 065016, 2022.
\newblock \doi{10.1103/PhysRevD.106.065016}.

\bibitem[Brown et~al.(2023)Brown, Freedman, Lin, and Susskind]{Brown:2021rmz}
Adam~R. Brown, Michael~H. Freedman, Henry~W. Lin, and Leonard Susskind.
\newblock {Universality in long-distance geometry and quantum complexity}.
\newblock \emph{Nature}, 622\penalty0 (7981):\penalty0 58--62, 2023.
\newblock \doi{10.1038/s41586-023-06460-3}.

\bibitem[Balasubramanian et~al.(2021)Balasubramanian, DeCross, Kar, Li, and Parrikar]{Balasubramanian:2021mxo}
Vijay Balasubramanian, Matthew DeCross, Arjun Kar, Yue~(Cathy) Li, and Onkar Parrikar.
\newblock {Complexity growth in integrable and chaotic models}.
\newblock \emph{JHEP}, 07:\penalty0 011, 2021.
\newblock \doi{10.1007/JHEP07(2021)011}.

\bibitem[Brown(2024)]{Brown:2022phc}
Adam~R. Brown.
\newblock {Polynomial Equivalence of Complexity Geometries}.
\newblock \emph{Quantum}, 8:\penalty0 1391, 2024.
\newblock \doi{10.22331/q-2024-07-02-1391}.

\bibitem[Craps et~al.(2024{\natexlab{a}})Craps, Clerck, Evnin, and Hacker]{10.21468/SciPostPhys.16.2.041}
Ben Craps, Marine~De Clerck, Oleg Evnin, and Philip Hacker.
\newblock {Integrability and complexity in quantum spin chains}.
\newblock \emph{SciPost Phys.}, 16:\penalty0 041, 2024{\natexlab{a}}.
\newblock \doi{10.21468/SciPostPhys.16.2.041}.
\newblock URL \url{https://scipost.org/10.21468/SciPostPhys.16.2.041}.

\bibitem[Craps et~al.(2022)Craps, De~Clerck, Evnin, Hacker, and Pavlov]{Craps:2022ese}
Ben Craps, Marine De~Clerck, Oleg Evnin, Philip Hacker, and Maxim Pavlov.
\newblock {Bounds on quantum evolution complexity via lattice cryptography}.
\newblock \emph{SciPost Phys.}, 13\penalty0 (4):\penalty0 090, 2022.
\newblock \doi{10.21468/SciPostPhys.13.4.090}.

\bibitem[Chapman and Policastro(2022)]{Chapman:2021jbh}
Shira Chapman and Giuseppe Policastro.
\newblock {Quantum computational complexity from quantum information to black holes and back}.
\newblock \emph{Eur. Phys. J. C}, 82\penalty0 (2):\penalty0 128, 2022.
\newblock \doi{10.1140/epjc/s10052-022-10037-1}.

\bibitem[Meter et~al.(2008)Meter, Munro, Nemoto, and Itoh]{meter2008arithmetic}
Rodney~Van Meter, W.~J. Munro, Kae Nemoto, and Kohei~M. Itoh.
\newblock Arithmetic on a distributed-memory quantum multicomputer.
\newblock \emph{{ACM} Journal on Emerging Technologies in Computing Systems}, 3\penalty0 (4):\penalty0 1--23, jan 2008.
\newblock \doi{10.1145/1324177.1324179}.
\newblock URL \url{https://doi.org/10.1145%2F1324177.1324179}.

\bibitem[Beals et~al.(2013)Beals, Brierley, Gray, Harrow, Kutin, Linden, Shepherd, and Stather]{beals2013efficient}
Robert Beals, Stephen Brierley, Oliver Gray, Aram~W. Harrow, Samuel Kutin, Noah Linden, Dan Shepherd, and Mark Stather.
\newblock Efficient distributed quantum computing.
\newblock \emph{Proceedings of the Royal Society A: Mathematical, Physical and Engineering Sciences}, 469\penalty0 (2153):\penalty0 20120686, may 2013.
\newblock \doi{10.1098/rspa.2012.0686}.
\newblock URL \url{https://doi.org/10.1098%2Frspa.2012.0686}.

\bibitem[Caleffi et~al.(2022)Caleffi, Amoretti, Ferrari, Cuomo, Illiano, Manzalini, and Cacciapuoti]{Caleffi:2022wxp}
Marcello Caleffi, Michele Amoretti, Davide Ferrari, Daniele Cuomo, Jessica Illiano, Antonio Manzalini, and Angela~Sara Cacciapuoti.
\newblock {Distributed Quantum Computing: a Survey}.
\newblock 12 2022.
\newblock \doi{10.48550/arXiv.2212.10609}.

\bibitem[Buhrman et~al.(2010)Buhrman, Cleve, Massar, and de~Wolf]{buhrman:2010}
Harry Buhrman, Richard Cleve, Serge Massar, and Ronald de~Wolf.
\newblock Nonlocality and communication complexity.
\newblock \emph{Reviews of Modern Physics}, 82\penalty0 (1):\penalty0 665--698, mar 2010.
\newblock \doi{10.1103/revmodphys.82.665}.
\newblock URL \url{https://doi.org/10.1103%2Frevmodphys.82.665}.

\bibitem[Rudnicki(2021)]{rudnicki:2021}
{\L}ukasz Rudnicki.
\newblock Quantum speed limit and geometric measure of entanglement.
\newblock \emph{Physical Review A}, 104\penalty0 (3), sep 2021.
\newblock \doi{10.1103/physreva.104.032417}.
\newblock URL \url{https://doi.org/10.1103%2Fphysreva.104.032417}.

\bibitem[Zolfi(2023)]{Zolfi:2023bdp}
Hamed Zolfi.
\newblock {Complexity and Multi-boundary Wormholes in 2 + 1 dimensions}.
\newblock \emph{JHEP}, 04:\penalty0 076, 2023.
\newblock \doi{10.1007/JHEP04(2023)076}.

\bibitem[Zhang(2022)]{Zhang:2021xwx}
Yuxuan Zhang.
\newblock {Straddling-gates problem in multipartite quantum systems}.
\newblock \emph{Phys. Rev. A}, 105\penalty0 (6):\penalty0 062430, 2022.
\newblock \doi{10.1103/PhysRevA.105.062430}.

\bibitem[Eisert(2021)]{Eisert:2021}
J.~Eisert.
\newblock {Entangling Power and Quantum Circuit Complexity}.
\newblock \emph{Phys. Rev. Lett.}, 127\penalty0 (2):\penalty0 020501, 2021.
\newblock \doi{10.1103/PhysRevLett.127.020501}.

\bibitem[Brown and Susskind(2018)]{Brown:2017jil}
Adam~R. Brown and Leonard Susskind.
\newblock {Second law of quantum complexity}.
\newblock \emph{Phys. Rev. D}, 97\penalty0 (8):\penalty0 086015, 2018.
\newblock \doi{10.1103/PhysRevD.97.086015}.

\bibitem[Brown et~al.(2017)Brown, Susskind, and Zhao]{Brown:2016wib}
Adam~R. Brown, Leonard Susskind, and Ying Zhao.
\newblock {Quantum Complexity and Negative Curvature}.
\newblock \emph{Phys. Rev. D}, 95\penalty0 (4):\penalty0 045010, 2017.
\newblock \doi{10.1103/PhysRevD.95.045010}.

\bibitem[Craps et~al.(2024{\natexlab{b}})Craps, Evnin, and Pascuzzi]{Craps:2023ivc}
Ben Craps, Oleg Evnin, and Gabriele Pascuzzi.
\newblock {A Relation between Krylov and Nielsen Complexity}.
\newblock \emph{Phys. Rev. Lett.}, 132\penalty0 (16):\penalty0 160402, 2024{\natexlab{b}}.
\newblock \doi{10.1103/PhysRevLett.132.160402}.

\bibitem[Wang et~al.(2015)Wang, Allegra, Jacobs, Lloyd, Lupo, and Mohseni]{Wang_2015}
Xiaoting Wang, Michele Allegra, Kurt Jacobs, Seth Lloyd, Cosmo Lupo, and Masoud Mohseni.
\newblock Quantum brachistochrone curves as geodesics: Obtaining accurate minimum-time protocols for the control of quantum systems.
\newblock \emph{Physical Review Letters}, 114\penalty0 (17), apr 2015.
\newblock \doi{10.1103/physrevlett.114.170501}.
\newblock URL \url{https://doi.org/10.1103%2Fphysrevlett.114.170501}.

\bibitem[Arnold(1966)]{AIF_1966__16_1_319_0}
Vladimir Arnold.
\newblock Sur la g\'eom\'etrie diff\'erentielle des groupes de {Lie} de dimension infinie et ses applications \`a l'hydrodynamique des fluides parfaits.
\newblock \emph{Annales de l'Institut Fourier}, 16\penalty0 (1):\penalty0 319--361, 1966.
\newblock \doi{10.5802/aif.233}.
\newblock URL \url{http://www.numdam.org/articles/10.5802/aif.233/}.

\bibitem[Vidal(2000)]{Vidal:1998re}
Guifre Vidal.
\newblock {On the characterization of entanglement}.
\newblock \emph{J. Mod. Opt.}, 47:\penalty0 355, 2000.
\newblock \doi{10.1080/09500340008244048}.

\bibitem[Dur et~al.(2001)Dur, Vidal, Cirac, Linden, and Popescu]{Cirac:2001}
W.~Dur, G.~Vidal, J.~I. Cirac, N.~Linden, and S.~Popescu.
\newblock {Entanglement capabilities of nonlocal Hamiltonians}.
\newblock \emph{Phys. Rev. Lett.}, 87:\penalty0 137901, 2001.
\newblock \doi{10.1103/PhysRevLett.87.137901}.

\bibitem[Wang et~al.(2020)Wang, Hu, Sanders, and Kais]{Wang:2020}
Yuchen Wang, Zixuan Hu, Barry~C. Sanders, and Sabre Kais.
\newblock Qudits and high-dimensional quantum computing.
\newblock \emph{Frontiers in Physics}, 8, nov 2020.
\newblock \doi{10.3389/fphy.2020.589504}.
\newblock URL \url{https://doi.org/10.3389%2Ffphy.2020.589504}.

\bibitem[Guo et~al.(2018)Guo, Hernandez, Myers, and Ruan]{Guo:2018kzl}
Minyong Guo, Juan Hernandez, Robert~C. Myers, and Shan-Ming Ruan.
\newblock {Circuit Complexity for Coherent States}.
\newblock \emph{JHEP}, 10:\penalty0 011, 2018.
\newblock \doi{10.1007/JHEP10(2018)011}.

\bibitem[Bravyi(2007)]{bravyi:2007}
Sergey Bravyi.
\newblock Upper bounds on entangling rates of bipartite hamiltonians.
\newblock \emph{Phys. Rev. A}, 76:\penalty0 052319, Nov 2007.
\newblock \doi{10.1103/PhysRevA.76.052319}.
\newblock URL \url{https://link.aps.org/doi/10.1103/PhysRevA.76.052319}.

\bibitem[Acoleyen et~al.(2013)Acoleyen, Mariën, and Verstraete]{van:2013}
Karel~Van Acoleyen, Michaël Mariën, and Frank Verstraete.
\newblock Entanglement rates and area laws.
\newblock \emph{Physical Review Letters}, 111\penalty0 (17), oct 2013.
\newblock \doi{10.1103/physrevlett.111.170501}.
\newblock URL \url{https://doi.org/10.1103%2Fphysrevlett.111.170501}.

\bibitem[Mari{\"e}n et~al.(2016)Mari{\"e}n, Audenaert, Van~Acoleyen, and Verstraete]{Verstraete:2016}
Micha{\"e}l Mari{\"e}n, Koenraad M.~R. Audenaert, Karel Van~Acoleyen, and Frank Verstraete.
\newblock Entanglement rates and the stability of the area law for the entanglement entropy.
\newblock \emph{Communications in Mathematical Physics}, 346\penalty0 (1):\penalty0 35--73, 2016.
\newblock \doi{10.1007/s00220-016-2709-5}.

\bibitem[Wei and Goldbart(2003)]{Chieh:2003}
Tzu-Chieh Wei and Paul~M. Goldbart.
\newblock Geometric measure of entanglement and applications to bipartite and multipartite quantum states.
\newblock \emph{Physical Review A}, 68\penalty0 (4), oct 2003.
\newblock \doi{10.1103/physreva.68.042307}.
\newblock URL \url{https://doi.org/10.1103%2Fphysreva.68.042307}.

\bibitem[Lloyd(1996)]{Lloyd:1996}
Seth Lloyd.
\newblock Universal quantum simulators.
\newblock \emph{Science}, 273\penalty0 (5278):\penalty0 1073--1078, 1996.
\newblock \doi{10.1126/science.273.5278.1073}.
\newblock URL \url{https://www.science.org/doi/abs/10.1126/science.273.5278.1073}.

\bibitem[Childs et~al.(2021)Childs, Su, Tran, Wiebe, and Zhu]{Childs:2021}
Andrew~M. Childs, Yuan Su, Minh~C. Tran, Nathan Wiebe, and Shuchen Zhu.
\newblock Theory of trotter error with commutator scaling.
\newblock \emph{Physical Review X}, 11\penalty0 (1), feb 2021.
\newblock \doi{10.1103/physrevx.11.011020}.
\newblock URL \url{https://doi.org/10.1103%2Fphysrevx.11.011020}.

\bibitem[Fu et~al.(2018)Fu, Maloney, Marolf, Maxfield, and Wang]{Fu:2018kcp}
Zicao Fu, Alexander Maloney, Donald Marolf, Henry Maxfield, and Zhencheng Wang.
\newblock {Holographic complexity is nonlocal}.
\newblock \emph{JHEP}, 02:\penalty0 072, 2018.
\newblock \doi{10.1007/JHEP02(2018)072}.

\bibitem[Castellani et~al.(1991)Castellani, D'Auria, and Fre]{Castellani:1991et}
L.~Castellani, R.~D'Auria, and P.~Fre.
\newblock \emph{{Supergravity and superstrings: A Geometric perspective. Vol. 1: Mathematical foundations}}.
\newblock 1991.
\newblock \doi{10.1142/0224}.

\bibitem[Skenderis and van Rees(2011)]{Skenderis:2009ju}
Kostas Skenderis and Balt~C. van Rees.
\newblock {Holography and wormholes in 2+1 dimensions}.
\newblock \emph{Commun. Math. Phys.}, 301:\penalty0 583--626, 2011.
\newblock \doi{10.1007/s00220-010-1163-z}.

\bibitem[Hung et~al.(2011)Hung, Myers, Smolkin, and Yale]{Hung:2011nu}
Ling-Yan Hung, Robert~C. Myers, Michael Smolkin, and Alexandre Yale.
\newblock {Holographic Calculations of Renyi Entropy}.
\newblock \emph{JHEP}, 12:\penalty0 047, 2011.
\newblock \doi{10.1007/JHEP12(2011)047}.

\bibitem[Dong(2016)]{Dong:2016}
Xi~Dong.
\newblock {The Gravity Dual of Renyi Entropy}.
\newblock \emph{Nature Commun.}, 7:\penalty0 12472, 2016.
\newblock \doi{10.1038/ncomms12472}.

\bibitem[Abt et~al.(2018)Abt, Erdmenger, Hinrichsen, Melby-Thompson, Meyer, Northe, and Reyes]{Abt:2017pmf}
Raimond Abt, Johanna Erdmenger, Haye Hinrichsen, Charles~M. Melby-Thompson, Ren\'e Meyer, Christian Northe, and Ignacio~A. Reyes.
\newblock {Topological Complexity in AdS$_3$/CFT$_2$}.
\newblock \emph{Fortsch. Phys.}, 66\penalty0 (6):\penalty0 1800034, 2018.
\newblock \doi{10.1002/prop.201800034}.

\bibitem[Abt et~al.(2019)Abt, Erdmenger, Gerbershagen, Melby-Thompson, and Northe]{Abt:2018ywl}
Raimond Abt, Johanna Erdmenger, Marius Gerbershagen, Charles~M. Melby-Thompson, and Christian Northe.
\newblock {Holographic Subregion Complexity from Kinematic Space}.
\newblock \emph{JHEP}, 01:\penalty0 012, 2019.
\newblock \doi{10.1007/JHEP01(2019)012}.

\bibitem[Orus et~al.(2006)Orus, Latorre, Eisert, and Cramer]{Orus:2005dpa}
R.~Orus, J.~I. Latorre, J.~Eisert, and M.~Cramer.
\newblock {Half the entanglement in critical systems is distillable from a single specimen}.
\newblock \emph{Phys. Rev. A}, 73:\penalty0 060303, 2006.
\newblock \doi{10.1103/PhysRevA.73.060303}.

\bibitem[Ag\'on et~al.(2019)Ag\'on, Headrick, and Swingle]{Agon:2018zso}
Cesar~A. Ag\'on, Matthew Headrick, and Brian Swingle.
\newblock {Subsystem Complexity and Holography}.
\newblock \emph{JHEP}, 02:\penalty0 145, 2019.
\newblock \doi{10.1007/JHEP02(2019)145}.

\bibitem[Alishahiha(2015)]{Alishahiha:2015rta}
Mohsen Alishahiha.
\newblock {Holographic Complexity}.
\newblock \emph{Phys. Rev. D}, 92\penalty0 (12):\penalty0 126009, 2015.
\newblock \doi{10.1103/PhysRevD.92.126009}.

\bibitem[Carmi et~al.(2017{\natexlab{b}})Carmi, Myers, and Rath]{Carmi:2016wjl}
Dean Carmi, Robert~C. Myers, and Pratik Rath.
\newblock {Comments on Holographic Complexity}.
\newblock \emph{JHEP}, 03:\penalty0 118, 2017{\natexlab{b}}.
\newblock \doi{10.1007/JHEP03(2017)118}.

\bibitem[Alishahiha et~al.(2019)Alishahiha, Babaei~Velni, and Mohammadi~Mozaffar]{Alishahiha:2018lfv}
Mohsen Alishahiha, Komeil Babaei~Velni, and M.~Reza Mohammadi~Mozaffar.
\newblock {Black hole subregion action and complexity}.
\newblock \emph{Phys. Rev. D}, 99\penalty0 (12):\penalty0 126016, 2019.
\newblock \doi{10.1103/PhysRevD.99.126016}.

\bibitem[C\'aceres et~al.(2019)C\'aceres, Couch, Eccles, and Fischler]{Caceres:2018blh}
Elena C\'aceres, Josiah Couch, Stefan Eccles, and Willy Fischler.
\newblock {Holographic Purification Complexity}.
\newblock \emph{Phys. Rev. D}, 99\penalty0 (8):\penalty0 086016, 2019.
\newblock \doi{10.1103/PhysRevD.99.086016}.

\bibitem[Caceres et~al.(2020)Caceres, Chapman, Couch, Hernandez, Myers, and Ruan]{Caceres:2019pgf}
Elena Caceres, Shira Chapman, Josiah~D. Couch, Juan~P. Hernandez, Robert~C. Myers, and Shan-Ming Ruan.
\newblock {Complexity of Mixed States in QFT and Holography}.
\newblock \emph{JHEP}, 03:\penalty0 012, 2020.
\newblock \doi{10.1007/JHEP03(2020)012}.

\bibitem[Auzzi et~al.(2020)Auzzi, Baiguera, Legramandi, Nardelli, Roy, and Zenoni]{Auzzi:2019vyh}
Roberto Auzzi, Stefano Baiguera, Andrea Legramandi, Giuseppe Nardelli, Pratim Roy, and Nicol\`o Zenoni.
\newblock {On subregion action complexity in AdS$_{3}$ and in the BTZ black hole}.
\newblock \emph{JHEP}, 01:\penalty0 066, 2020.
\newblock \doi{10.1007/JHEP01(2020)066}.

\bibitem[Auzzi et~al.(2019)Auzzi, Nardelli, Schaposnik~Massolo, Tallarita, and Zenoni]{Auzzi:2019mah}
Roberto Auzzi, Giuseppe Nardelli, Fidel~I. Schaposnik~Massolo, Gianni Tallarita, and Nicol\`o Zenoni.
\newblock {On volume subregion complexity in Vaidya spacetime}.
\newblock \emph{JHEP}, 11:\penalty0 098, 2019.
\newblock \doi{10.1007/JHEP11(2019)098}.

\bibitem[Di~Giulio and Tonni(2020)]{DiGiulio:2020hlz}
Giuseppe Di~Giulio and Erik Tonni.
\newblock {Complexity of mixed Gaussian states from Fisher information geometry}.
\newblock \emph{JHEP}, 12:\penalty0 101, 2020.
\newblock \doi{10.1007/JHEP12(2020)101}.

\bibitem[Jiang and Liu(2019)]{Jiang:2018nzg}
Jie Jiang and Xiangjing Liu.
\newblock {Circuit Complexity for Fermionic Thermofield Double states}.
\newblock \emph{Phys. Rev. D}, 99\penalty0 (2):\penalty0 026011, 2019.
\newblock \doi{10.1103/PhysRevD.99.026011}.

\bibitem[Bhatia(2013)]{bhatia2013matrix}
Rajendra Bhatia.
\newblock \emph{Matrix analysis}, volume 169.
\newblock Springer Science \& Business Media, 2013.
\newblock \doi{10.1007/978-1-4612-0653-8}.
\newblock URL \url{https://link.springer.com/book/10.1007/978-1-4612-0653-8}.

\bibitem[Horn and Johnson(1991)]{horn_johnson_1991}
Roger~A. Horn and Charles~R. Johnson.
\newblock \emph{Topics in Matrix Analysis}.
\newblock Cambridge University Press, 1991.
\newblock \doi{10.1017/CBO9780511840371}.

\bibitem[Chen and Wong(1992)]{CHEN1992109}
Ling Chen and Chi~Song Wong.
\newblock Inequalities for singular values and traces.
\newblock \emph{Linear Algebra and its Applications}, 171:\penalty0 109--120, 1992.
\newblock ISSN 0024-3795.
\newblock \doi{https://doi.org/10.1016/0024-3795(92)90253-7}.
\newblock URL \url{https://www.sciencedirect.com/science/article/pii/0024379592902537}.

\bibitem[Calabrese and Lefevre(2008)]{calabrese2008entanglement}
Pasquale Calabrese and Alexandre Lefevre.
\newblock Entanglement spectrum in one-dimensional systems.
\newblock \emph{Physical Review A}, 78\penalty0 (3), sep 2008.
\newblock \doi{10.1103/physreva.78.032329}.
\newblock URL \url{https://doi.org/10.1103%2Fphysreva.78.032329}.

\bibitem[Fubini(1904)]{Fubini}
Guido Fubini.
\newblock {Sulle metriche definite da una forma Hermitiana}.
\newblock \emph{Atti del Reale Istituto Veneto di Scienze, Lettere ed Arti}, 63:\penalty0 502--513, 1904.

\bibitem[Study(1905)]{Study}
Eduard Study.
\newblock {K\"{u}rzeste Wege im komplexen Gebiet}.
\newblock \emph{Mathematische Annalen}, 60:\penalty0 3, 1905.
\newblock \doi{10.1007/BF01457616}.
\newblock URL \url{https://link.springer.com/article/10.1007/BF01457616}.

\end{thebibliography}

\end{document}